\pdfoutput=1

\documentclass[11pt,twoside,a4paper,cmspaper,final,collab]{cms-tdr}
\def\svnVersion{488487P}\def\cmsCernNoTag{CERN-EP-2018-177}\def\cmsCernDate{\today}\def\cmsMessage{Published in the European Physical Journal C as \href{http://dx.doi.org/10.1140/epjc/s10052-019-6620-z}{\doi{10.1140/epjc/s10052-019-6620-z.}}}

\begin{document}\cmsNoteHeader{TOP-17-015}

\hyphenation{had-ron-i-za-tion}
\hyphenation{cal-or-i-me-ter}
\hyphenation{de-vices}
\RCS$HeadURL: svn+ssh://svn.cern.ch/reps/tdr2/papers/TOP-17-015/trunk/TOP-17-015.tex $
\RCS$Id: TOP-17-015.tex 488485 2019-02-07 20:31:33Z efe $
\newlength\cmsFigWidth
\newlength\cmsFigWidthMed
\newlength\cmsFigWidthii\setlength\cmsFigWidthii{0.70\textwidth}
\newlength\cmsFigWidthi\setlength\cmsFigWidthi{0.85\textwidth}
\newlength\cmsFigWidthiv\setlength\cmsFigWidthiv{0.80\textwidth}
\newlength\cmsTabSkip\setlength{\cmsTabSkip}{1ex}
\ifthenelse{\boolean{cms@external}}{\setlength\cmsFigWidth{0.49\textwidth}}{\setlength\cmsFigWidth{0.9\textwidth}}
\ifthenelse{\boolean{cms@external}}{\setlength\cmsFigWidthMed{0.49\textwidth}}{\setlength\cmsFigWidthMed{0.7\textwidth}}
\ifthenelse{\boolean{cms@external}}{\providecommand{\cmsLeft}{top\xspace}}{\providecommand{\cmsLeft}{left\xspace}}
\ifthenelse{\boolean{cms@external}}{\providecommand{\cmsRight}{bottom\xspace}}{\providecommand{\cmsRight}{right\xspace}}
\ifthenelse{\boolean{cms@external}}{\providecommand{\cmsTable}[1]{#1}}{\providecommand{\cmsTable}[1]{\resizebox{\textwidth}{!}{#1}}}
\newcommand{\RIVET}{\textsc{rivet}}
\newcommand{\nch}{\ensuremath{N_\text{ch}}\xspace}
\newcommand{\sumpt}{\ensuremath{\sum\pt}\xspace}
\newcommand{\sumpz}{\ensuremath{\sum p_{z}}\xspace}
\newcommand{\avgpt}{\ensuremath{\overline{\pt}}\xspace}
\newcommand{\avgpz}{\ensuremath{\overline{p_z}}\xspace}
\newcommand{\chpt}{\ensuremath{\abs{\ptvec}}\xspace}
\newcommand{\ptll}{\ensuremath{\pt(\ell\ell)}\xspace}
\newcommand{\ptllvec}{\ensuremath{\ptvec(\ell\ell)}\xspace}
\newcommand{\mll}{\ensuremath{m(\ell\ell)}}
\newcommand{\pwpy}{\textsc{Pw+Py8}}
\newcommand{\pwhwpp}{\textsc{Pw+Hw++}}
\newcommand{\pwhw}{\textsc{Pw+Hw7}}
\newcommand{\amcpy}{\textsc{MG5}\_a\textsc{MC}}
\cmsNoteHeader{TOP-17-015}
\title{Study of the underlying event in top quark pair production in $\Pp\Pp$ collisions at 13\TeV}

\date{\today}
\abstract{
Measurements of normalized differential cross sections
as functions of the multiplicity and kinematic variables of
charged-particle tracks from the underlying event
in top quark and antiquark pair production are presented.
The measurements are performed in proton-proton collisions at a center-of-mass energy of 13\TeV,
and are based on data collected by the CMS experiment at the LHC in 2016
corresponding to an integrated luminosity of 35.9\fbinv.
Events containing one electron, one muon, and two jets from the hadronization and fragmentation of \cPqb\ quarks are used.
These measurements characterize, for the first time,
properties of the underlying event in top quark pair production
and show no deviation from the universality hypothesis
at energy scales typically above twice the top quark mass.
}

\hypersetup{
pdfauthor={CMS Collaboration},
pdftitle={Study of the underlying event in top quark pair production in pp collisions at 13 TeV},
pdfsubject={CMS},
pdfkeywords={top quark, underlying event}}

\maketitle
\section{Introduction}
\label{sec:introduction}
At the LHC, top quark and antiquark pairs (\ttbar) are
dominantly produced
in the scattering of the proton constituents via quantum chromodynamics (QCD)
at an energy scale ($Q$) of about
two times the \cPqt{} quark mass ($m_\cPqt$).
The properties of the \cPqt{} quark can be studied directly from its decay
products, as it decays before hadronizing.
Mediated by the electroweak interaction, the \cPqt{} quark decay yields a {\PW} boson and a quark,
the latter carrying the QCD color charge of the mother particle.
Given the large branching fraction for the decay into a bottom quark,
$\mathcal{B}(\cPqt\to\PW\cPqb)=0.957\pm0.034$~\cite{Patrignani:2016xqp},
in this analysis we assume that each \cPqt{} or \cPaqt{} quark yields
a corresponding bottom (\cPqb) or antibottom (\cPaqb) quark in its decay.
Other quarks may also be produced, in a color-singlet state, if a $\PW\to\qqbar^\prime$ decay occurs.
Being colored, these quarks will fragment and hadronize giving rise to an experimental signature with jets.
Thus, when performing precision measurements of \cPqt{} quark properties at hadron colliders,
an accurate description of the fragmentation and hadronization
of the quarks from the hard scatter process as well as of the ``underlying event'' (UE),
defined below, is essential.
First studies of the fragmentation and hadronization of the {\cPqb} quarks in \ttbar events
have been reported in Ref.~\cite{Khachatryan:2016wqo,Aad:2013fba}.
In this paper, we present the first measurement of the properties of the UE in \ttbar events
at a scale $Q\geq 2m_\cPqt$.

The UE is defined as any hadronic activity that cannot be attributed to the particles stemming from the hard scatter, and in this case from \ttbar decays.
Because of energy-momentum conservation, the UE constitutes the recoil against the \ttbar system.
In this study, the hadronization products of initial- and final-state radiation (ISR and FSR)
that cannot be associated to the particles from the \ttbar decays are
probed as part of the UE, even if they can be partially modeled by perturbative QCD.
The main contribution to the UE comes from the color exchanges between the beam particles
and is modeled in terms of multiparton interactions (MPI),
color reconnection (CR), and beam-beam remnants (BBR),
whose model parameters can be tuned to minimum bias and Drell--Yan (DY) data.

The study of the UE in \ttbar events provides a direct test of its universality
at higher energy scales than those probed in minimum bias or DY events.
This is relevant as a direct probe of CR,
which is needed to confine the initial QCD color charge of the {\cPqt} quark into color-neutral states.
The CR mainly occurs between one of the products of the fragmentation of the {\cPqb}
quark from the \cPqt{} quark decay and the proton remnants.
This is expected to induce an ambiguity in the origin of some of the final states present in a bottom quark jet~\cite{Sjostrand:2013cya,Argyropoulos:2014zoa,Corcella:2015kth}.
The impact of these ambiguities in the measurement of \cPqt{} quark properties
is evaluated through phenomenological models that need to be tuned to the data.
Recent examples of the impact that different model parameters have on $m_\cPqt$
can be found in Ref.~\cite{Corcella:2017rpt,Ravasio:2018lzi}.

The analysis is performed using final states where both of the {\PW} bosons decay to leptons,
yielding one electron and one muon with opposite charge sign, and the corresponding neutrinos.
In addition, two {\cPqb} jets are required in the selection,
as expected from the $\ttbar\to(\Pe\nu\cPqb)(\mu\nu\cPqb)$ decay.
This final state is chosen because of its expected high purity
and because the products of the hard process can be distinguished with high efficiency
and small contamination from objects not associated with \cPqt{} quark decays,
\eg, jets from ISR.

After discussing the experimental setup in Section~\ref{sec:thecmsdetector},
and the signal and background modeling in Section~\ref{sec:dataandsimulatedsamples},
we present the strategy employed to select the events
in Section~\ref{sec:evsel} and to measure the UE contribution in each selected event in Section~\ref{sec:uechar}.
The measurements are corrected to a particle-level definition
using the method described in Section~\ref{sec:correctionstoparticlelevel}
and the associated systematic uncertainties are discussed
in Section~\ref{sec:systematic uncertainties}.
Finally, in Section~\ref{sec:results},
the results are discussed and compared to predictions from different Monte Carlo (MC) simulations.
The measurements are summarized in Section~\ref{sec:summary}.

\section{The CMS detector}
\label{sec:thecmsdetector}
The central feature of the CMS apparatus is a superconducting solenoid of 6\unit{m} internal diameter, providing a
magnetic field of 3.8\unit{T} parallel to the beam direction.

Within the solenoid volume are a silicon pixel and strip tracker,
a lead tungstate crystal electromagnetic calorimeter (ECAL),
and a brass and scintillator hadron calorimeter (HCAL), each composed of a barrel and two endcap sections.
A preshower detector, consisting of two planes of silicon sensors interleaved with about three radiation lengths of lead,
is located in front of the endcap regions of the ECAL.
Hadron forward calorimeters, using steel as an absorber and quartz fibers as the sensitive material,
extend the pseudorapidity coverage provided by the barrel and endcap detectors from $\abs{\eta} = 3.0$ to 5.2. Muons are detected in the window $\abs{\eta}<2.4$ in gas-ionization detectors embedded in the steel flux-return yoke outside the solenoid.

Charged-particle trajectories with $\abs{\eta}<2.5$ are measured by the tracker system.
The particle-flow algorithm~\cite{Sirunyan:2017ulk} is used to reconstruct and identify individual particles in an event, with an optimized combination of information from the various elements of the CMS detector. The energy of the photons is directly obtained from the ECAL measurement, corrected for zero-suppression effects. The energy of the electrons is determined from a combination of the electron momentum at the primary interaction vertex as determined by the tracker, the energy of the corresponding ECAL cluster, and the energy sum of all bremsstrahlung photons spatially compatible with originating from the electron track. The energy of the muons is obtained from the curvature of the corresponding track. The energy of charged hadrons is determined from a combination of their momentum measured in the tracker and the matching ECAL and HCAL energy deposits, corrected for zero-suppression effects and for the response function of the calorimeters to hadronic showers. Finally, the energy of neutral hadrons is obtained from the corresponding corrected ECAL and HCAL energies.

Events of interest are selected using a two-tiered trigger system~\cite{Khachatryan:2016bia}.
The first level, composed of custom hardware processors, uses information from the calorimeters
and muon detectors to select events at a rate of around 100\unit{kHz} within a time interval of less than 4\mus.
The second level, known as the high-level trigger, consists of a farm of processors
running a version of the full event reconstruction software optimized for fast processing,
and reduces the event rate to around 1\unit{kHz} before data storage.

A more detailed description of the CMS detector,
together with a definition of the coordinate system used and the  relevant kinematic variables,
can be found in Ref.~\cite{Chatrchyan:2008zzk}.

\section{Signal and background modeling}
\label{sec:dataandsimulatedsamples}
This analysis is based on proton-proton (\Pp\Pp) collision data at a center-of-mass energy $\sqrt{s}=13\TeV$,
collected by the CMS detector in 2016 and corresponds to an integrated luminosity of 35.9$\fbinv$~\cite{CMS-PAS-LUM-17-001}.

{\tolerance=8000
The \ttbar process is simulated with the \POWHEG (v2) generator in the heavy quark production (hvq) mode~\cite{powhegv21,powhegv22,powhegv23}.
The NNPDF3.0 next-to-leading-order (NLO) parton distribution functions (PDFs)
with the strong coupling parameter $\alpS=0.118$ at the \cPZ{} boson mass scale ($M_\cPZ$)~\cite{Ball:2014uwa}
are utilized in the matrix-element (ME) calculation.
The renormalization and factorization scales, $\mu_\mathrm{R}$ and $\mu_\mathrm{F}$,
are set
to $\mT=\sqrt{\smash[b]{m_\cPqt^2+\pt^2}}$,
where $m_\cPqt=172.5\GeV$
and \pt is the transverse momentum in the \ttbar rest frame.
Parton showering is simulated using {\PYTHIA}8 (v8.219)~\cite{Sjostrand:2014zea}
and the CUETP8M2T4 UE tune \cite{CMS-PAS-TOP-16-021}.
The CUETP8M2T4 tune is based on the CUETP8M1 tune~\cite{Khachatryan:2015pea}
but uses a lower value of $\alpS^\text{ISR}(M_\cPZ)=0.1108$ in the parton shower (PS);
this value leads to a better description of jet multiplicities in \ttbar events at $\sqrt{s}=8\TeV$~\cite{Khachatryan:2015mva}.
The leading-order (LO) version of the same NNPDF3.0 is used in the PS and MPI simulation in the CUETP8M2T4 tune.
The cross section used for the \ttbar simulation is
832$^{+20}_{-29}~(\text{scale})\pm35(\text{PDF}+\alpS)$\unit{pb},
computed at the next-to-next-to-leading-order (NNLO) plus
next-to-next-to-leading-logarithmic accuracy ~\cite{CZAKON20142930}.
\par}

Throughout this paper, data are compared to the predictions of different generator settings for the \ttbar process.
Table~\ref{tab:mcsetups} summarizes the main characteristics of the setups
and abbreviations used in the paper.
Among other UE properties, CR and MPI are modeled differently in the alternative setups considered,
hence the interest in comparing them to the data.
Three different signal ME generators are used: \POWHEG,
\MGvATNLO{} (v2.2.2) with the FxFx merging scheme~\cite{Alwall:2014hca,Frederix:2012ps} for jets from the ME calculations and PS,
and \SHERPA (v2.2.4)~\cite{Gleisberg:2008ta}.
The latter is used in combination with \textsc{OpenLoops} (v1.3.1)~\cite{Cascioli:2011va},
and with the CS parton shower based on the Catani--Seymour dipole subtraction scheme~\cite{Schumann:2007mg}.
In addition, two different \HERWIG PS versions are used and interfaced with \POWHEG:
\HERWIGpp~\cite{Bahr:2008pv} with the EE5C UE tune~\cite{Seymour:2013qka} and the
CTEQ6 (L1)~\cite{Pumplin:2002vw} PDF set,
and \HERWIG7~\cite{Bahr:2008pv,Bellm:2015jjp} with its default tune
and the MMHT2014 (LO)~\cite{Harland-Lang:2014zoa} PDF set.

\begin{table*}[ht]
\centering
\topcaption{MC simulation settings used for the comparisons with the differential cross section measurements of the UE.
The table lists the main characteristics and values used for the most relevant parameters of the generators.
The row labeled ``Setup designation'' shows the definitions of the  abbreviations used throughout this paper.
}
\cmsTable{
\begin{tabular}{lccc}
Event generator                    & \POWHEG (v2)      &  \MGvATNLO~(v2.2.2)                      & \SHERPA(v2.2.4) \\
\hline
\multicolumn{4}{l}{\textit{ Matrix element characteristics}} \\[\cmsTabSkip]
Mode                             &  hvq            & FxFx Merging             & \textsc{OpenLoops} \\
Scales ($\mu_\mathrm{R},\mu_\mathrm{F}$)     & \mT  & $\sum_{\cPqt,\cPaqt}\mT/2$ & METS~+~QCD\\
$\alpS(M_\cPZ)$                       & 0.118           & 0.118                    & 0.118 \\
PDF                              & NNPDF3.0 NLO    & NNPDF3.0 NLO             & NNPDF3.0 NNLO \\
Accuracy                    & \ttbar [NLO]    & \ttbar~+~0,1,2 jets [NLO]  & \ttbar [NLO] \\
                                   & 1 jet [LO]      & 3 jets [LO]              & \\[\cmsTabSkip]
\multicolumn{4}{l}{\textit{ Parton shower}} \\[\cmsTabSkip]
Setup designation                    & \pwpy          & \amcpy               & \SHERPA \\
~~~~PS                                 & \multicolumn{2}{c}{\PYTHIA (v8.219)}    & CS   \\
~~~~Tune                               & \multicolumn{2}{c}{CUETP8M2T4}             & default \\
~~~~PDF                                & \multicolumn{2}{c}{NNPDF2.3 LO}     & NNPDF3.0 NNLO \\
~~~~($\alpS^\text{ISR}(M_\cPZ),\alpS^\text{FSR}(M_\cPZ)$)  & \multicolumn{2}{c}{(0.1108, 0.1365)} & (0.118, 0.118) \\
~~~~ME Corrections                     & \multicolumn{2}{c}{on}              & \NA \\[\cmsTabSkip]
Setup designation                      & \pwhwpp          &      \pwhw           & \\
~~~~PS                                 &  \HERWIGpp       &    \HERWIG7            & \\
~~~~Tune                               &  EE5C            &      Default          & \\
~~~~PDF                                &  CTEQ6 (L1)      &      MMHT2014 LO         & \\
~~~~($\alpS^\text{ISR}(M_\cPZ),\alpS^\text{FSR}(M_\cPZ)$)  & (0.1262, 0.1262)  &     (0.1262, 0.1262)           & \\
~~~~ME Corrections                     &  off             &      on          & \\
\end{tabular}
}
\label{tab:mcsetups}
\end{table*}

{\tolerance=8000
Additional variations of the {\pwpy} sample are used to illustrate the sensitivity of the measurements
to different parameters of the UE model.
A supplementary table, presented in the appendix,
details the parameters that have been changed with respect to the CUETP8M2T4 tune in these additional variations.
The variations include extreme models that highlight separately the contributions of MPI and CR to the UE,
fine-grained variations of different CR models~\cite{Argyropoulos:2014zoa,Christiansen:2015yqa},
an alternative MPI model based on the ``Rope hadronization'' framework
describing Lund color strings overlapping in the same area~\cite{Bierlich:2014xba,Bierlich:2015rha},
variations of the choice of $\alpS(M_\cPZ)$ in the parton shower,
and a variation of the values that constitute the
CUETP8M2T4 tune, according to their uncertainties.
\par}

{\tolerance=8000
Background processes are simulated with several generators.
The \PW\cPZ{}, \PW+jets, and $\cPZ\cPZ\to 2\ell 2q$
(where $\ell$ denotes any of the charged leptons e/$\mu$/$\tau$)
processes are simulated at NLO, using {\MGvATNLO} with the FxFx merging.
Drell--Yan production,
with dilepton invariant mass, \mll, greater than 50\GeV,
is simulated at LO with {\MGvATNLO}
using the so-called MLM matching scheme~\cite{Alwall:2007fs} for jet merging.
The \POWHEG (v2) program is furthermore used to simulate the \PW\PW{}, and $\cPZ\cPZ\to 2\ell 2\nu$ processes~\cite{Melia:2011tj,Nason:2013ydw}, while \POWHEG (v1) is used to simulate the \cPqt\PW{} process~\cite{Re:2010bp}.
The single top quark $t$-channel background is simulated at NLO using \POWHEG (v2) and \textsc{MadSpin} contained in \MGvATNLO{} (v2.2.2)~\cite{Alioli:2009je,Artoisenet:2012st}.
The residual \cPqt\cPaqt+V backgrounds, where $\mathrm{V}=\PW{}$ or \cPZ{}, are generated at NLO using \MGvATNLO.
The cross sections of the DY and \PW+jets processes are normalized to the NNLO prediction, computed using {\FEWZ} (v3.1.b2)~\cite{fewz},
and single top quark processes are normalized to the approximate NNLO prediction~\cite{Kidonakis:2013zqa}.
Processes containing two vector bosons (hereafter referred to as dibosons)
are normalized to the NLO predictions computed with \MGvATNLO,
with the exception of the \PW\PW{} process, for which the NNLO prediction~\cite{Gehrmann:2014fva} is used.
\par}

{\tolerance=8000
All generated events are processed through the \GEANTfour-based~\cite{Agostinelli:2002hh,Allison:2006ve,ALLISON2016186}
CMS detector simulation and the standard CMS event reconstruction.
Additional $\Pp\Pp$ collisions per bunch crossing (pileup) are included in the simulations.
These simulate the effect of
pileup in the events,
with the same multiplicity distribution as that observed in data,
\ie, about 23 simultaneous interactions, on average, per bunch crossing.
\par}

\section{Event reconstruction and selection}
\label{sec:evsel}
The selection targets events in which each {\PW} boson decays to a charged lepton and a neutrino.
Data are selected online with single-lepton and dilepton triggers.
The particle flow (PF) algorithm~\cite{Sirunyan:2017ulk} is used for the reconstruction of final-state objects.
The offline event selection is similar to the one described in Ref.~\cite{Khachatryan:2016kzg}.
At least one PF charged lepton candidate with $\pt>25\GeV$ and another one with $\pt>20\GeV$,
both having  $\abs{\eta}<2.5$, are required.
The two leptons must have opposite charges and an invariant mass $m(\ell^\pm\ell^\mp)>12\GeV$.
When extra leptons are present in the event, the dilepton candidate is built from the highest \pt leptons in the event.
Events with $\Pe^\pm\mu^\mp$ in the final state are used for the main analysis,
while $\Pe^\pm\Pe^\mp$ and $\mu^\pm\mu^\mp$ events are used
to derive the normalization of the DY background.
The simulated events are corrected for the differences between data and simulation
in the efficiencies of the trigger, lepton identification, and lepton isolation criteria.
The corrections are derived with
$\cPZ\to\Pe^\pm\Pe^\mp$ and $\cPZ\to\mu^\pm\mu^\mp$
events using the ``tag-and-probe" method~\cite{Khachatryan:2010xn}
and are parameterized as functions of the \pt and $\eta$ of the leptons.

Jets are clustered using the anti-\kt jet finding algorithm~\cite{Cacciari:2008gp,Cacciari:2011ma}
with a distance parameter of 0.4 and all the reconstructed PF candidates in the event.
The charged hadron subtraction algorithm is used to mitigate the contribution from pileup to the jets~\cite{Khachatryan:2016kdb}.
At least two jets with $\pt>30\GeV$, $\abs{\eta}<2.5$
and identified by a \cPqb-tagging algorithm are required.
The \cPqb-tagging is based on a ``combined secondary vertex'' algorithm ~\cite{Sirunyan:2017ezt}
characterized by an efficiency of about 66\%, corresponding to misidentification probabilities for light quark and \cPqc{} quark jets
of 1.5 and 18\%, respectively.
A \pt-dependent scale factor is applied to the simulations in order to reproduce the efficiency of this algorithm,
as measured in data.

{\tolerance=8000
The reconstructed vertex with the largest value of summed physics-object $\pt^2$ is taken to be the primary $\Pp\Pp$ interaction vertex.
The physics objects are the jets, clustered using the jet finding algorithm~\cite{Cacciari:2008gp,Cacciari:2011ma}
with the tracks assigned to the vertex as inputs,
and the associated missing transverse momentum, \ptmiss, taken as the negative vector sum of the \pt of those jets.
The latter is defined as the magnitude of the negative vector sum of the momenta of all reconstructed PF candidates in an event,
projected onto the plane perpendicular to the direction of the proton beams.
\par}

All backgrounds are estimated from simulation, with the exception of the DY background normalization.
The latter is estimated making use of the so-called $R_\text{out/in}$ method~\cite{Khachatryan:2010ez},
in which events with same-flavor leptons are used to normalize the yield of $\Pe\mu$ pairs from DY production
of $\tau$ lepton pairs.
The normalization of the simulation is estimated from the number of events in the data within a 15\GeV window around the {\PZ} boson mass \cite{Khachatryan:2010ez}.
For $\Pe\mu$ events,
we use the geometric mean of the scale factors determined for $\Pe\Pe$
and $\mu\mu$ events.
With respect to the simulated predictions,
a scale factor $1.3\pm0.4$ is obtained from this method, with statistical and systematic
uncertainties added in quadrature.
The systematic uncertainty is estimated from the differences found in the scale factor for
events with 0 or 1 \cPqb-tagged jets, in the same-flavor channels.

We select a total of 52\,645 $\Pe\mu$ events with an expected purity of 96\%.
The data agree with the expected yields within 2.2\%,
a value smaller than the uncertainty
in the integrated luminosity alone, 2.5\%~\cite{CMS-PAS-LUM-17-001}.
The \cPqt\PW{} events are expected to constitute 90\% of the total background.

In the simulation, the selection is mimicked at the particle level with the techniques described in Ref.~\cite{Collaboration:2267573}.
Jets and leptons are defined at the particle level with the same conventions as adopted by the {\RIVET} framework~\cite{Buckley:2013}.
The procedure ensures that the selections and definitions of the objects at particle level are consistent with those used in the {\RIVET} routines.
A brief description of the particle-level definitions follows:

\begin{itemize}

\item prompt charged leptons (i.e., not produced as a result of hadron decays)
      are reconstructed as ``dressed'' leptons with nearby photon recombination
      using the anti-\kt algorithm with a distance parameter of 0.1;

\item jets are clustered with the anti-\kt algorithm with a distance parameter of 0.4
      using all particles remaining after removing both the leptons from the hard process and the neutrinos;

\item the flavor of a jet is identified by including {\PB} hadrons in the clustering.

\end{itemize}

{\tolerance=8000
Using these definitions, the fiducial region of this analysis is specified by the same
requirements that are applied offline (reconstruction level)  for leptons and jets.
Simulated events are categorized as fulfilling only the reconstruction-based, only the particle-based, or both selection requirements.
If a simulated event passes only the reconstruction-level selection, it is considered in the ``misidentified signal'' category,
\ie, it does not contribute to the fiducial region defined in the analysis and thus is considered as a background process.
In the majority of the bins of each of the distributions analyzed,
the fraction of signal events passing both the reconstruction- and particle-level selections is estimated to be about 80\%,
while the fraction of misidentified signal events is estimated to be less than 10\%.
\par}

\section{Characterization of the underlying event}
\label{sec:uechar}
In order to isolate the UE activity in data,
the contribution from both pileup and the hard process itself must be identified and excluded from the analysis.
The contamination from pileup events is expected to yield soft particles in time with the hard process,
as well as tails in the energy deposits from out-of-time interactions.
The contamination from the hard process is expected to be associated with the
two charged leptons and two {\cPqb} jets originating from the \ttbar decay chain.

{\tolerance=8000
In order to minimize the contribution from these sources, we use the properties of the reconstructed PF candidates in each event.
The track associated to the charged PF candidate is required to be compatible with originating from the primary vertex.
This condition reduces to a negligible amount the contamination from pileup in the charged particle collection.
A simple association by proximity in $z$ with respect to the primary vertex of the event
is expected to yield a pileup-robust, high-purity selection.
For the purpose of this analysis all charged PF candidates are required to satisfy the following requirements:
\par}

\begin{itemize}
\item $\pt>900\MeV$ and $\abs{\eta}<2.1$;
\item the associated track needs to be either used in the fit of the primary vertex
      or to be closer to it in $z$ than with respect to other reconstructed vertices in the event.
\end{itemize}

\begin{figure}[!htb]
\centering
\includegraphics[width=0.5\textwidth]{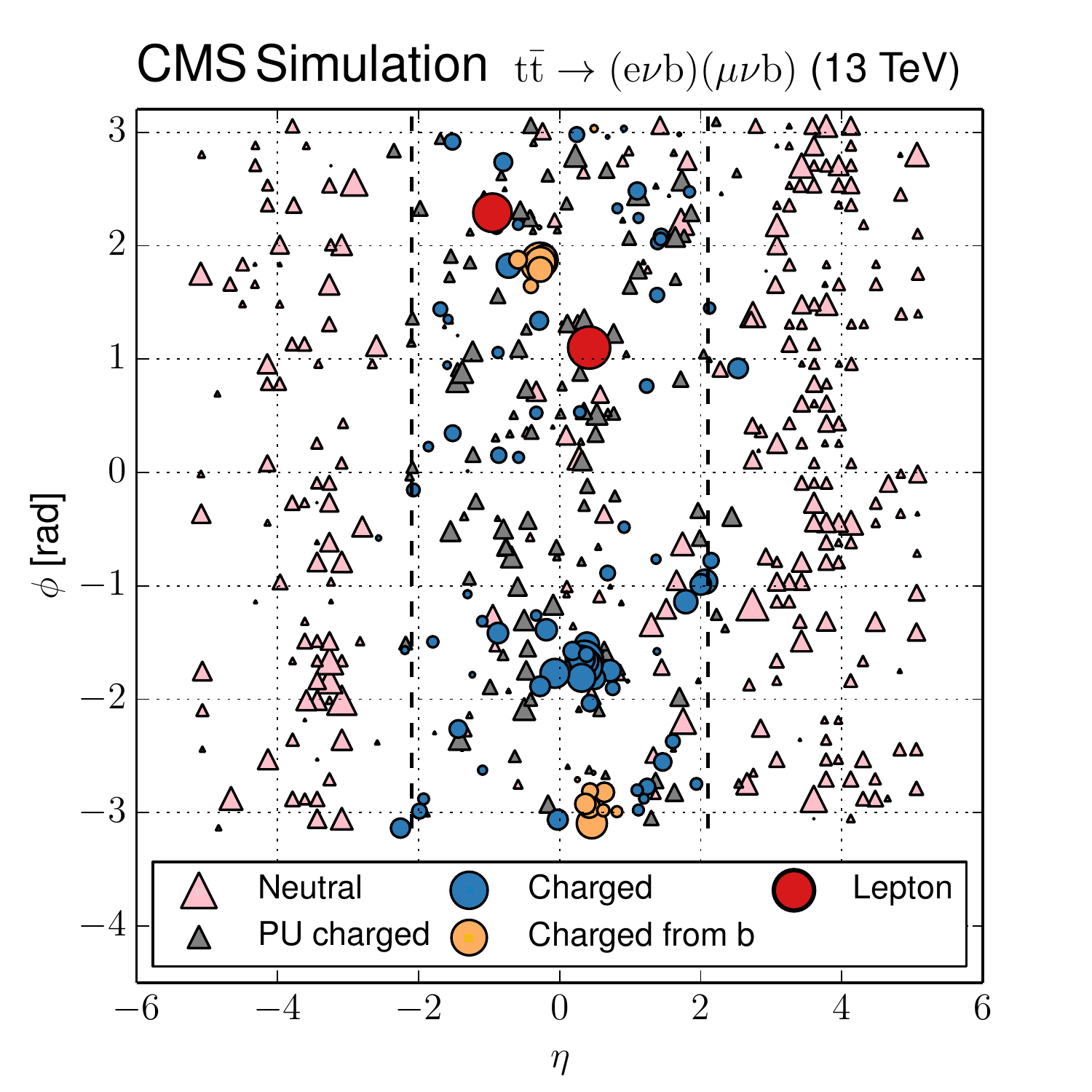}
\caption{
Distribution of all PF candidates reconstructed in a {\pwpy} simulated \ttbar event in the $\eta$--$\phi$ plane.
Only particles with $\pt>900\MeV$ are shown, with a marker whose area is proportional to the particle \pt.
The fiducial region in $\eta$ is represented by the dashed lines.
}
\label{fig:uesel}
\end{figure}

{\tolerance=8000
After performing the selection of the charged PF candidates we check which ones have been used in the clustering of the two \cPqb-tagged jets and which ones match the two charged lepton candidates within a $\Delta R=\sqrt{\smash[b]{(\Delta\eta)^2+(\Delta\phi)^2}}=0.05$ cone,
where $\phi$ is the azimuthal angle in radians.
All PF candidates failing the kinematic requirements, being matched to another primary vertex in the event,
or being matched to the charged leptons and \cPqb-tagged jets, are removed from the analysis.
The UE analysis proceeds by using the remaining charged PF candidates.
Figure~\ref{fig:uesel} shows, in a simulated \ttbar event,
the contribution from charged and neutral PF candidates, the charged component of the pileup, and the hard process.
The charged PF candidates that are used in the study of the UE are represented after applying the selection described above.
\par}

Various characteristics, such as the multiplicity of the selected charged particles, the flux of momentum, and the topology or shape of the event
have different sensitivity to the modeling of the recoil, the contribution from MPI and CR, and other parameters.

The first set of observables chosen in this analysis is related to the multiplicity and momentum flux in the event:
\begin{itemize}
\item charged-particle multiplicity: \nch;
\item magnitude of the \pt of the charged particle recoil system: $\abs{\ptvec}=\abs{{\sum_{i=1}^{N_\text{ch}}} \vec{p}_{\mathrm{T},i}}$;
\item scalar sum of the \pt (or $p_z$) of charged particles: $\sum p_\mathrm{k}={\sum_{i=1}^{N_\text{ch}}} \abs{\vec{p}_{\mathrm{k},i}}$, where $\mathrm{k=T}$ or $z$;
\item average \pt (or $p_z$) per charged particle: computed from the ratio between the scalar sum and the charged multiplicity: \avgpt (or \avgpz).
\end{itemize}
The second set of observables characterizes the UE shape and it is computed from the so-called linearized sphericity tensor~\cite{Parisi:1978eg,Donoghue:1979vi}:
\begin{equation}
S^{\mu\nu}= {\sum_{i=1}^{N_\text{ch}}} p_i^\mu p_i^\nu / \abs{p_i} \Big/~ {\sum_{i=1}^{N_\text{ch}}} \abs{p_i},
\label{eq:sphertens}
\end{equation}
where the $i$ index runs over the particles associated with the UE, as for the previous variables,
and the $\mu$ and $\nu$ indices refer to one of the $(x,y,z)$ components of the momentum of the particles.
The eigenvalues ($\lambda_i$) of $S^{\mu\nu}$ are in decreasing order,
\ie, with $\lambda_1$ the largest one,
and are used to compute the following observables~\cite{Ellis:1980wv}:

\begin{itemize}
\item{{Aplanarity:}} $A=\frac{3}{2}\lambda_3$ measures the $\pt$  component out of the event plane, defined by the two leading eigenvectors.
                        Isotropic (planar) events are expected to have $A=1/2\,(0)$.
\item{{Sphericity:}} $S=\frac{3}{2}(\lambda_2+\lambda_3)$ measures the $\pt^2$ with respect to the axis of the event.
                        An isotropic (dijet) event is expected to have $S=1\,(0)$.
\item ${C}=3(\lambda_1\lambda_2+\lambda_1\lambda_3+\lambda_2\lambda_3)$ identifies 3 jet events (tends to be 0 for dijet events).
\item ${D}=27\lambda_1\lambda_2\lambda_3$ identifies 4 jet events (tends to be 0 otherwise).
\end{itemize}

Further insight can be gained by studying the evolution of the two sets of observables
in different categories of the \ttbar system kinematic quantities.
The categories chosen below are
sensitive to the recoil or the scale of the energy of the hard process,
and are expected to be reconstructed with very good resolution.
Additionally, these variables minimize the effect of migration of events
between categories due to resolution effects.

\begin{figure*}[!htp]
\centering
\includegraphics[width=0.45\textwidth]{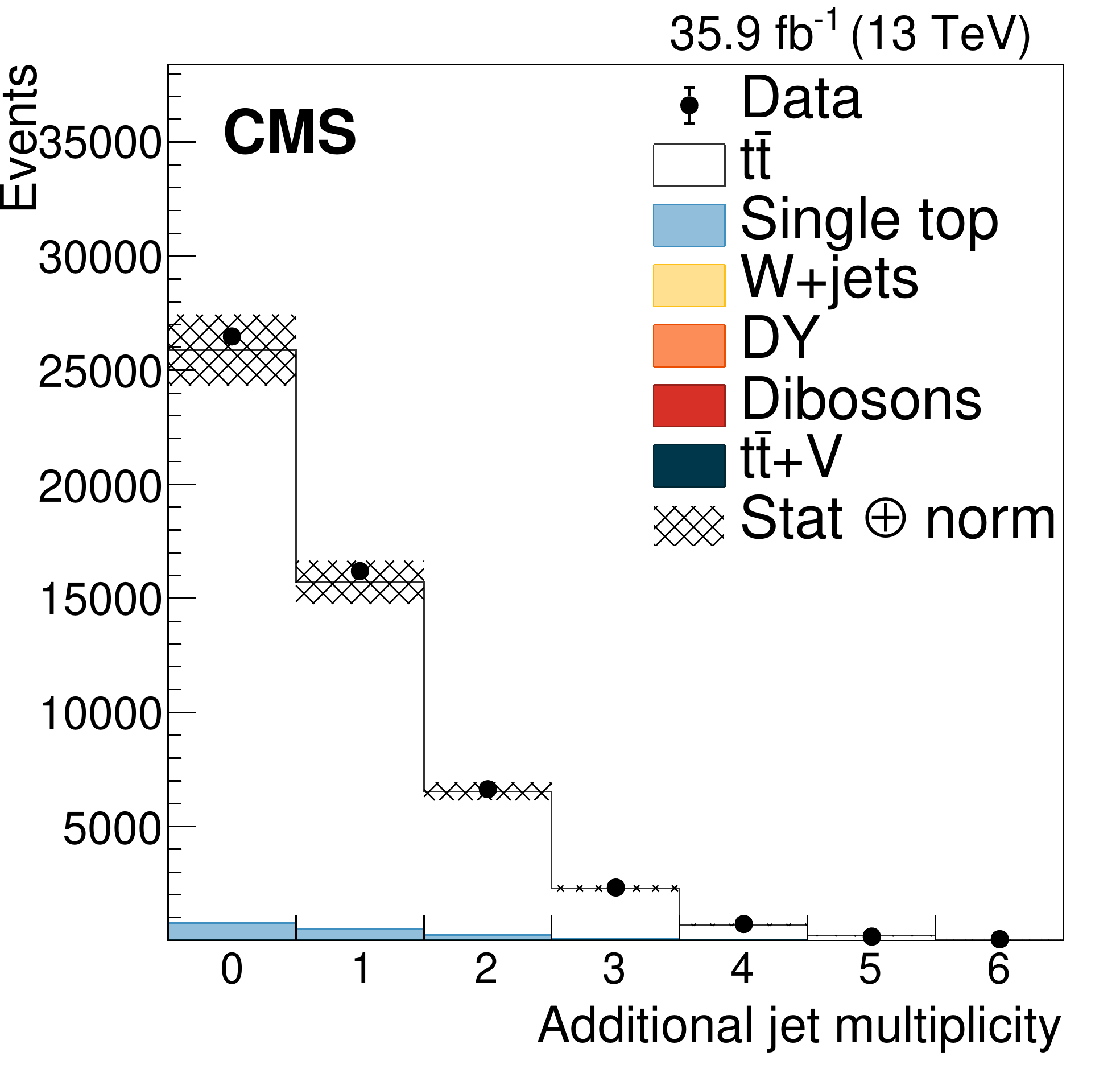}
\includegraphics[width=0.45\textwidth]{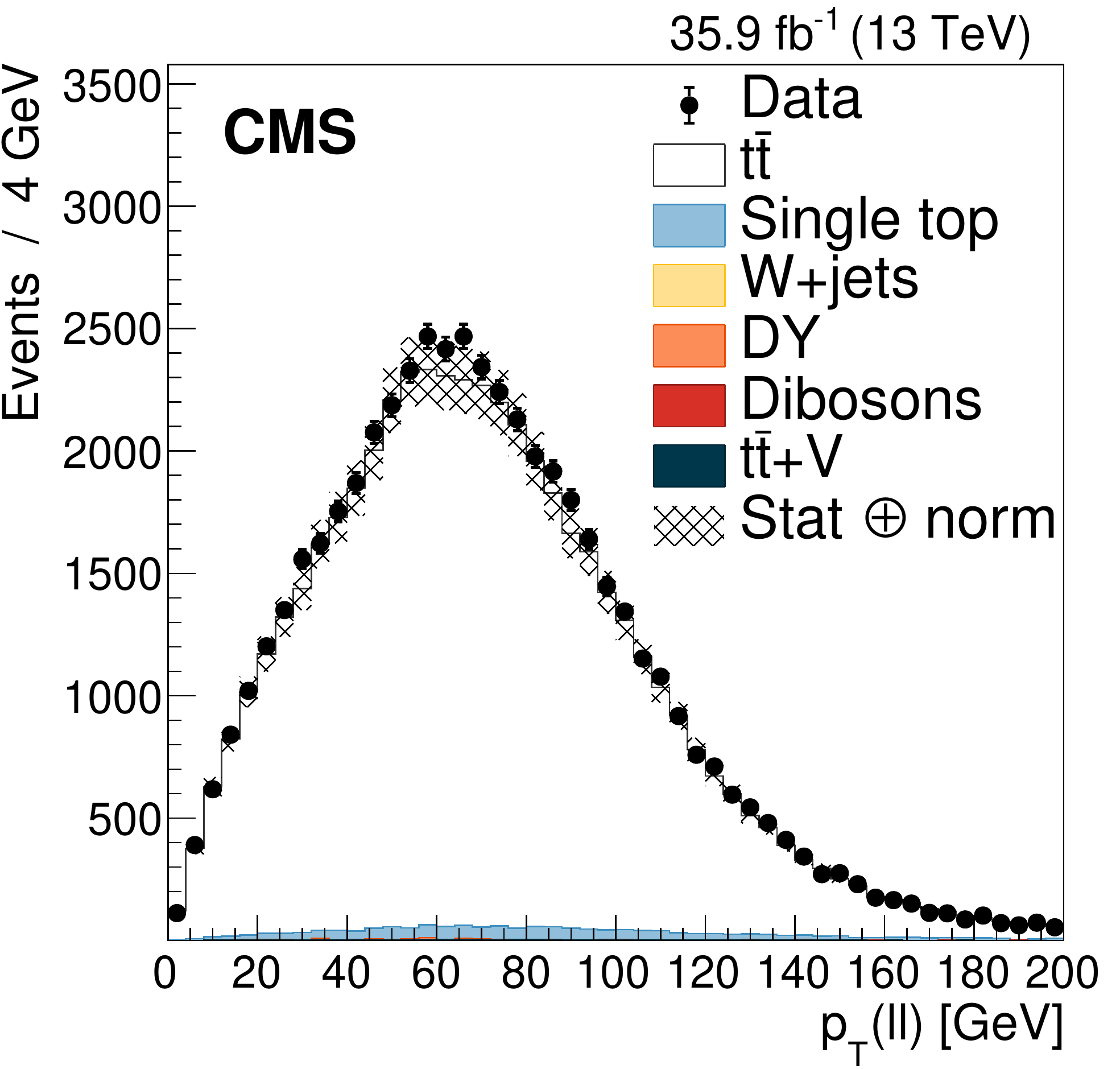}
\includegraphics[width=0.45\textwidth]{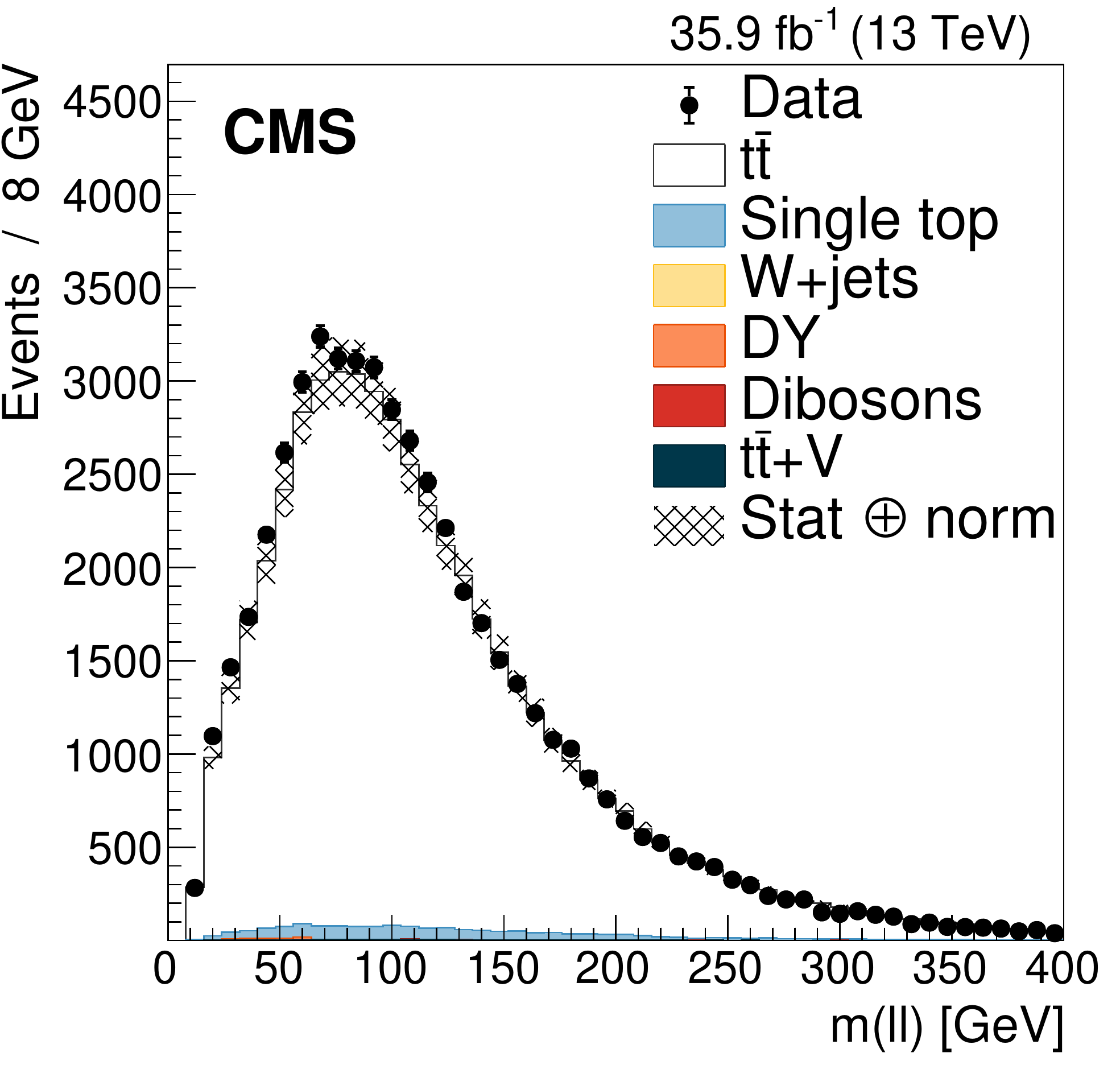}
\caption{
Distributions of the variables used to categorize the study of the UE.
Upper left: multiplicity of additional jets ($\pt>30\GeV$).
Upper right: \ptll.
Lower: \mll.
The distributions in data are compared to the sum of the expectations for the signal and backgrounds.
The shaded band represents
the uncertainty associated to the integrated luminosity
and the theoretical value of the \ttbar cross section.
}
\label{fig:category_vars}
\end{figure*}

The dependence of the UE on the recoil system is studied in categories that are defined
according to the multiplicity of additional jets with $\pt>30\GeV$ and $\abs{\eta}<2.5$,
excluding the two selected \cPqb-tagged jets.
The categories with 0, 1, or more than 1 additional jet are used for this purpose.
The additional jet multiplicity is partially correlated with the charged-particle multiplicity and helps to factorize the contribution from ISR.
The distribution of the number of additional jets is shown in Fig.~\ref{fig:category_vars} (upper left).

\begin{figure}[!htp]
\centering
\includegraphics[width=0.5\textwidth]{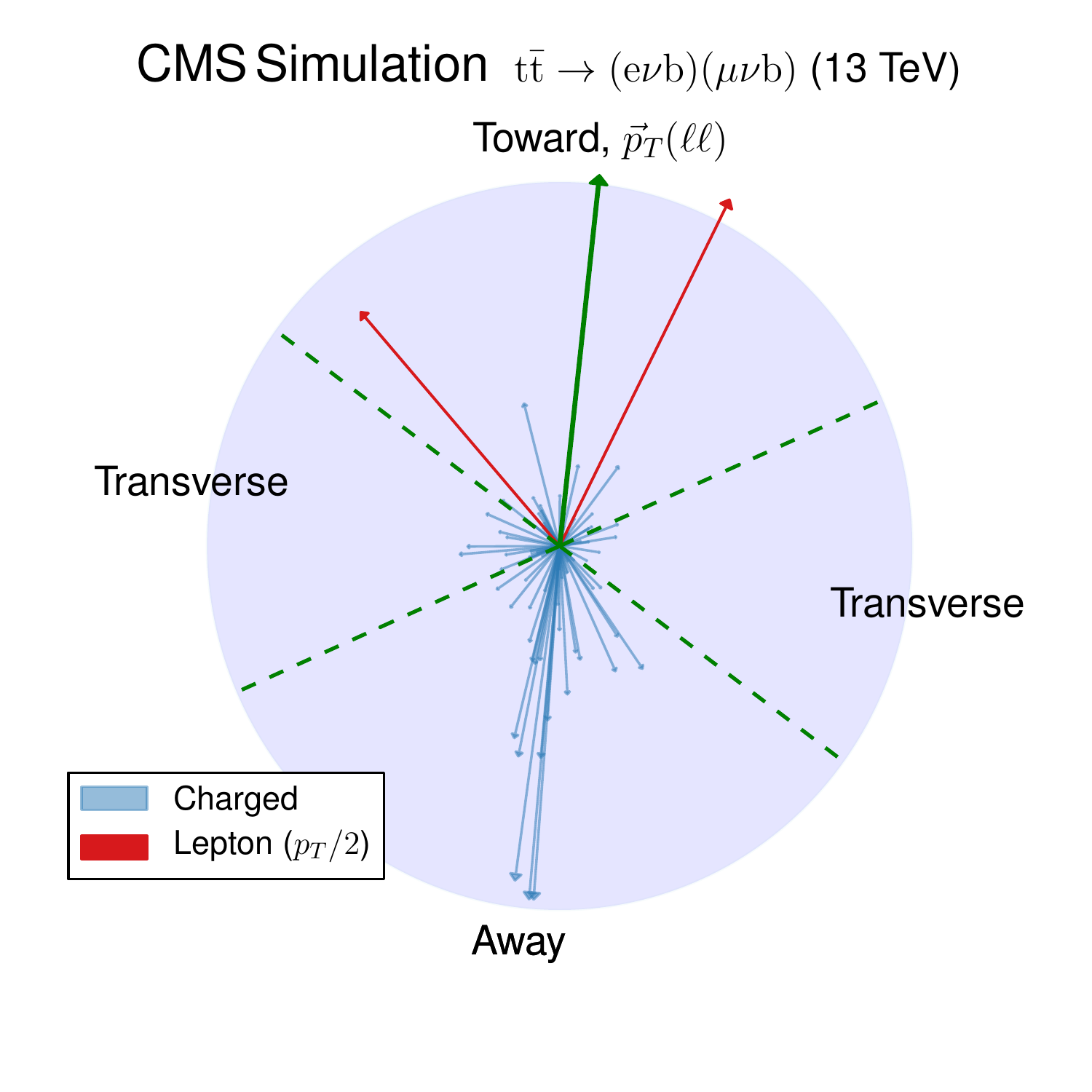}
\caption{
Display of the transverse momentum of the selected charged particles, the two leptons, and the dilepton pair in the transverse plane
corresponding to the same event as in Fig.~\ref{fig:uesel}.
The \pt of the particles is proportional to the length of the arrows and
the dashed lines represent the regions that are defined relative to the \ptllvec{} direction.
For clarity, the \pt of the leptons has been rescaled by a factor of 0.5.
}
\label{fig:ueregions}
\end{figure}

In addition to these categories, the transverse momentum of the dilepton system, \ptllvec{}, is used
as it preserves some correlation with the transverse momentum of the \ttbar system and, consequently, with the recoil of the system.
The \ptllvec{} direction is used to define three regions in the transverse plane of each event.
The regions are denoted as ``transverse" ($60^\circ<\abs{\Delta\phi}<120^\circ$),
``away" ($\abs{\Delta\phi}>120^\circ$), and ``toward" ($\abs{\Delta\phi}<60^\circ$).
Each reconstructed particle in an event is assigned to one of these regions,
depending on the difference of their azimuthal angle with respect to the \ptllvec{} vector.
Figure~\ref{fig:ueregions} illustrates how this classification is performed on a typical event.
This classification is expected to enhance the sensitivity of the measurements to the contributions from ISR,
MPI and CR in different regions.
In addition, the magnitude, \ptll{}, is used to categorize the events
and its distribution is shown in Fig.~\ref{fig:category_vars} (upper right).
The \ptll{} variable is estimated with a resolution better than 3\%.

Lastly, the dependence of the UE on the energy scale of the hard process is characterized
by measuring it in different categories of the \mll{} variable.
This variable is correlated with the invariant mass of the \ttbar system, but not with its \pt.
The \mll~distribution is shown in Fig.~\ref{fig:category_vars} (lower).
A resolution better than 2\%  is expected in the measurement of \mll.

Although both \ptll{} and \mll{} are only partially correlated with the \ttbar kinematic quantities,
they are expected to be reconstructed with very good resolution.
Because of the two escaping neutrinos, the kinematics of the \ttbar pair can only
be reconstructed by using the \ptmiss measurement,
which has poorer experimental resolution when compared to the leptons.
In addition, given that \ptmiss is correlated with the UE activity,
as it stems from the balance of all PF candidates in the transverse plane,
it could introduce a bias in the definition of the categories and the observables studied to characterize the UE.
Hence the choice to use only dilepton-related variables.

\section{Corrections to the particle level}
\label{sec:correctionstoparticlelevel}
Inefficiencies of the track reconstruction due to the residual contamination from pileup,
nuclear interactions in the tracker material,
and accidental splittings of the primary vertex~\cite{Chatrchyan:2014fea} are expected to cause a slight bias in the observables described above.
The correction for these biases is estimated from simulation and applied to the data by means of an unfolding procedure.

At particle (generator) level, the distributions of the observables of interest are binned according to the resolutions expected  from simulation.
Furthermore, we require that each bin contains at least 2\% of the total number of events.
The migration matrix ($K$), used to map the reconstruction- to particle-level distributions,
is constructed using twice the number of bins at the reconstruction level than the ones used at particle level.
This procedure ensures almost diagonal matrices, which have a numerically stable inverse.
The matrix is extended with an additional row that
is used to count the events failing the reconstruction-level requirements,
but found in the fiducial region of the analysis, \ie, passing the particle-level requirements.
The inversion of the migration matrix is made using a Tikhonov regularization procedure~\cite{Tikhonov}, as implemented in the \textsc{TUnfoldDensity} package~\cite{Schmitt:2012kp}.
The unfolded distribution is found by minimizing a $\chi^2$ function
\begin{equation}
\chi^2=(y-K\lambda)^T V_{yy}^{-1} (y-K\lambda) + \tau^2 ||L(\lambda-\lambda_0)||^2,
\label{eq:unfoldchi2}
\end{equation}
where $y$ are the observations,
$V_{yy}$ is an estimate of the covariance of $y$
(calculated using the simulated signal sample),
$\lambda$ is the particle-level expectation,
$||L(\lambda-\lambda_0)||^2$ is a penalty function
(with $\lambda_0$ being estimated from the simulated samples),
and $\tau>0$ is the so-called regularization parameter.
The latter regulates how strongly the penalty term should contribute to the minimization of $\chi^2$.
In our setup we choose the function $L$ to be the curvature, \ie, the second derivative, of the output distribution.
The chosen value of the $\tau$ parameter is optimized for each distribution by minimizing its average global correlation coefficient~\cite{Schmitt:2012kp}.
Small values, \ie, $\tau<10^{-3}$, are found for all the distributions;
the global correlation coefficients are around 50\%.
After unfolding, the distributions are normalized to unity.

The statistical coverage of the unfolding procedure is checked by means of
pseudo-experiments based on independent {\pwpy} samples.
The pull of each bin in each distribution is found to be consistent with that of a standard normal distribution.
The effect of the regularization term in the unfolding is checked in the data by
folding the measured distributions and comparing the outcome to the originally-reconstructed data.
In general the folded and the original distributions agree within 1--5\% in each bin, with the largest differences observed in bins with low yield.

\section{Systematic uncertainties}
\label{sec:systematic uncertainties}
The impact of different sources of uncertainty is evaluated
by unfolding the data with alternative migration matrices,
which are obtained after changing the settings in the simulations as explained below.
The effect of a source of uncertainty in non-fiducial \ttbar events
is included in this estimate, by updating the background prediction.
The observed bin-by-bin differences are used as estimates of the uncertainty.
The impact of the uncertainty in the background normalization is the only exception to this procedure,
as detailed below.
The covariance matrices associated to each source of uncertainty are built using the procedure described in detail in~\cite{Sirunyan:2017lvd}. In case several sub-contributions are used to estimate a source of uncertainty, the corresponding differences in each bin are treated independently, symmetrized, and used to compute individual covariance matrices, which preserve the normalization. Variations on the event yields are fully absorbed by normalizing the measured cross sections. Thus, only the sources of uncertainty that yield variations in the shapes have a non-negligible impact.

\subsection{Experimental uncertainties}
\label{subsubsec:expuncs}

The following experimental sources of uncertainty are considered:

\begin{description}

\item[Pileup:] Although pileup is included in the simulation, there
  is an intrinsic uncertainty in modeling its multiplicity.
  An uncertainty of $\pm$4.6\% in the inelastic $\Pp\Pp$ cross section is used and propagated to the event weights~\cite{Sirunyan:2018nqx}.

\item[Trigger and selection efficiency:] The scale factors used to correct the simulation
  for different trigger and lepton selection efficiencies in data and simulation are
  varied up or down, according to their uncertainty.
  The uncertainties in the muon track and electron reconstruction efficiencies
  are included in this category and added in quadrature.

\item[Lepton energy scale:] The corrections applied to the electron energy and muon momentum scales
  are varied separately, according to their uncertainties.
  The corrections and uncertainties are obtained using  methods similar to those described in Refs.~\cite{Khachatryan:2015hwa,Chatrchyan:2012xi}.
  These variations lead to a small migration of events
  between the different \ptll{} or \mll{} categories used in the analysis.

\item[Jet energy scale:] A \pt- and $\eta$-dependent parameterization of
  the jet energy scale is used to vary the calibration of the jets in
  the simulation. The corrections and uncertainties are obtained using methods similar to those described in Ref.~\cite{Khachatryan:2016kdb}.
  The effect of these variations is similar to that described
  for the lepton energy scale uncertainty; in this case the migration of events
  occurs between different jet multiplicity categories.

\item[Jet energy resolution:] Each jet is further smeared up or down
  depending on its \pt and $\eta$, with respect to the central value measured in data.
  The difference with respect to data is measured using methods similar to those described in Ref.~\cite{Khachatryan:2016kdb}.
  The main effect induced in the analysis from altering the jet energy resolution is similar to that described for the jet energy scale uncertainty.

\item[{\cPqb} tagging and misidentification efficiencies:] The scale factors
  used to correct for the difference in performance between data and simulation
  are varied according to their uncertainties and depending on the flavor of the jet~\cite{Sirunyan:2017ezt}.
  The main effect of this variation is to move jets into the candidate {\cPqb} jets sample
  or remove them from it.

\item[Background normalization:] The impact of the uncertainty in the normalization of the backgrounds
 is estimated by computing the difference obtained with respect to the nominal result when these contributions are not subtracted from data.
 This difference is expected to cover the uncertainty in the normalization of the main backgrounds, \ie, DY and the \cPqt\PW{} process,
 and the uncertainty in the normalization of the \ttbar events
 that are expected to pass the reconstruction-level requirements but fail the
 generator-level ones.
 The total expected background contribution is at the level of 8--10\%, depending on the bin.
 The impact from this uncertainty is estimated to be $<$5\%.

\item[Tracking reconstruction efficiency:] The efficiency of track reconstruction is found to be more than 90\%.
  It is monitored using muon candidates from $\cPZ\to\PGmp\PGmm$ decays,
  and the ratio of the four-body final $\PDz \to \PK^-\PGpp\PGpm\PGpp$ decay
  to the two-body $\PDz\to \PK^-\PGpp$ decay.
  The latter is used to determine a data-to-simulation scale factor ($SF_\text{trk}$) as a function of the pseudorapidity of the tracks,
  and for different periods of the data taking used in this analysis.
  The envelope of the $SF_\text{trk}$ values, with uncertainties included, ranges from 0.89 to 1.17~\cite{CMS-DP-2016-012},
  and it provides an adequate coverage for the residual variations observed in the charged-particle multiplicity between different
  data taking periods.
  The impact of the variation of $SF_\text{trk}$ by its uncertainty is estimated
  by using the value of $\abs{1-SF_\text{trk}}$ for the probability to remove a reconstructed track from the event
  or to promote an unmatched generator-level charged particle to a reconstructed track,
  depending on whether $SF_\text{trk}<1$ or $>$1, respectively.
  Different migration matrices,
  reflecting the different tracking efficiencies obtained from varying the uncertainty in $SF_\text{trk}$,
  are obtained by this method and used to unfold the data.
  Although the impact is nonnegligible on variables such as \nch{} or \sumpt{}, it has very small impact ($<$1\%) on variables such as {\avgpt} and \avgpz.

\end{description}

\subsection{Theoretical uncertainties}
\label{subsec:thuncs}

The following theoretical uncertainties are considered:

\begin{description}

\item[Scale choices:]
$\mu_\mathrm{R}$ and $\mu_\mathrm{F}$ are varied individually in the ME by
factors between 0.5 and 2, excluding the extreme cases $\mu_\mathrm{R}/\mu_\mathrm{F}=\mu(2,0.5)$ and $\mu(0.5,2)$,
according to the prescription described in Refs.~\cite{Cacciari:2003fi,Catani:2003zt}.

\item[Resummation scale and \alpS used in the parton shower:]
In \POWHEG, the real emission cross section is scaled by a damping function,
parameterized by the so-called $h_\text{damp}$ variable~\cite{powhegv21,powhegv22,powhegv23}.
This parameter controls the ME-PS matching and regulates the high-\pt radiation by reducing
real emissions generated by \POWHEG with a factor of $h_\text{damp}^2/(\pt^2+h_\text{damp}^2)$.
In the simulation used to derive the migration matrices, $h_\text{damp}=1.58~m_\cPqt$
and the uncertainty in this value is evaluated by changing it by +42 or -37\%,
a range that is determined from the jet multiplicity measurements in \ttbar at $\sqrt{s}=8\TeV$~\cite{Khachatryan:2015mva}.
Likewise, the uncertainty associated with the choice of
$\alpS^\text{ISR}(M_\cPZ)=0.1108$ for space-like and $\alpS^\text{FSR}(M_\cPZ)=0.1365$
for time-like showers
in the CUETP8M2T4 tune is evaluated by varying the scale at which it is computed, $M_\cPZ$, by a factor of 2 or 1/2.

\item[UE model:]
The dependence of the migration matrix on the UE model assumed in the simulation is tested
by varying the parameters that model the MPI and CR
in the range of values corresponding to the uncertainty envelope associated to the CUETP8M2T4 tune.
The uncertainty envelope has been determined using the same methods as described in Ref.~\cite{Khachatryan:2015pea}.
In the following, these will be referred to as UE up/down variations.
The dependence on the CR model is furthermore tested using other models besides the nominal one,
which is the MPI-based CR model where the \ttbar decay products are excluded from reconnections to the UE.
A dedicated sample where the reconnections to resonant decay products are enabled (hereafter designated as ERDon)
is used to evaluate possible differences in the unfolded results.
In addition, alternative models for the CR are tested.
One sample utilizing the ``gluon move'' model~\cite{Argyropoulos:2014zoa}, in which gluons can be moved to another string,
and another utilizing the ``QCD-based'' model with string formation beyond LO~\cite{Christiansen:2015yqa}  are used for this purpose.
In both samples, the reconnections to the decay resonant processes are enabled.
The envelope of the differences is considered as a systematic uncertainty.

\item[\cPqt{} quark $\textbf{\pt}$:]
The effect of reweighting of the simulated \cPqt{} quark \pt ($\pt(\cPqt)$) distribution
to match the one reconstructed from data~\cite{Khachatryan:2016mnb,Sirunyan:2017mzl} is added as an additional uncertainty.
This has the most noticeable effect
on the fraction of events that do not pass the reconstruction-level
requirements and migrate out of the fiducial phase space.

\item[\cPqt{} quark mass:]
An additional uncertainty is considered, related to the value of $m_\cPqt=172.5\GeV$ used in the simulations,
by varying this parameter by $\pm 0.5\GeV$~\cite{Khachatryan:2015hba}.

\end{description}

Any possible uncertainty from the choice of the hadronization model is expected to be significantly smaller than the theory uncertainties described above. This has been explicitly tested by comparing the results at reconstruction level and after unfolding the data with the \pwpy{} and \pwhwpp{} migration matrices. The latter relies on a different hadronization model, but it folds other modelling differences such as the underlying event tune or the parton shower as well. Thus it can only be used as a test setup to validate the measurement.

\subsection{Summary of systematic uncertainties}
\label{subsec:systsummary}

The uncertainties on the measurement of the normalized differential cross sections are dominated by the systematic uncertainties,
although in some bins of the distributions the statistical uncertainties are a large
component.
The experimental uncertainties have, in general, small impact; the most relevant
are the tracking reconstruction efficiency for the \nch, \sumpt{}, \sumpz{}, and \chpt{} observables.
Other observables are affected at a sub-percent level by this uncertainty.
Theory uncertainties affect the measurements more significantly,
a fact that underlines the need of better tuning of the model parameters.

Event shape observables are found to be the most robust against this uncertainty,
while \sumpt{}, \sumpz{}, and \chpt{} are the ones that suffer more from it.
Other sources of theoretical uncertainty typically have a smaller effect.

To further illustrate the impact of different sources on the observables considered,
we list in Table~\ref{tab:uncmean}
the uncertainties on the average of each observable.
In the table, only systematic uncertainties that impact the average of one of the observables
by at least 0.5\% are included.
The total uncertainty on the average of a given quantity ranges from 1 to 8\%,
and hence the comparison with model predictions can be carried out in a discrete manner.

\begin{table*}[htb]
\centering
\topcaption{Uncertainties affecting the measurement of the average of the UE observables.
The values are expressed in \% and
the last row reports the quadratic sum of the individual contributions.}
\label{tab:uncmean}
\begin{tabular}{lcccccccccc}
\multirow{2}{*}{Source} & \multicolumn{10}{c}{\% Uncertainty} \\
                        & \nch & \sumpt & \sumpz & \avgpt & \avgpz & \chpt & $S$ & $A$ & $C$ & $D$\\
\hline
Statistical                        & 0.1 & 0.2 & 0.3 & 0.2 & 0.2 & 0.3 & 0.1 & 0.1 & 0.1 & 0.1 \\[\cmsTabSkip]
\multicolumn{2}{c}{~} & \multicolumn{8}{c}{Experimental} & \\[\cmsTabSkip]
Background                       & 1.2 & 1.6 & 1.8 & 0.4 & 0.7 & 1.6 & 0.4 & 0.7 & 0.3 & 0.7 \\
Tracking eff.                    & 4.4 & 4.2 & 4.9 & 0.8 & 0.4 & 4.0 & 0.4 & 0.6 & 0.2 & 0.6 \\[\cmsTabSkip]
\multicolumn{2}{c}{~} & \multicolumn{8}{c}{Theory} & \\[\cmsTabSkip]
$\mu_\mathrm{R}/\mu_\mathrm{F}$        & 0.5 & 0.8 & 1.0 & 0.3 & 0.3 & 1.0 & 0.1 & 0.1 & 0.1 & 0.2 \\
Resummation scale                & 0.2 & 0.8 & 0.5 & 1.1 & 0.2 & 1.6 & 0.8 & 0.4 & 0.2 & 0.7 \\
$\alpS^\text{FSR}(M_\cPZ)$     & 0.5 & 0.7 & 0.7 & 0.8 & 1.7 & 0.7 & 0.2 & 1.0 & 0.2 & 1.2 \\
$\alpS^\text{ISR}(M_\cPZ)$     & 0.1 & 0.3 & 1.1 & 1.2 & 0.7 & 0.4 & 0.2 & 0.5 & 0.1 & 1.3 \\
UE model                         & 0.1 & 0.1 & 0.2 & 1.0 & 0.4 & 0.5 & 0.2 & 0.2 & 0.1 & 0.9 \\
$m_\cPqt$                        & 0.4 & 0.7 & 1.5 & 0.6 & 0.9 & 0.5 & 0.1 & 0.1 & 0.1 & 0.7 \\
$\pt(\cPqt)$                     & 1.4 & 4.4 & 4.5 & 2.8 & 2.1 & 6.7 & 0.2 & 0.5 & 0.2 & 0.3 \\[\cmsTabSkip]
Total                              & 4.9 & 6.5 & 7.3 & 3.7 & 3.1 & 8.2 & 1.1 & 1.6 & 0.6 & 2.4 \\
\end{tabular}
\end{table*}

\section{Results}
\label{sec:results}

\subsection{Inclusive distributions}
\label{subsec:incdistr}

\begin{figure*}[!htp]
\centering
\includegraphics[width=\cmsFigWidthii]{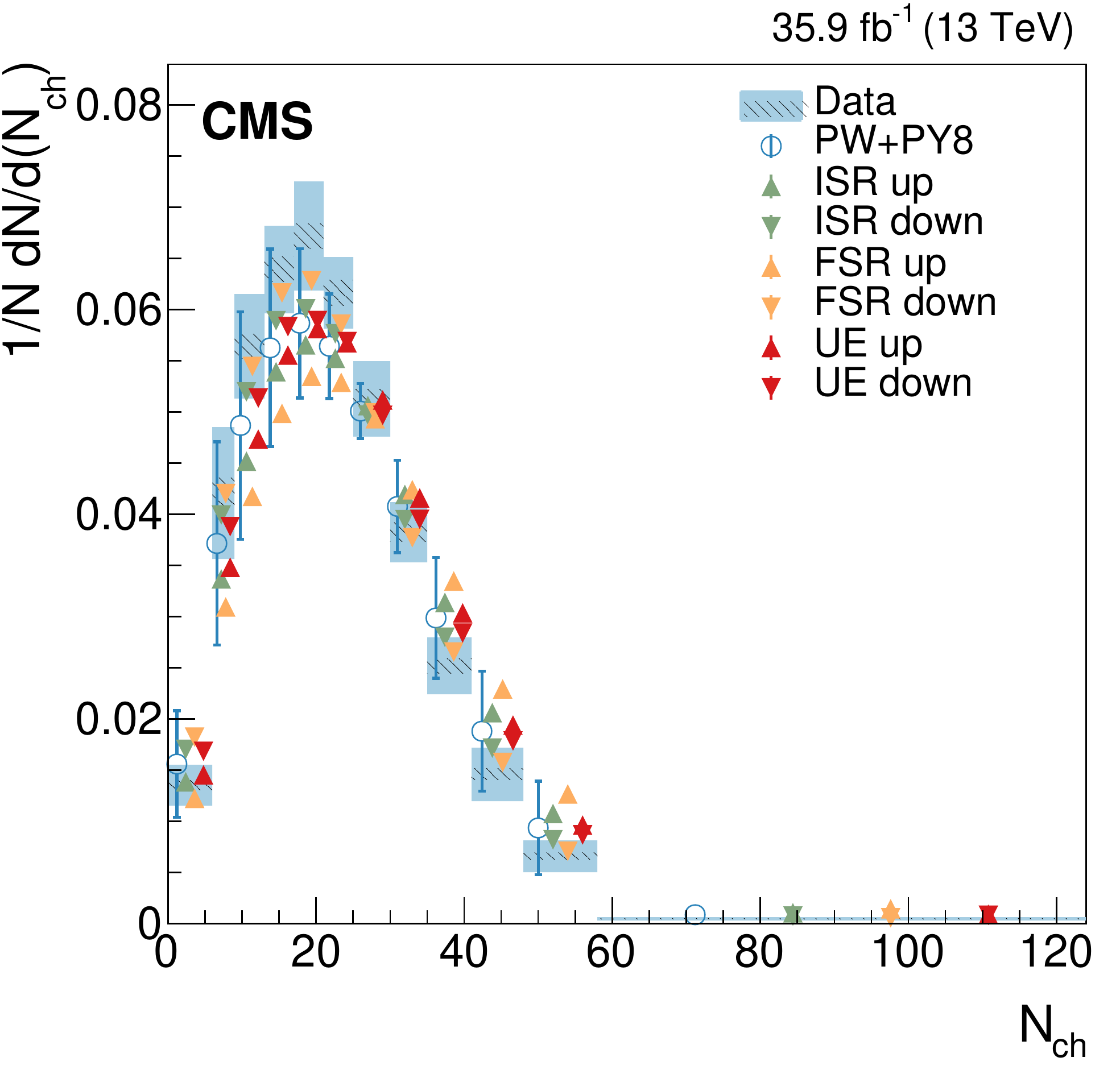}
\includegraphics[width=\cmsFigWidthi]{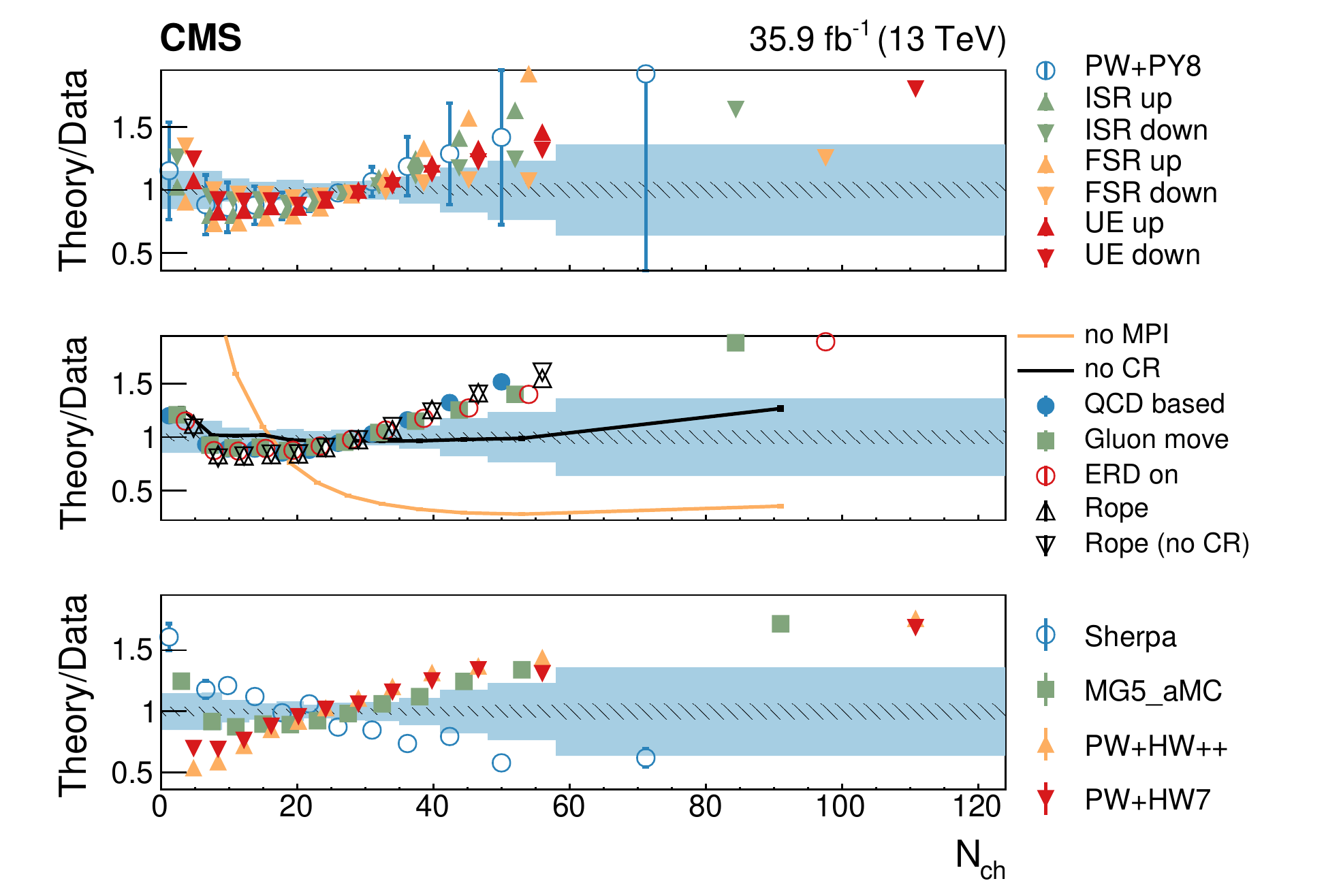}
\caption{
The normalized differential cross section as a function of \nch{} is shown on the upper panel.
The data (colored boxes) are compared to the nominal {\pwpy} predictions
and to the expectations obtained
from varied $\alpS^\text{ISR}(M_\cPZ)$ or $\alpS^\text{FSR}(M_\cPZ)$ {\pwpy} setups (markers).
The different panels on the lower display show
the ratio between each model tested (see text) and the data.
In both cases the shaded (hatched) band represents the total (statistical) uncertainty of the data,
while the error bars represent either the total uncertainty of the {\pwpy} setup,
computed as described in the text,
or the statistical uncertainty of the other MC simulation setups.
}
\label{fig:chmult_unfolded}
\end{figure*}

\begin{figure*}[!htp]
\centering
\includegraphics[width=\cmsFigWidthii]{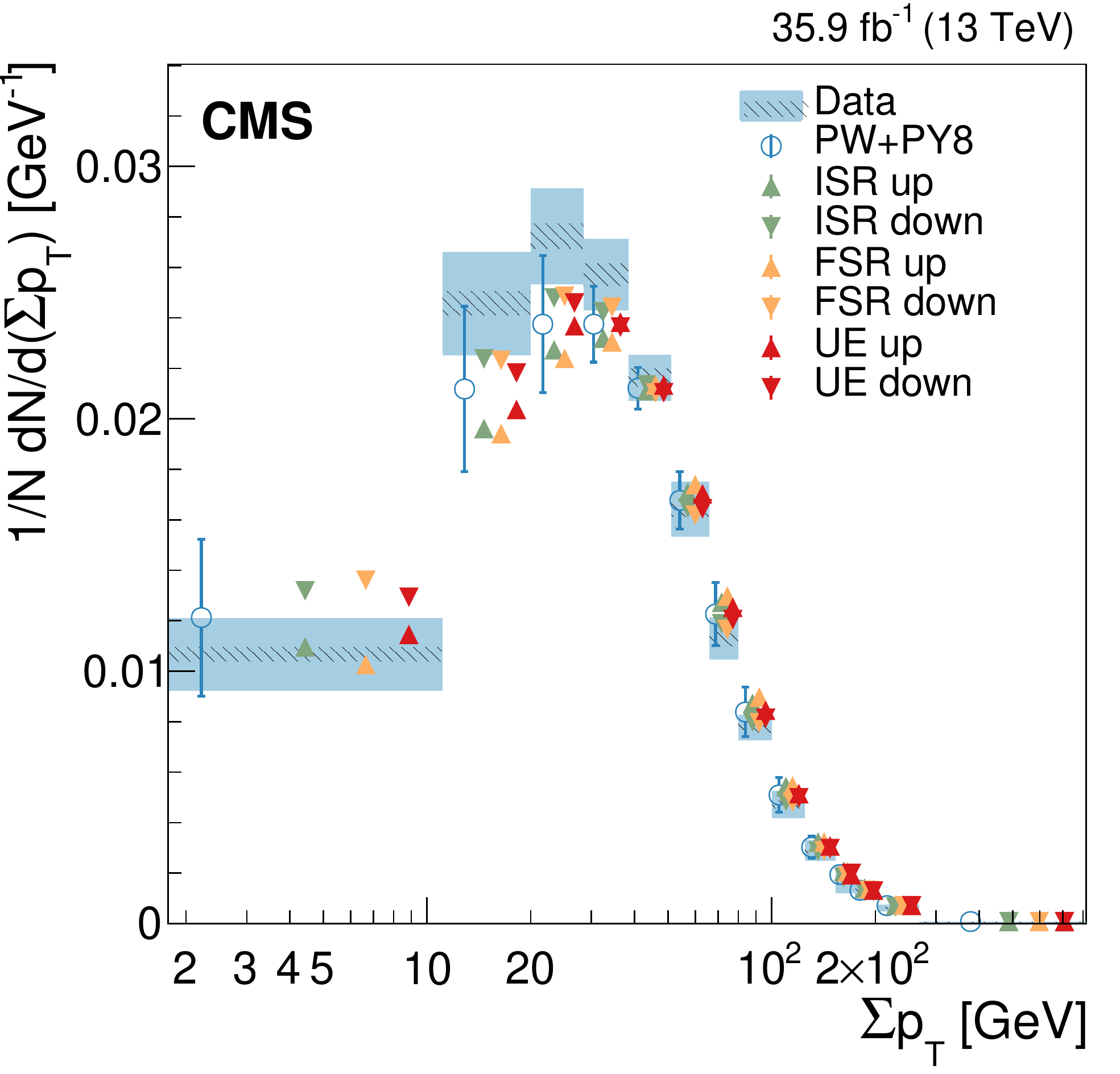}
\includegraphics[width=\cmsFigWidthi]{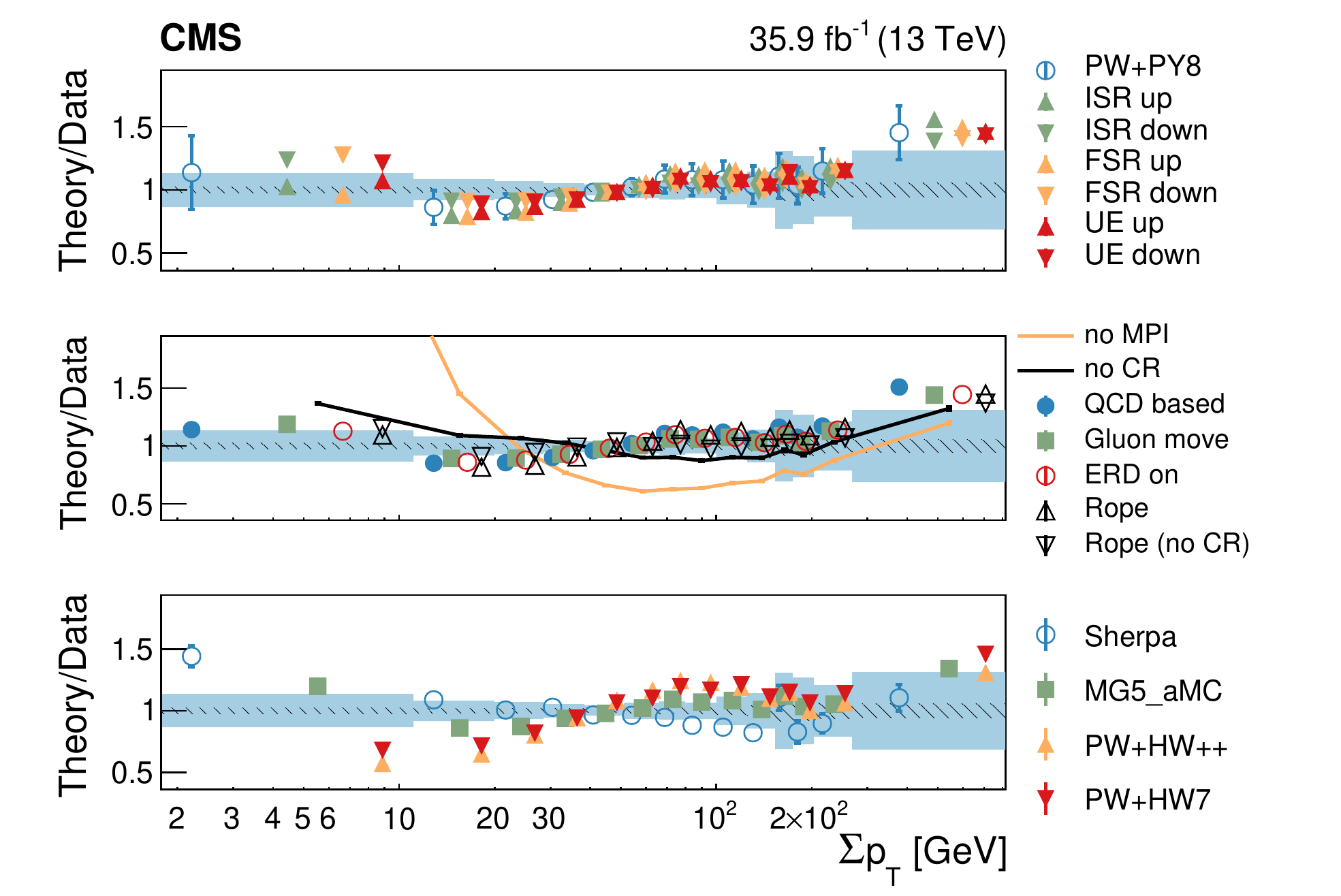}
\caption{
Normalized differential cross section as function of \sumpt{},
compared to the predictions of different models.
The conventions of Fig.~\ref{fig:chmult_unfolded} are used.
}
\label{fig:chflux_unfolded}
\end{figure*}

\begin{figure*}[!htp]
\centering
\includegraphics[width=\cmsFigWidthii]{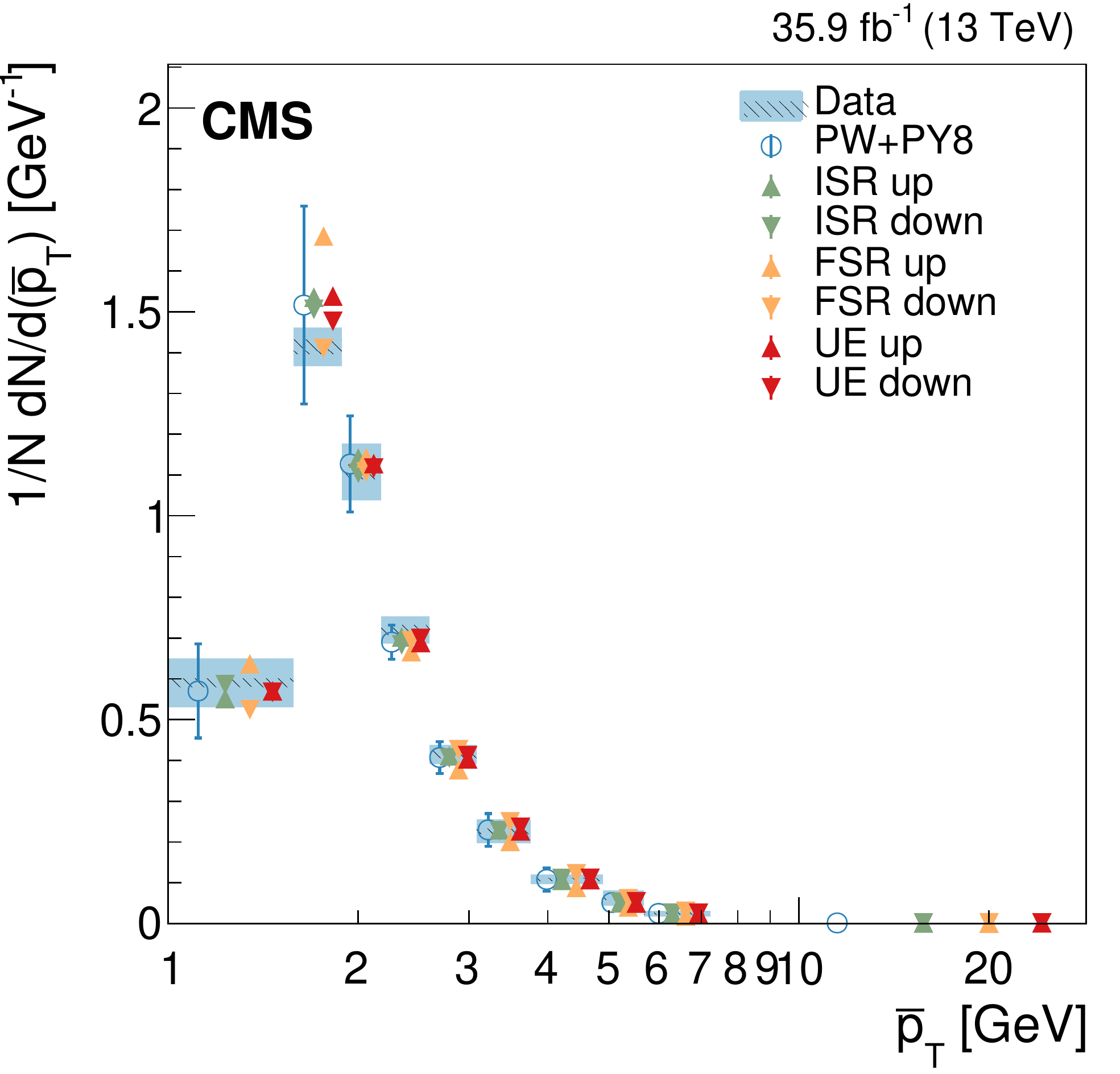}
\includegraphics[width=\cmsFigWidthi]{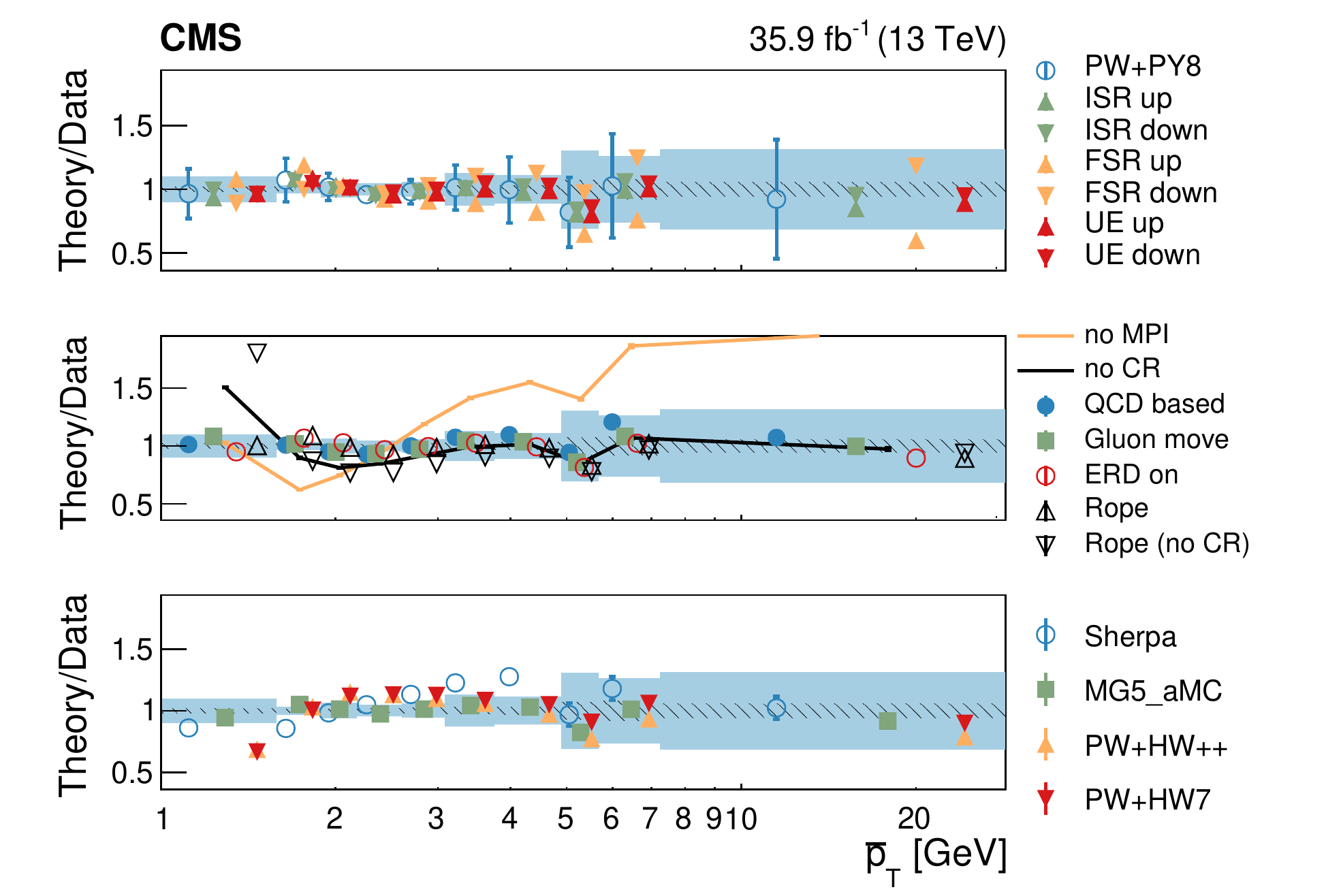}
\caption{
Normalized differential cross section as function of \avgpt{},
compared to the predictions of different models.
The conventions of Fig.~\ref{fig:chmult_unfolded} are used.
}
\label{fig:chavgpt_unfolded}
\end{figure*}

\begin{figure*}[!htp]
\centering
\includegraphics[width=\cmsFigWidthii]{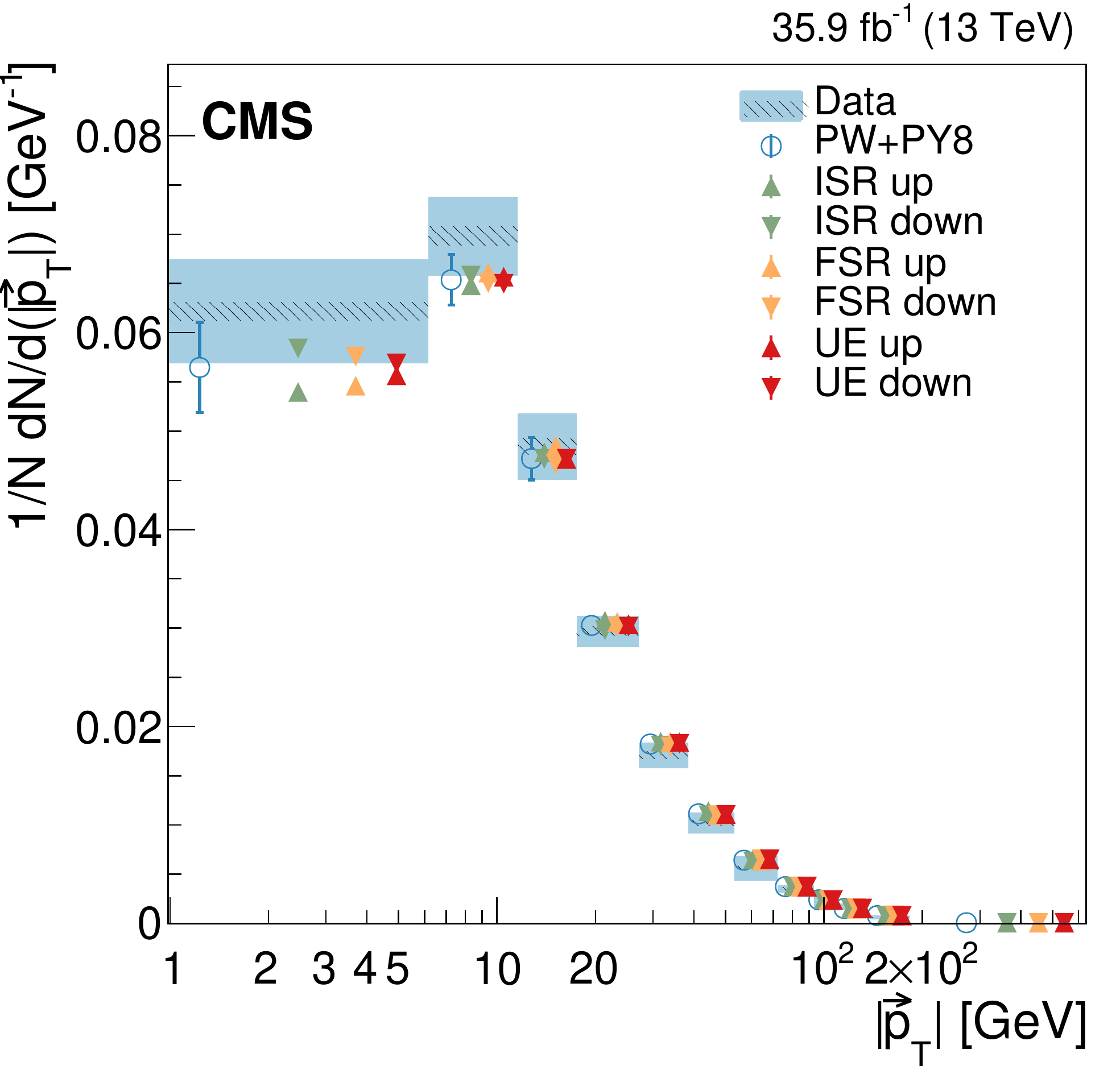}
\includegraphics[width=\cmsFigWidthi]{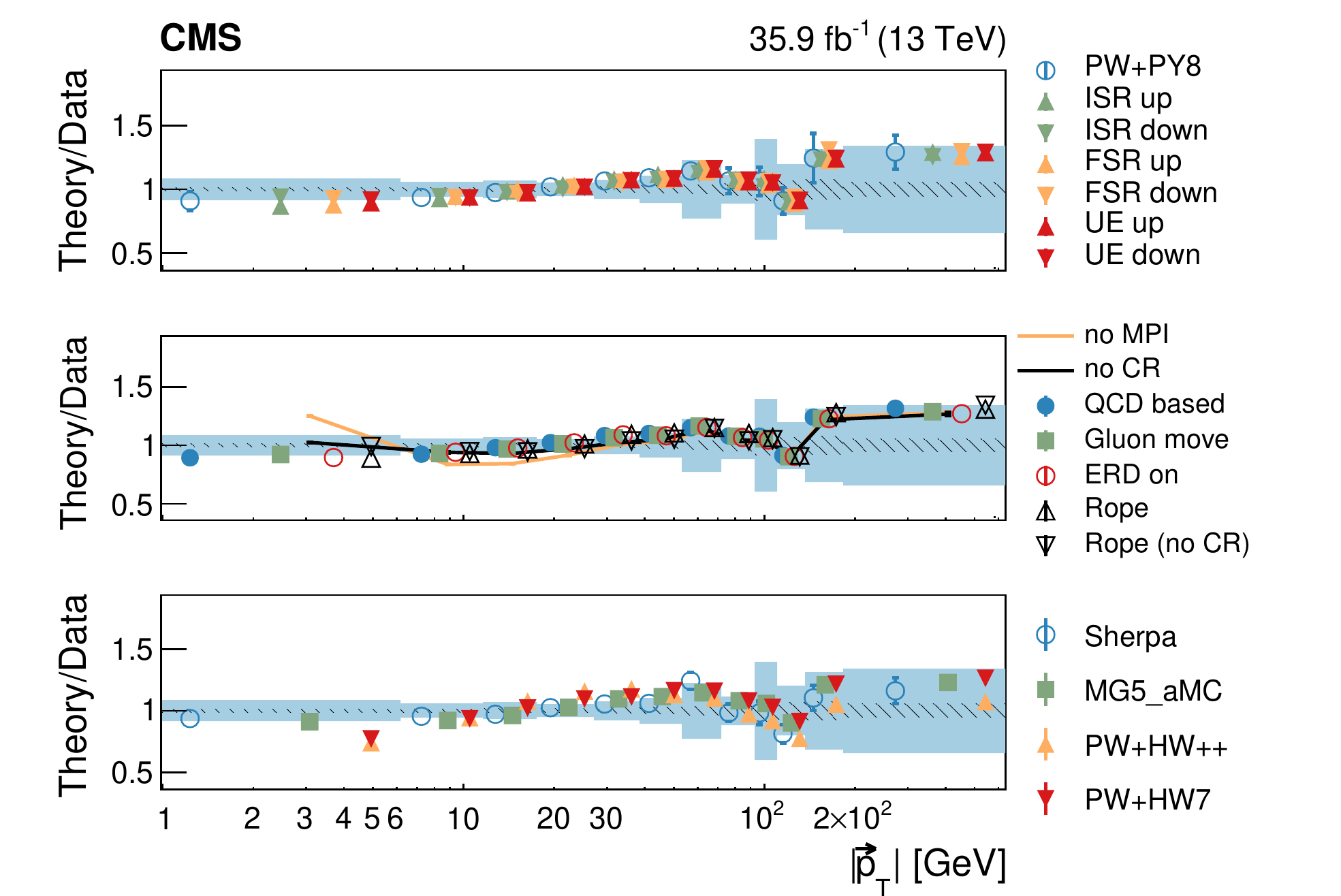}
\caption{
Normalized differential cross section as function of \chpt{},
compared to the predictions of different models.
The conventions of Fig.~\ref{fig:chmult_unfolded} are used.
}
\label{fig:chrecoil_unfolded}
\end{figure*}

\begin{figure*}[!htp]
\centering
\includegraphics[width=\cmsFigWidthii]{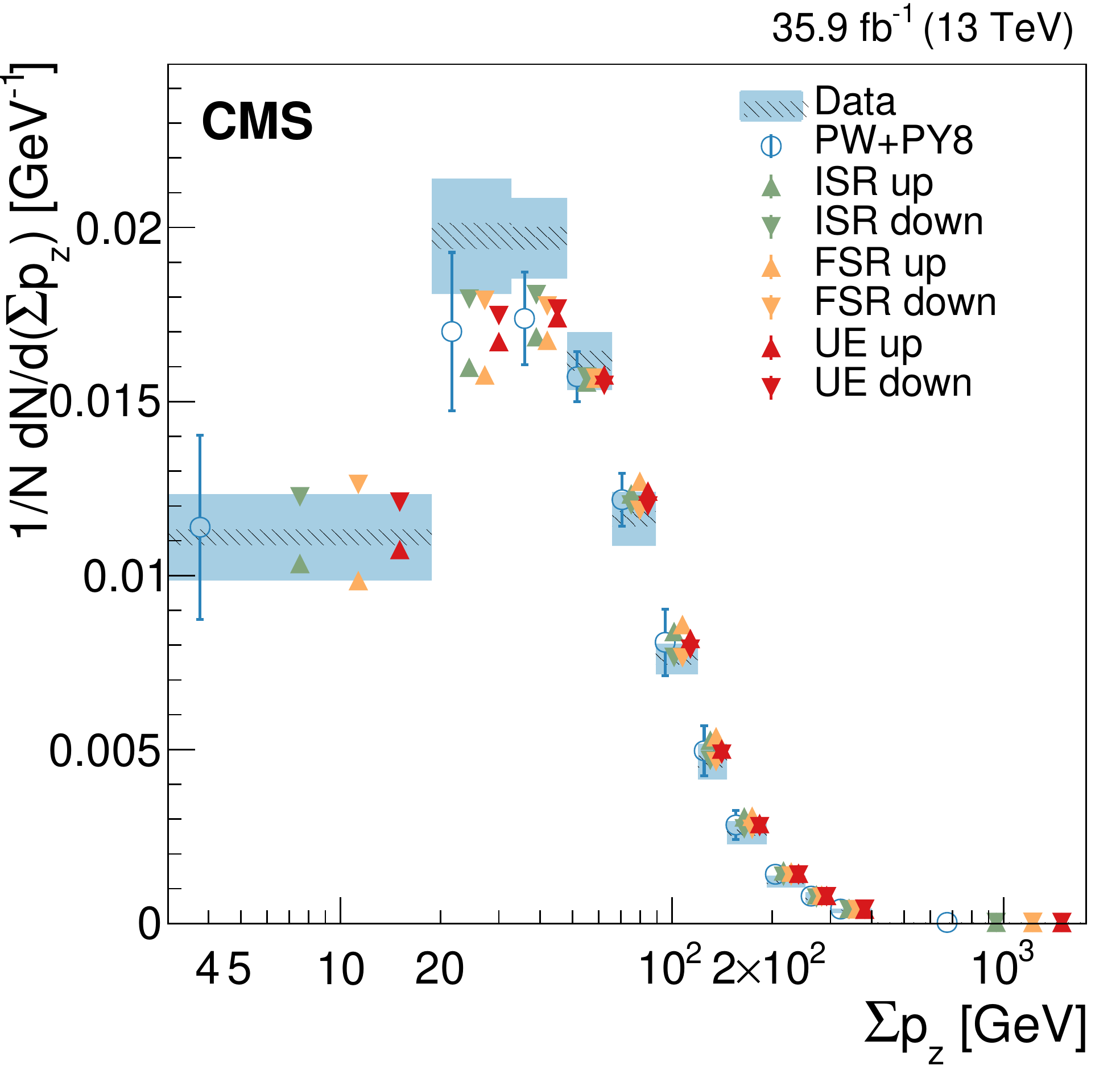}
\includegraphics[width=\cmsFigWidthi]{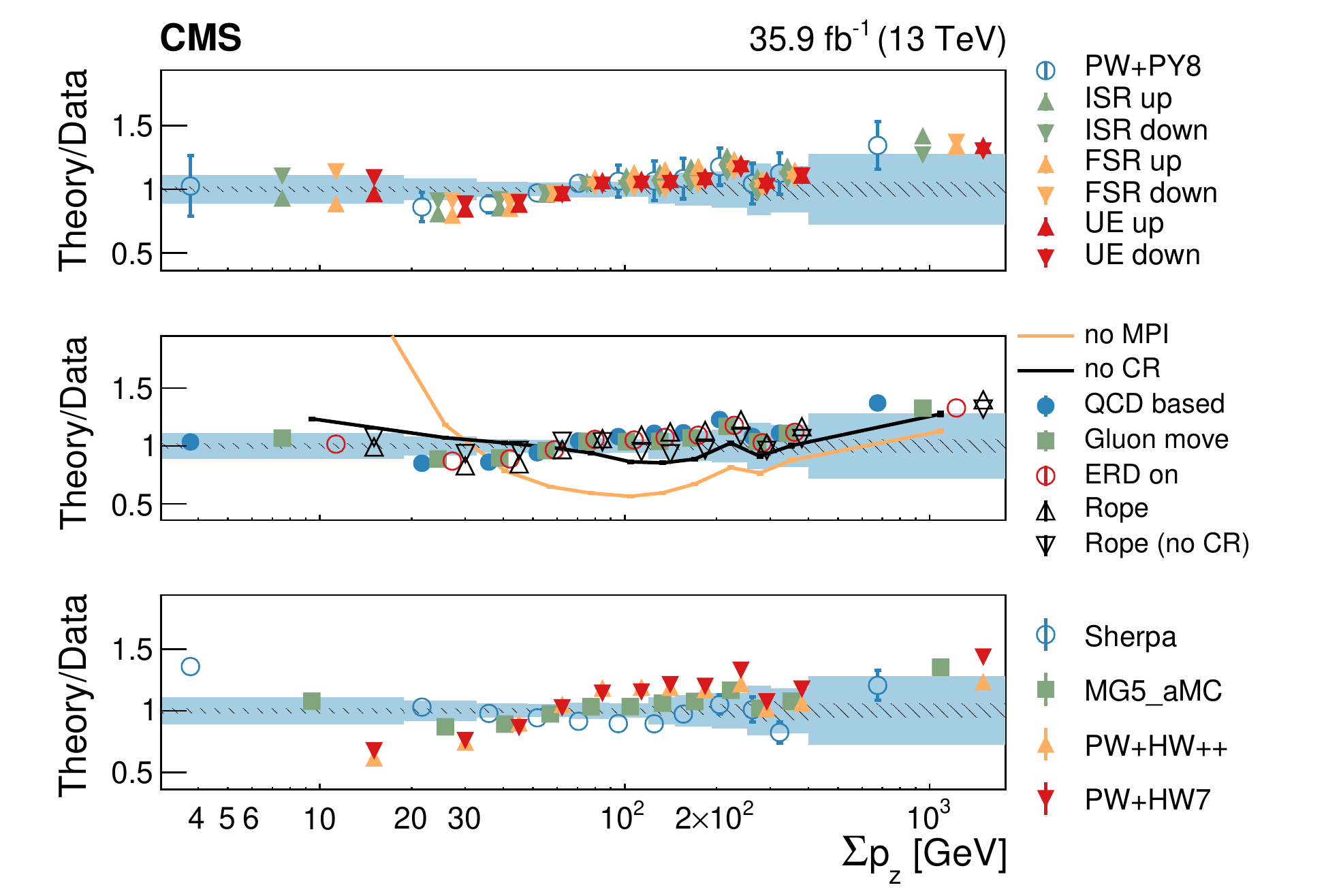}
\caption{
Normalized differential cross section as function of \sumpz{},
compared to the predictions of different models.
The conventions of Fig.~\ref{fig:chmult_unfolded} are used.
}
\label{fig:chfluxz_unfolded}
\end{figure*}

\begin{figure*}[!htp]
\centering
\includegraphics[width=\cmsFigWidthii]{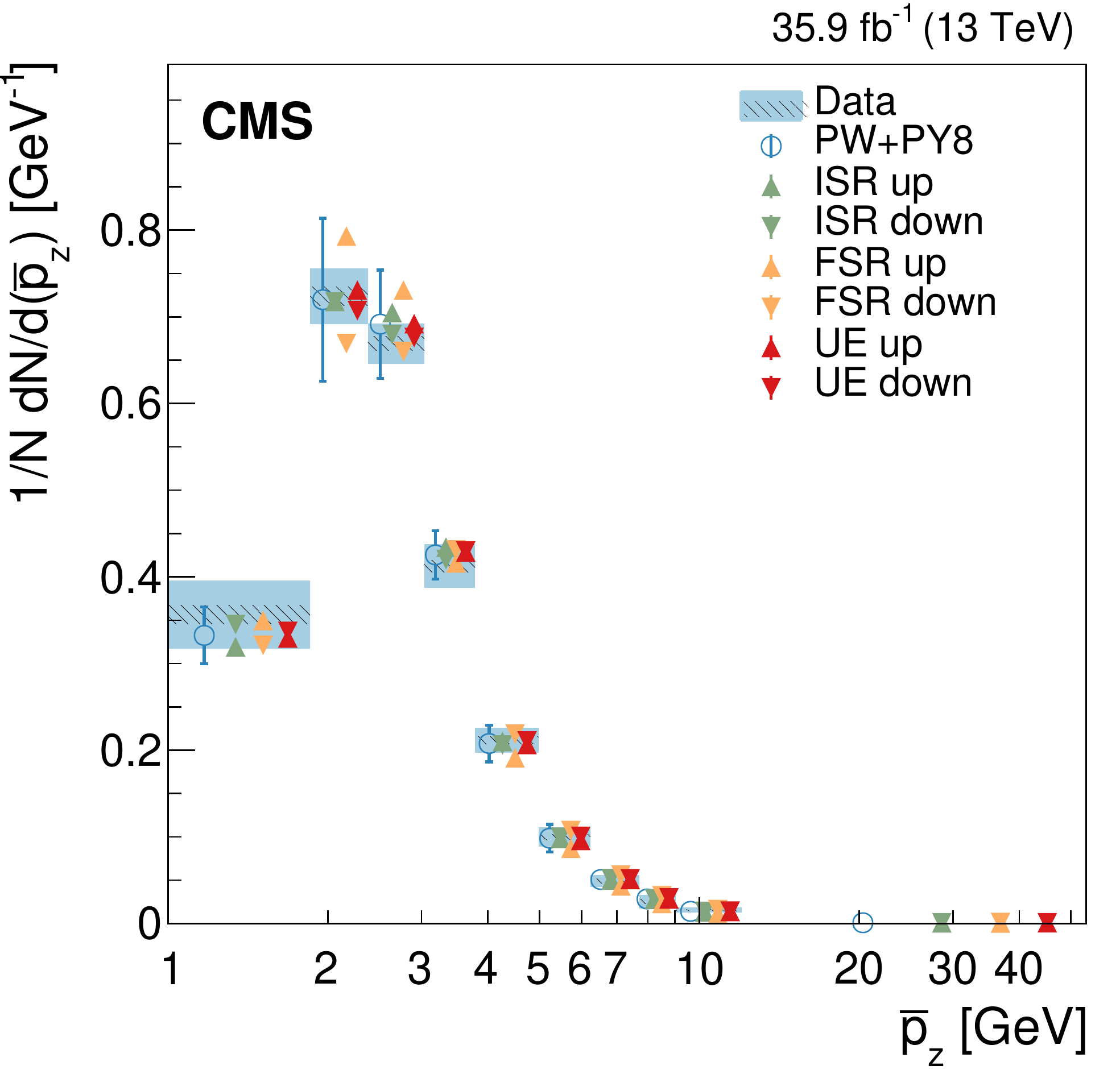}
\includegraphics[width=\cmsFigWidthi]{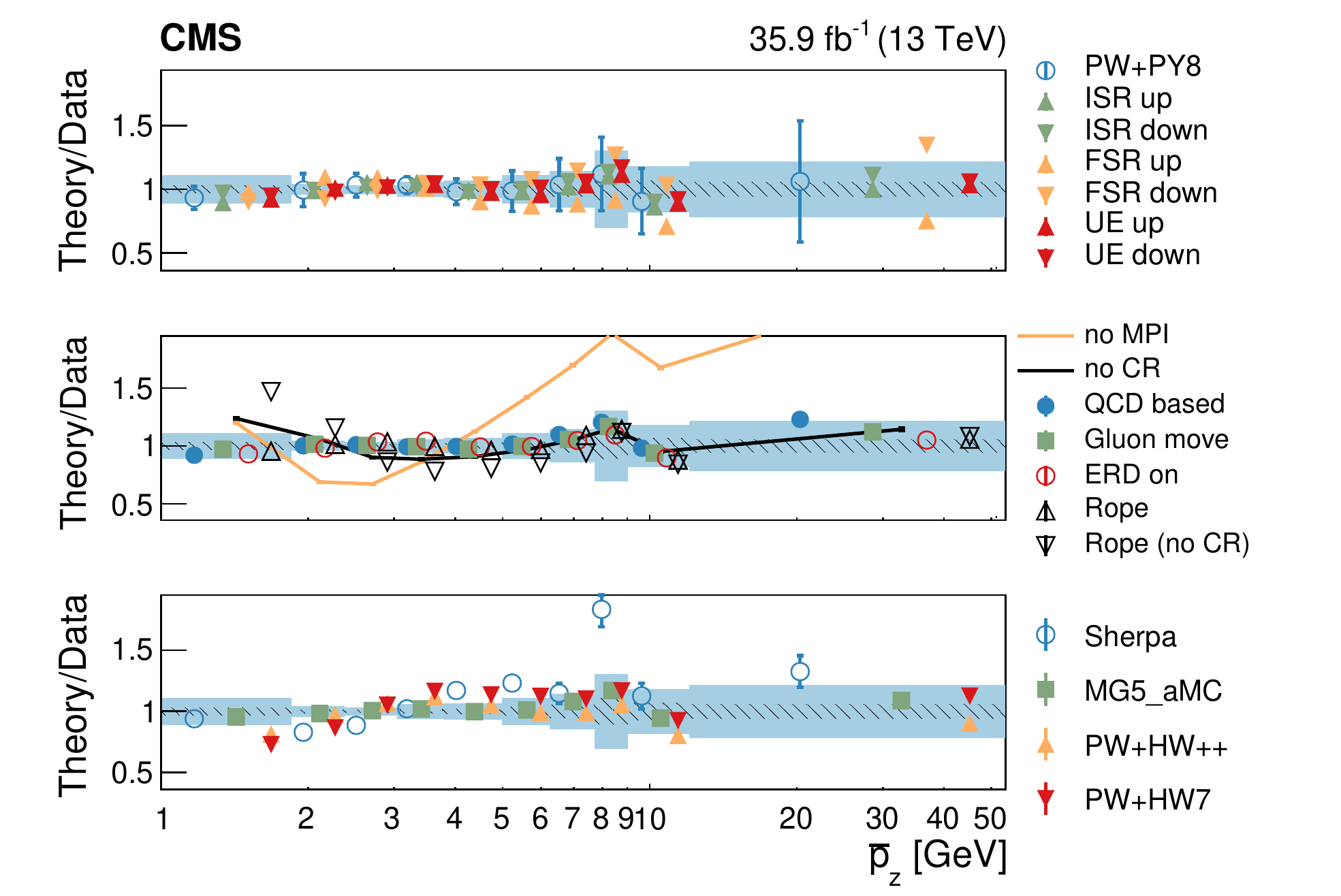}
\caption{
Normalized differential cross section as function of \avgpz{},
compared to the predictions of different models.
The conventions of Fig.~\ref{fig:chmult_unfolded} are used.
}
\label{fig:chavgpz_unfolded}
\end{figure*}

\begin{figure*}[!htp]
\centering
\includegraphics[width=\cmsFigWidthii]{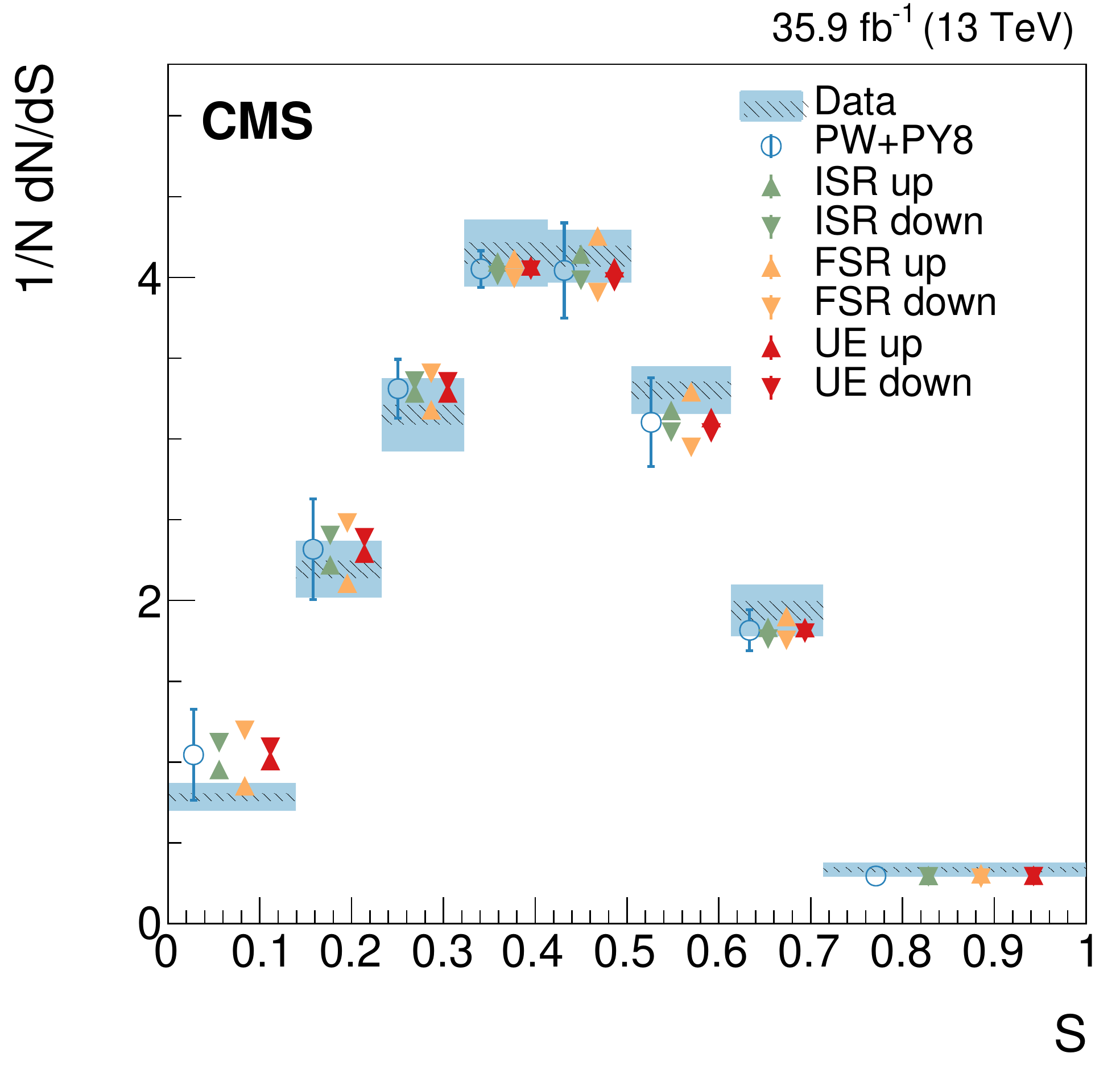}
\includegraphics[width=\cmsFigWidthi]{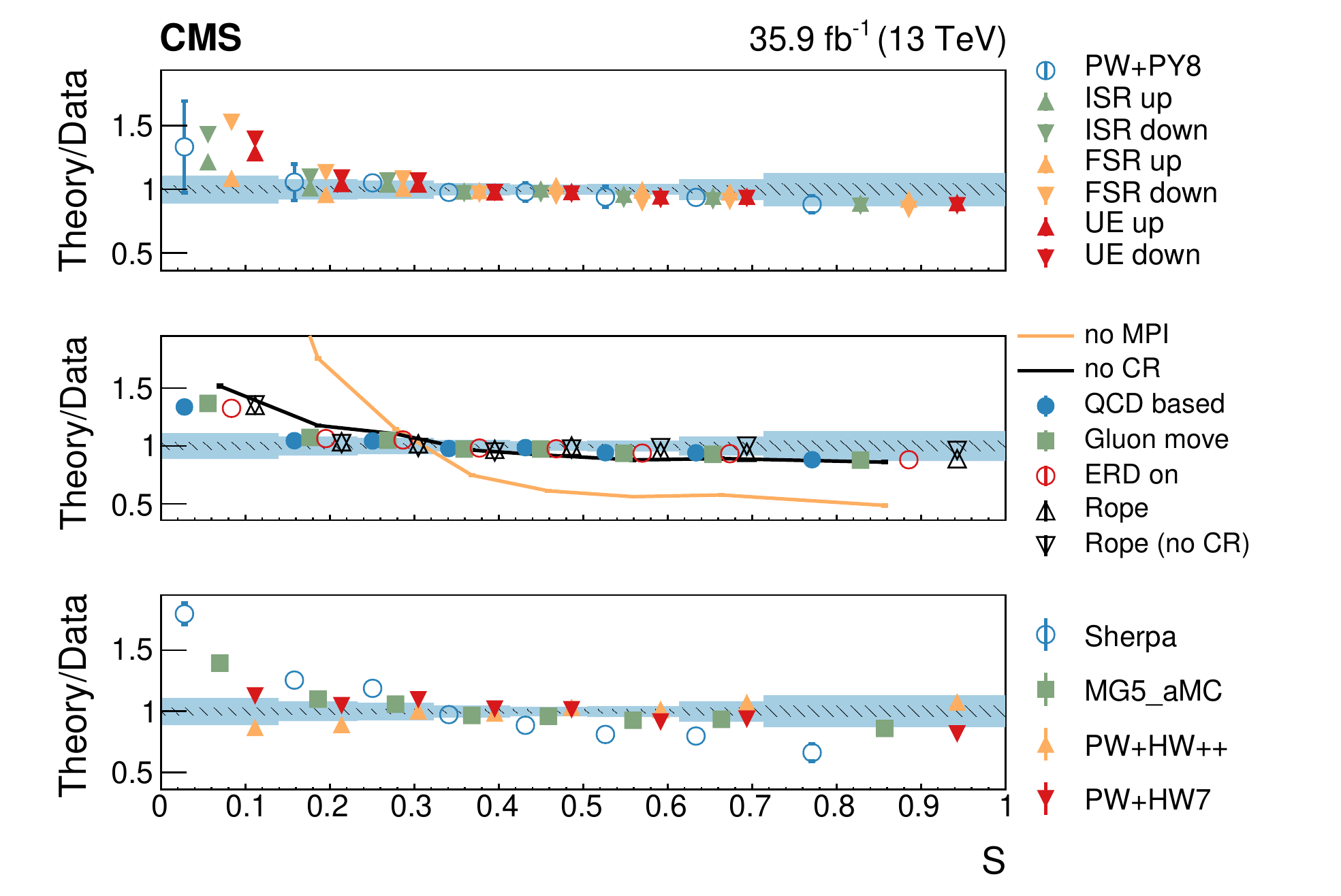}
\caption{
Normalized differential cross section as function of the sphericity variable,
compared to the predictions of different models.
The conventions of Fig.~\ref{fig:chmult_unfolded} are used.
}
\label{fig:sphericity_unfolded}
\end{figure*}

\begin{figure*}[!htp]
\centering
\includegraphics[width=\cmsFigWidthii]{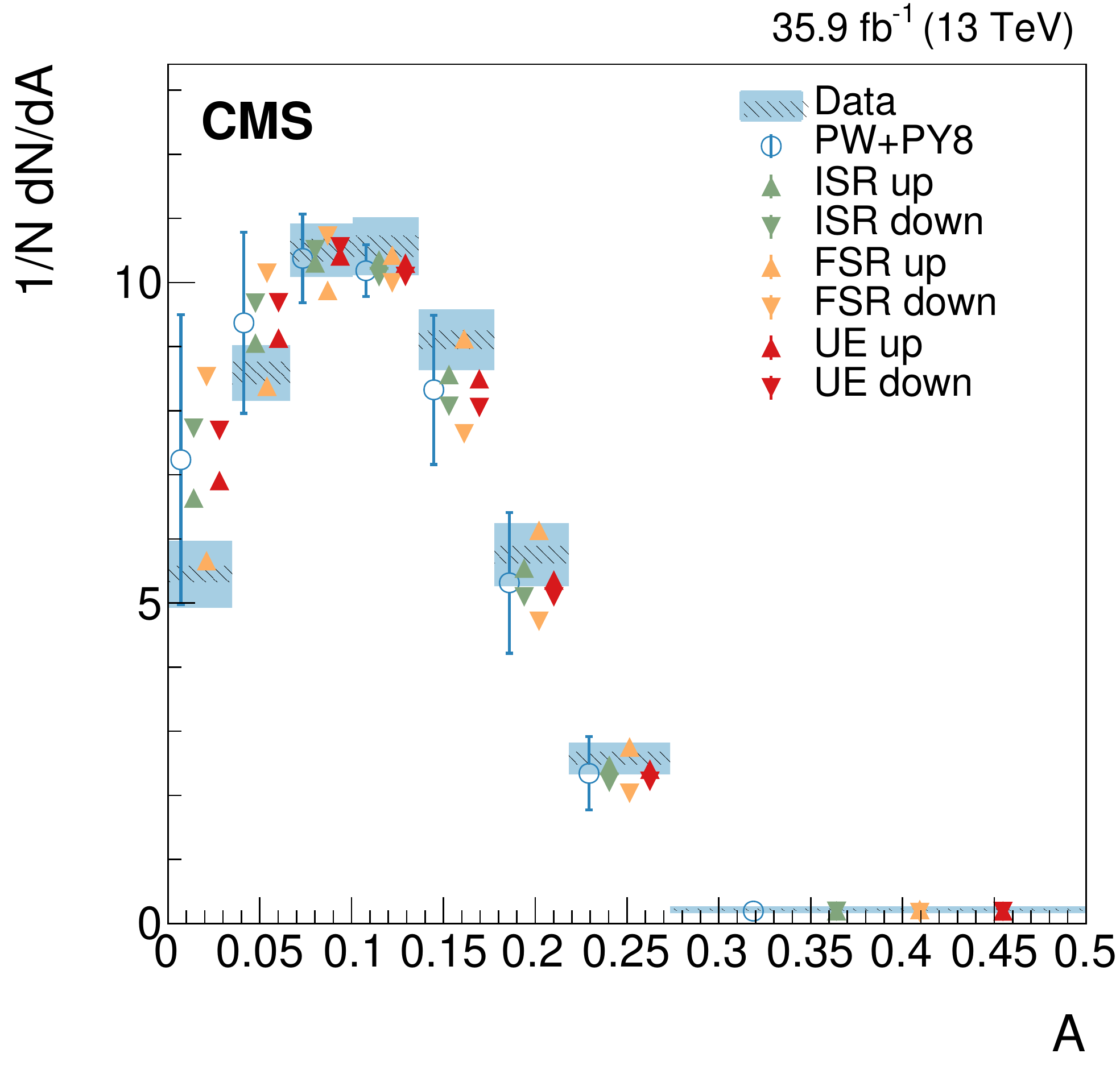}
\includegraphics[width=\cmsFigWidthi]{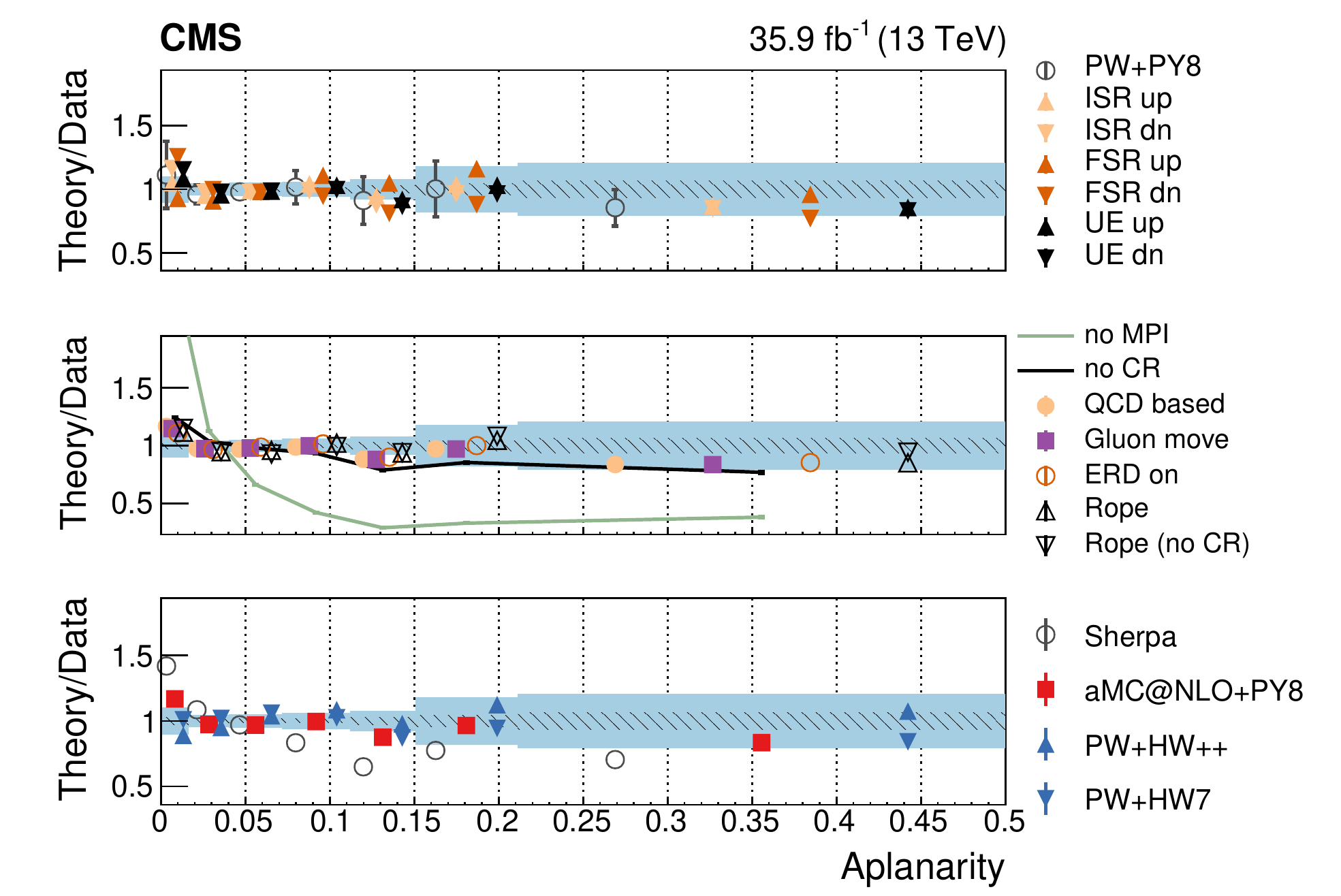}
\caption{
Normalized differential cross section as function of the aplanarity variable,
compared to the predictions of different models.
The conventions of Fig.~\ref{fig:chmult_unfolded} are used.
}
\label{fig:aplanarity_unfolded}
\end{figure*}

\begin{figure*}[!htp]
\centering
\includegraphics[width=\cmsFigWidthii]{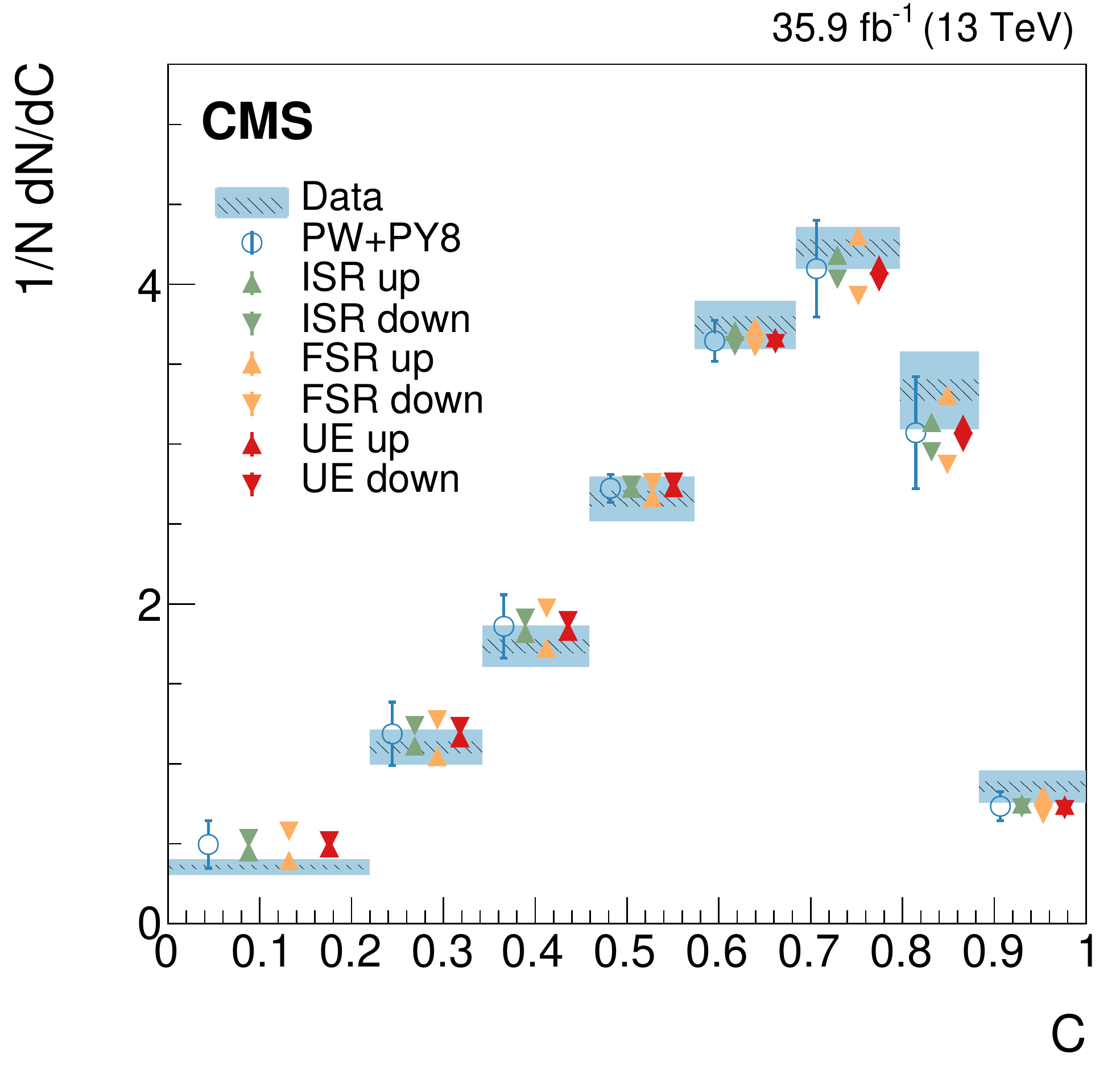}
\includegraphics[width=\cmsFigWidthi]{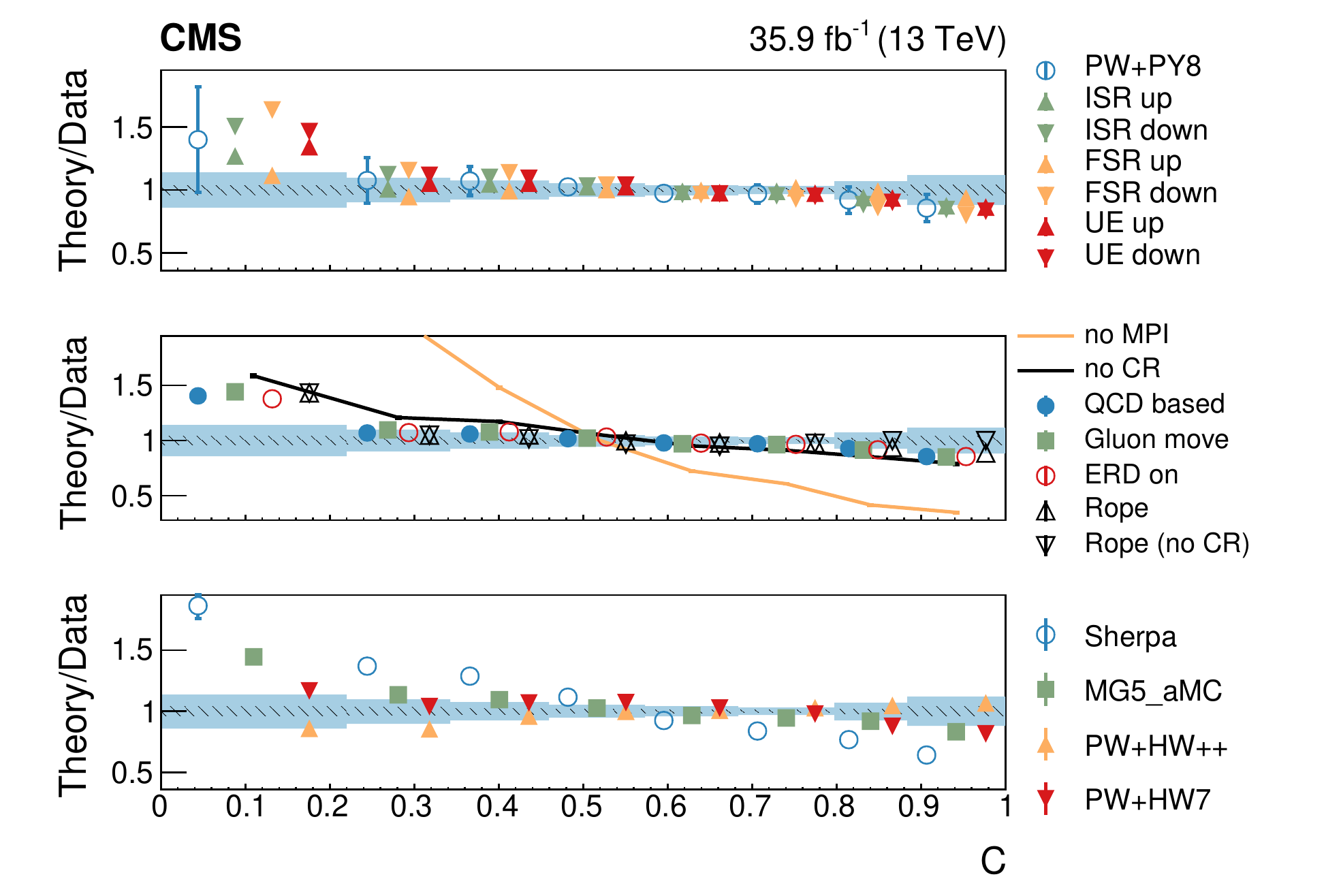}
\caption{
Normalized differential cross section as function of the $C$ variable,
compared to the predictions of different models.
The conventions of Fig.~\ref{fig:chmult_unfolded} are used.
}
\label{fig:C_unfolded}
\end{figure*}

\begin{figure*}[!htp]
\centering
\includegraphics[width=\cmsFigWidthii]{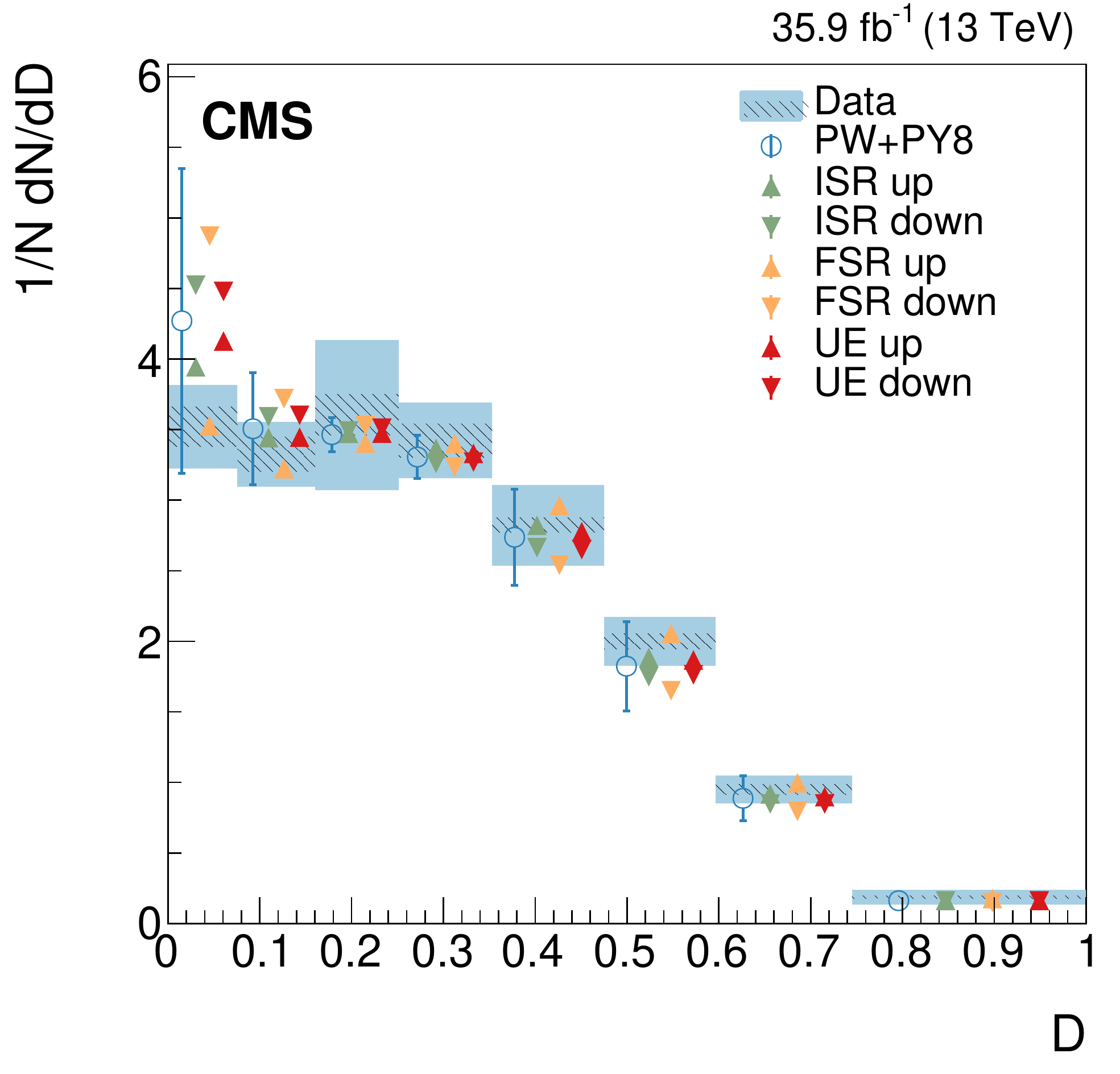}
\includegraphics[width=\cmsFigWidthi]{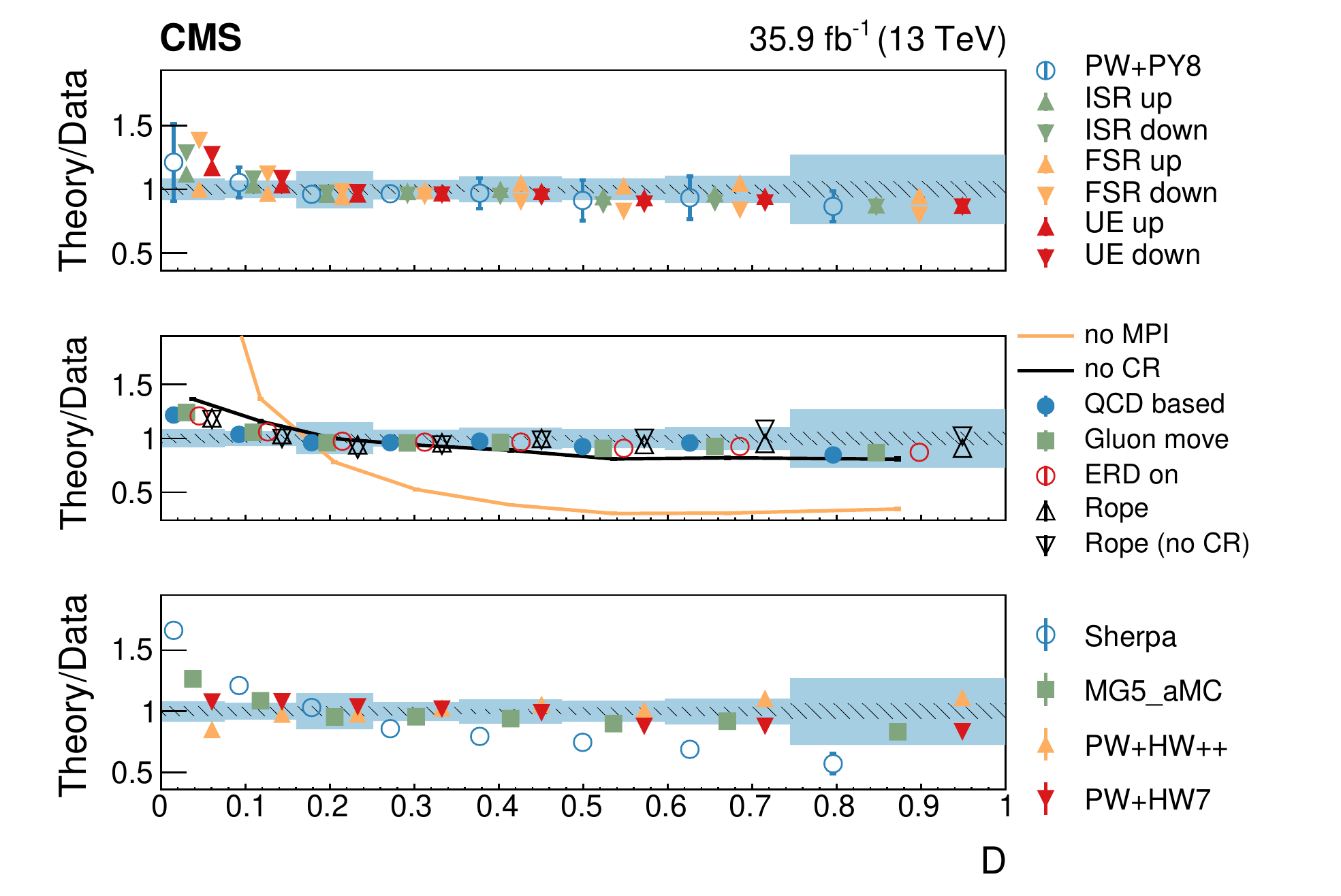}
\caption{
Normalized differential cross section as function of the $D$ variable,
compared to the predictions of different models.
The conventions of Fig.~\ref{fig:chmult_unfolded} are used.
}
\label{fig:D_unfolded}
\end{figure*}

The normalized differential cross sections measured as functions of
\nch, \sumpt, \avgpt, \chpt, \sumpz, \avgpz, sphericity, aplanarity, $C$, and $D$
are shown in Figs.~\ref{fig:chmult_unfolded}--\ref{fig:D_unfolded}, respectively.
The distributions are obtained after unfolding the background-subtracted data and normalizing the result to unity.
The result is compared to the simulations,
whose settings are summarized in Table~\ref{tab:mcsetups} and in the appendix.
For the predictions, the statistical uncertainty is represented as an error bar.
In the specific case of the {\pwpy} setup, the error bar represents the envelope
obtained by varying the main parameters of the CUETP8M2T4 tune, according to their uncertainties.
The envelope includes the variation of the CR model, $\alpS^\text{ISR}(M_\cPZ)$,
$\alpS^\text{FSR}(M_\cPZ)$, the $h_\text{damp}$ parameter,
and the $\mu_\mathrm{R}/\mu_\mathrm{F}$ scales at the ME level.
Thus, the uncertainty band represented for the {\pwpy} setup
should be interpreted as the theory uncertainty in that prediction.
For each distribution we give, in addition,
the ratio between different predictions and the data.

In \ttbar events the UE contribution is determined to have
typically $\mathcal{O}(20)$ charged particles with
$\avgpt\sim\avgpz\approx 2\GeV$, vectorially summing to a
recoil of about 10\GeV.
The distribution of the UE activity is anisotropic (as the sphericity is $<1$),
close to planar (as the aplanarity peaks at low values of $\approx$0.1),
and peaks at around 0.75 in the $C$ variable, which identifies three-jet topologies.
The $D$ variable, which identifies the four-jet topology, is instead found to have values closer to 0.
The three-prong configuration in the energy flux of the UE described by the $C$ variable
can be identified with two of the eigenvectors of the linearized sphericity tensor
being correlated with the direction of the \cPqb-tagged jets,
and the third one being determined by energy conservation.
When an extra jet with $\pt>30\GeV$ is selected, we measure a change
in the profile of the event shape variables, with average values lower by 20--40\%
with respect to the distributions in which no extra jet is found.
Thus when an extra jet is present,
the event has a dijet-like topology instead of an isotropic shape.

The results obtained with {\PYTHIA}8 for the parton shower simulation show negligible dependence
on the ME generator with which it is interfaced,
\ie, {\pwpy} and \amcpy{} yield similar results.
In all distributions the contribution from MPI is strong:
switching off this component in the simulation
has a drastic effect on the predictions of all the variables analyzed.
Color reconnection effects are more subtle to identify in the data.
In the inclusive distributions, CR effects are needed to improve the theory accuracy
for $\avgpt<3\GeV$ or $\avgpz<5\GeV$.
The differences between the CR models tested
(as discussed in detail in Section~\ref{sec:dataandsimulatedsamples}) are nevertheless small
and almost indistinguishable in the inclusive distributions.
In general the {\pwpy} setup is found to be in agreement with the data,
when the total theory uncertainty is taken into account.
In most of the distributions it is the variation of $\alpS^\text{FSR}(M_\cPZ)$ that
dominates the theory uncertainty, as this variation
leads to the most visible changes in the UE.
The other parton shower setups tested do not describe the data as accurately, but they were not tuned to the same level of detail as {\pwpy}.
The {\pwhwpp} and {\pwhw}-based setups show distinct trends with respect to the data
from those observed in any of the {\PYTHIA}8-based setups.
While describing fairly well the UE event shape variables, \HERWIGpp and \HERWIG7
disagree with the \nch, \avgpt{}, and {\avgpz} measurements.
The \SHERPA predictions disagree with data in most of the observables.

For each distribution the level of agreement between theory predictions and data
is quantified by means of a $\chi^2$ variable defined as:
\begin{equation}
\chi^2= \sum_{i,j=1}^{n}\delta y_i\,(\text{Cov}^{-1})_{ij}\,\delta y_j,
\label{eq:chi2}
\end{equation}
where $\delta y_i$ ($\delta y_j$) are the differences between the data and the model in the $i$-th ($j$-th) bin;
here $n$ represents the total number of bins, and $\text{Cov}^{-1}$ is the inverse of the covariance matrix.
Given that the distributions are normalized,
the covariance matrix becomes invertible after removing its first row and column.
In the calculation of Eq.~\ref{eq:chi2} we assume that the theory uncertainties
are uncorrelated with those assigned to the measurements.
Table~\ref{tab:chi2_summary} summarizes the values obtained for the main models
with respect to each of the distributions
shown in Figs.~\ref{fig:chmult_unfolded}--\ref{fig:D_unfolded}.
The values presented in the table quantify the level of agreement of each model with the measurements.
Low $\chi^2$ per number of degrees of freedom (dof) values are obtained for the  {\pwpy} setup
when all the theory uncertainties of the model are taken into account,
in particular for the event shape variables.
This indicates that the theory uncertainty envelope is conservative.

\begin{table*}[ht]
\centering
\topcaption{
Comparison between the measured distributions at particle level and the predictions of different generator setups.
We list the results of the $\chi^2$ tests together with dof.
For the comparison no uncertainties in the predictions are taken into account,
except for the {\pwpy} setup for which the comparison including the theoretical uncertainties is quoted separately in parenthesis.
}
\label{tab:chi2_summary}
\begin{tabular}{lccccc}
\multirow{2}{*}{Observable} & \multicolumn{5}{c}{$\chi^2/$dof} \\
           & \pwpy        & \pwhwpp & \pwhw & \amcpy & \SHERPA \\
\hline
\nch       & 30/11 (15/11) & 33/11  & 17/11 & 34/11 & 95/11 \\
\sumpt     & 24/13 (13/13) & 129/13 & 56/13 & 30/13 & 37/13 \\
\sumpz     & 8/11 (4/11)   & 34/11  & 20/11 & 9/11  & 18/11 \\
\avgpt     & 12/9 (1/9)    & 40/9   & 56/9  & 6/9   & 56/9 \\
\avgpz     & 2/9 (1/9)     & 9/9    & 32/9  & 1/9   & 36/9 \\
\chpt      & 17/11 (7/11)  & 102/11 & 49/11 & 20/11 & 34/11 \\
$S$        & 29/7 (3/7)    & 7/7    & 17/7  & 36/7  & 194/7 \\
$A$        & 18/7 (1/7)    & 8/7    & 13/7  & 26/7  & 167/7 \\
$C$        & 34/7 (4/7)    & 7/7    & 27/7  & 38/7  & 187/7 \\
$D$        & 7/7 (1/7)     & 5/7    & 8/7   & 11/7  & 83/7 \\
\end{tabular}
\end{table*}

\subsection{Profile of the UE in different categories}
\label{subsec:ueprofiles}

\begin{figure*}[!htp]
\centering
\includegraphics[width=\cmsFigWidthiv]{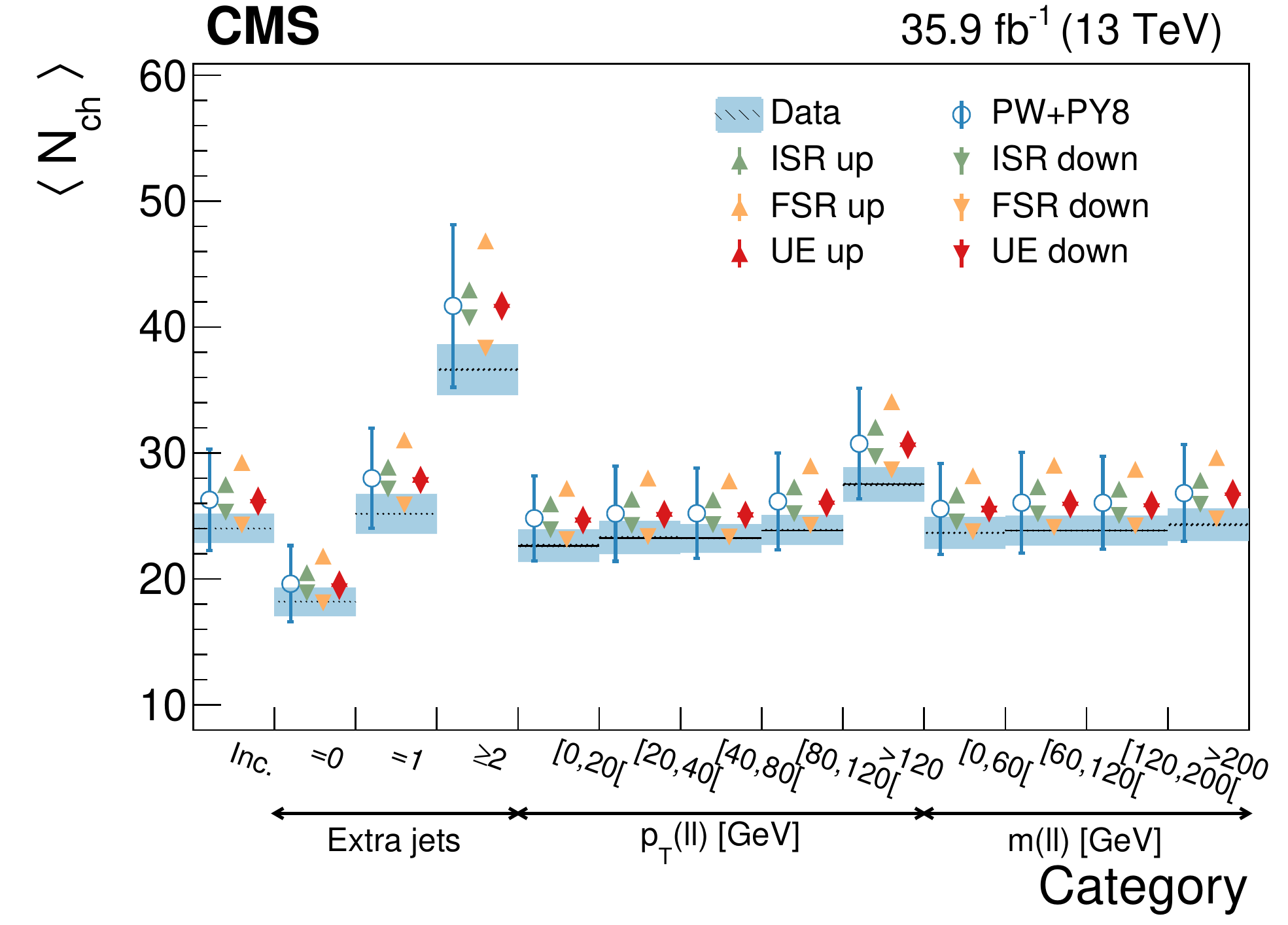}\\
\includegraphics[width=\cmsFigWidthi]{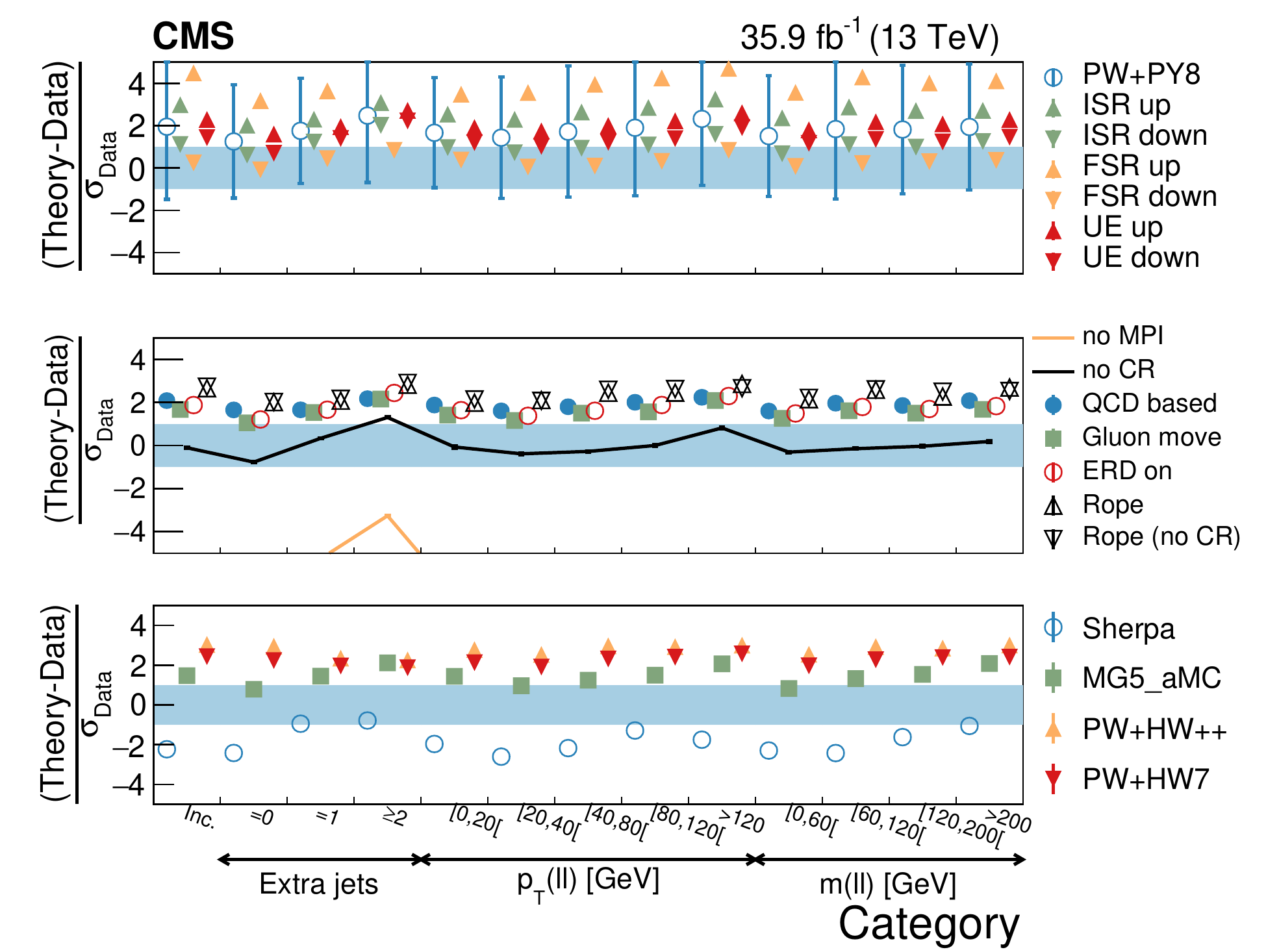}
\caption{
Average \nch{} in different event categories.
The mean observed in data (boxes) is compared to
the predictions from different models (markers), which are superimposed in the upper figure.
The total (statistical) uncertainty of the data is represented by a shaded (hatched) area
and the statistical uncertainty of the models is represented with error bars.
In the specific case of the {\pwpy} model the error bars represent the total uncertainty (see text).
The lower figure displays the pull between different models and the data,
with the different panels corresponding to different sets of models.
The bands represent the interval where $\abs{\text{pull}}<1$.
The error bar for the {\pwpy} model represents the range of variation of the pull
for the different configurations described in the text.
}
\label{fig:ueprofile_chmult}
\end{figure*}

\begin{figure*}[!htp]
\centering
\includegraphics[width=\cmsFigWidthiv]{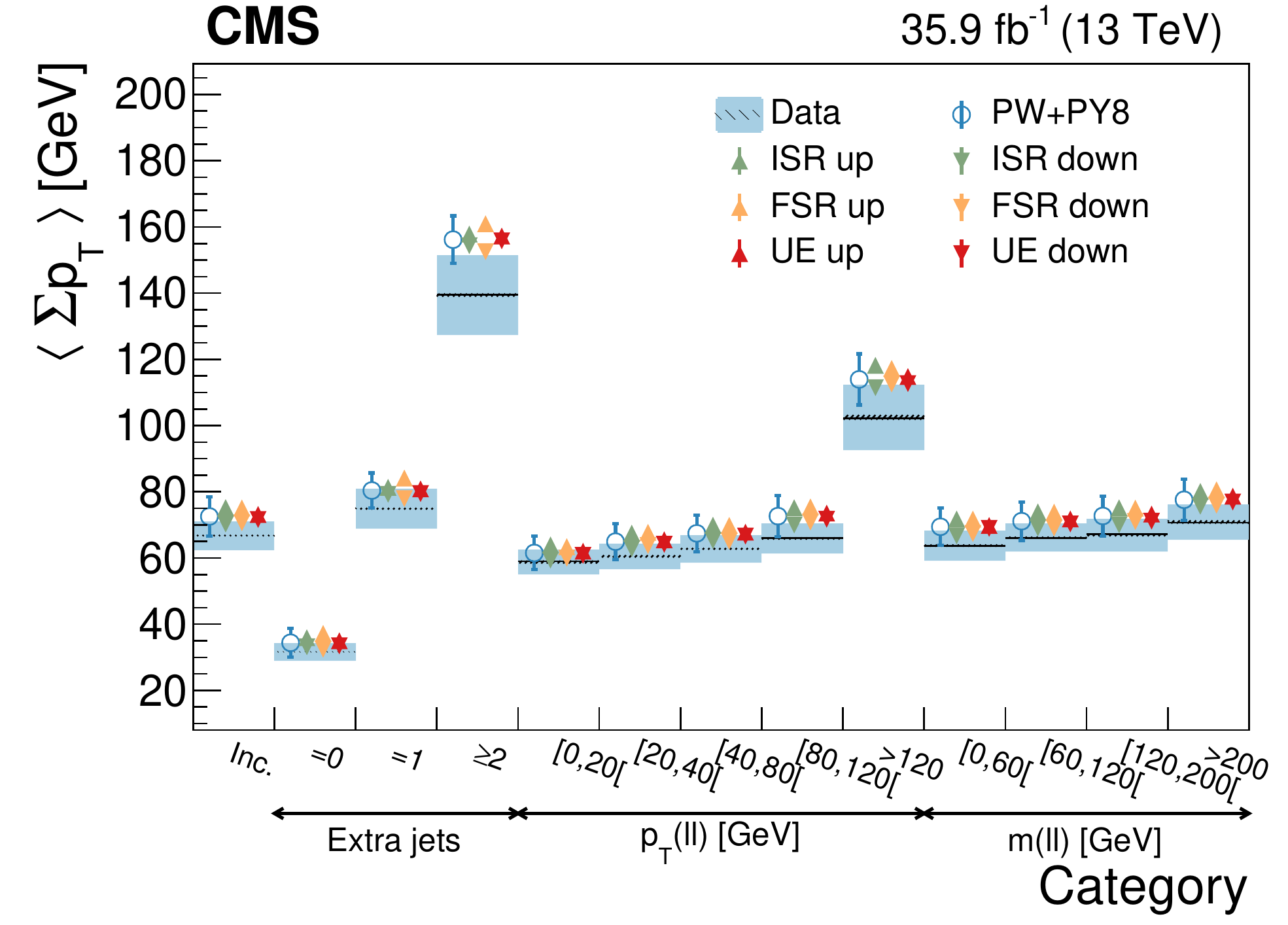}\\
\includegraphics[width=\cmsFigWidthi]{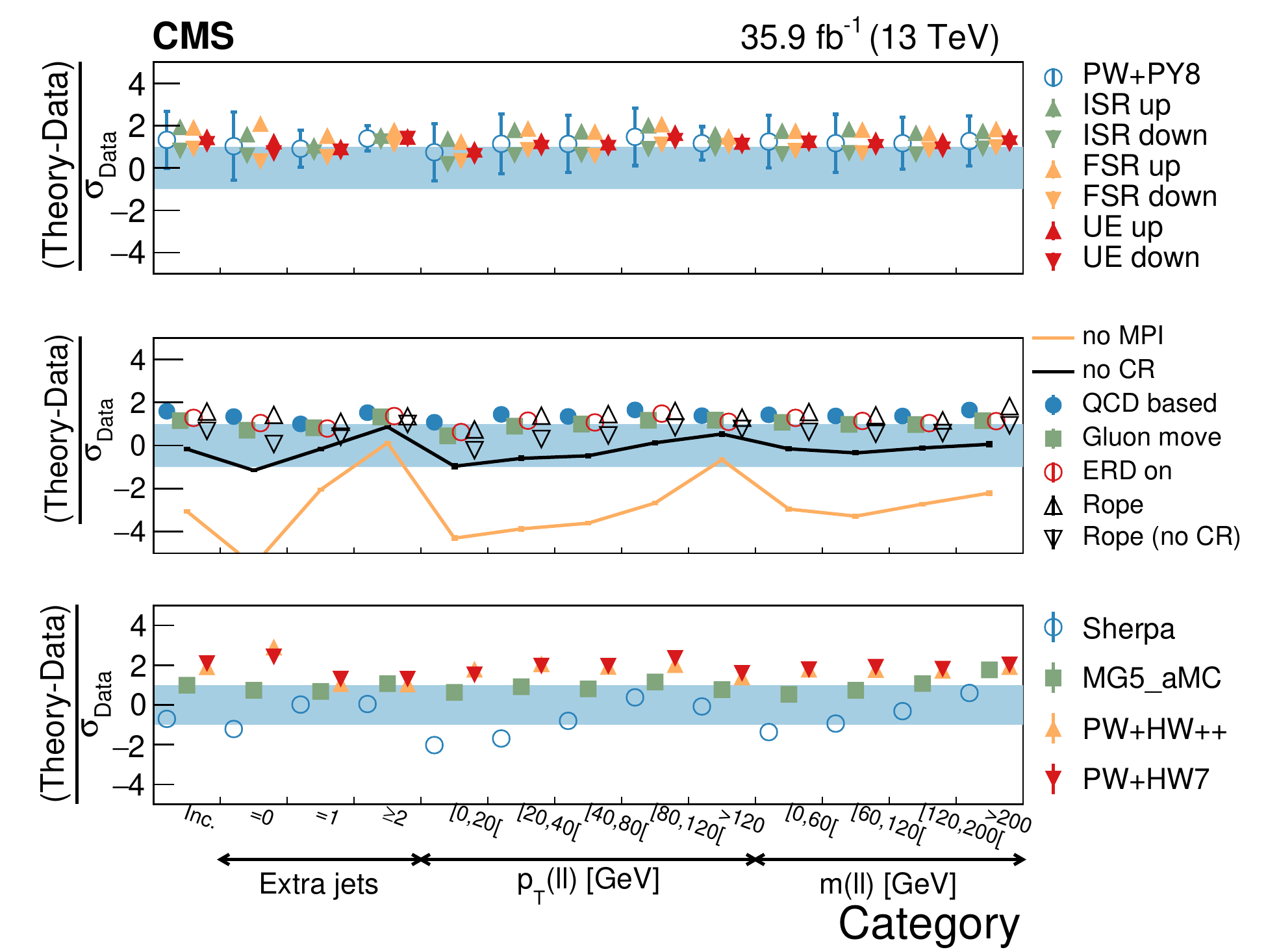}
\caption{
Average \sumpt{} in different event categories.
The conventions of Fig.~\ref{fig:ueprofile_chmult} are used.
}
\label{fig:ueprofile_chflux}
\end{figure*}

\begin{figure*}[!htp]
\centering
\includegraphics[width=\cmsFigWidthiv]{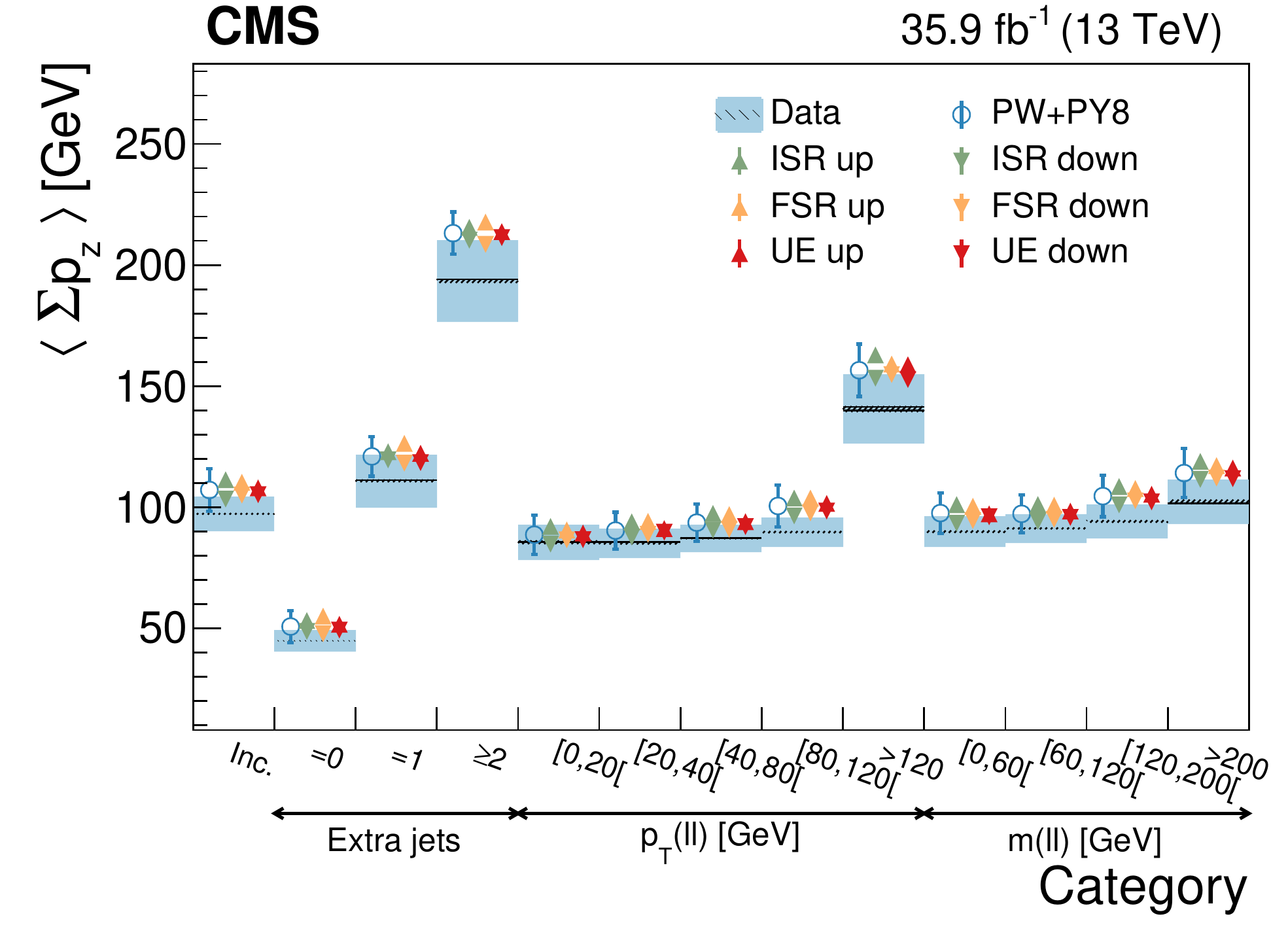}\\
\includegraphics[width=\cmsFigWidthi]{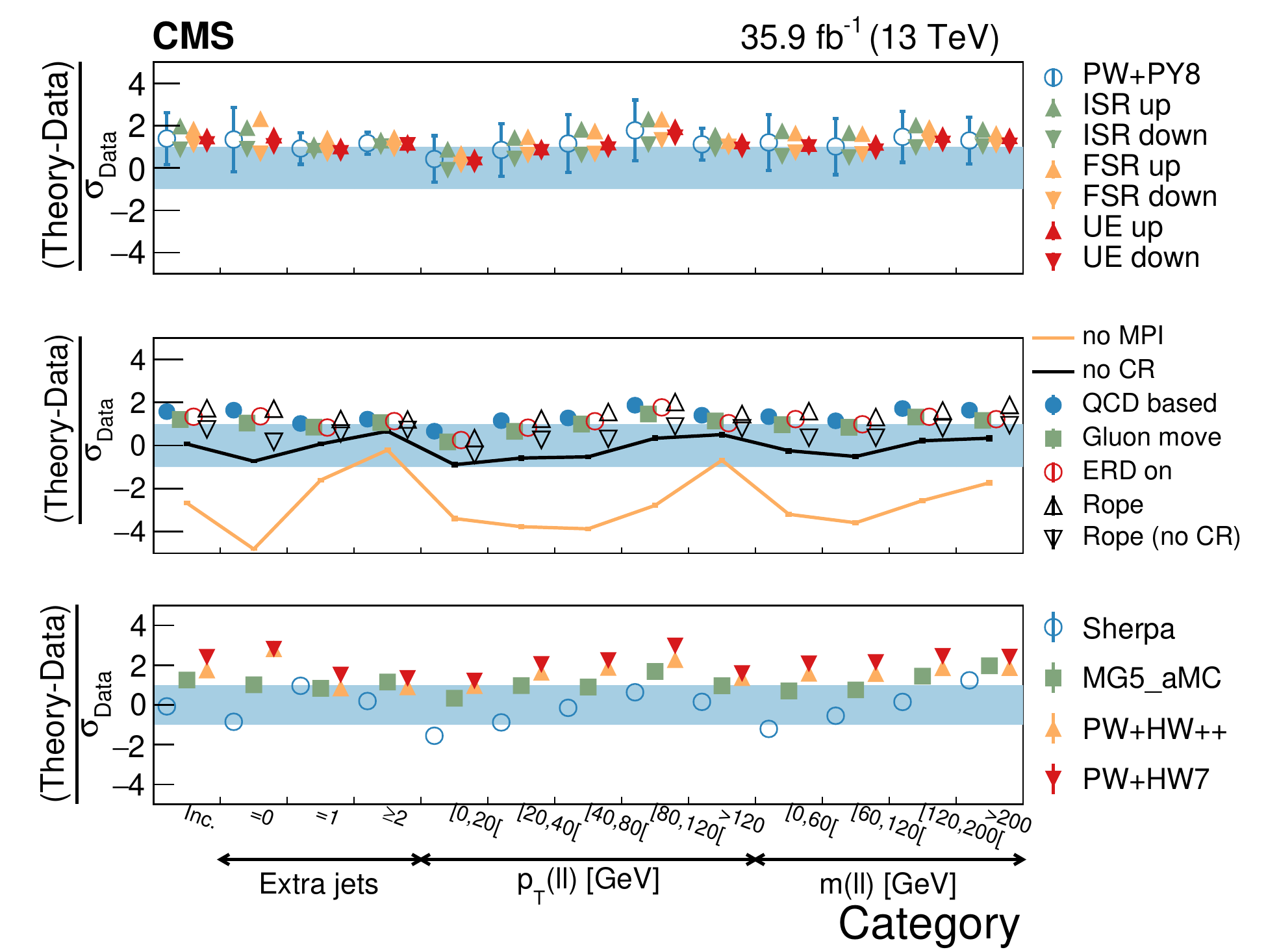}
\caption{
Average \sumpz{} in different categories.
The conventions of Fig.~\ref{fig:ueprofile_chmult} are used.
}
\label{fig:ueprofile_chfluxz}
\end{figure*}

\begin{figure*}[!htp]
\centering
\includegraphics[width=\cmsFigWidthiv]{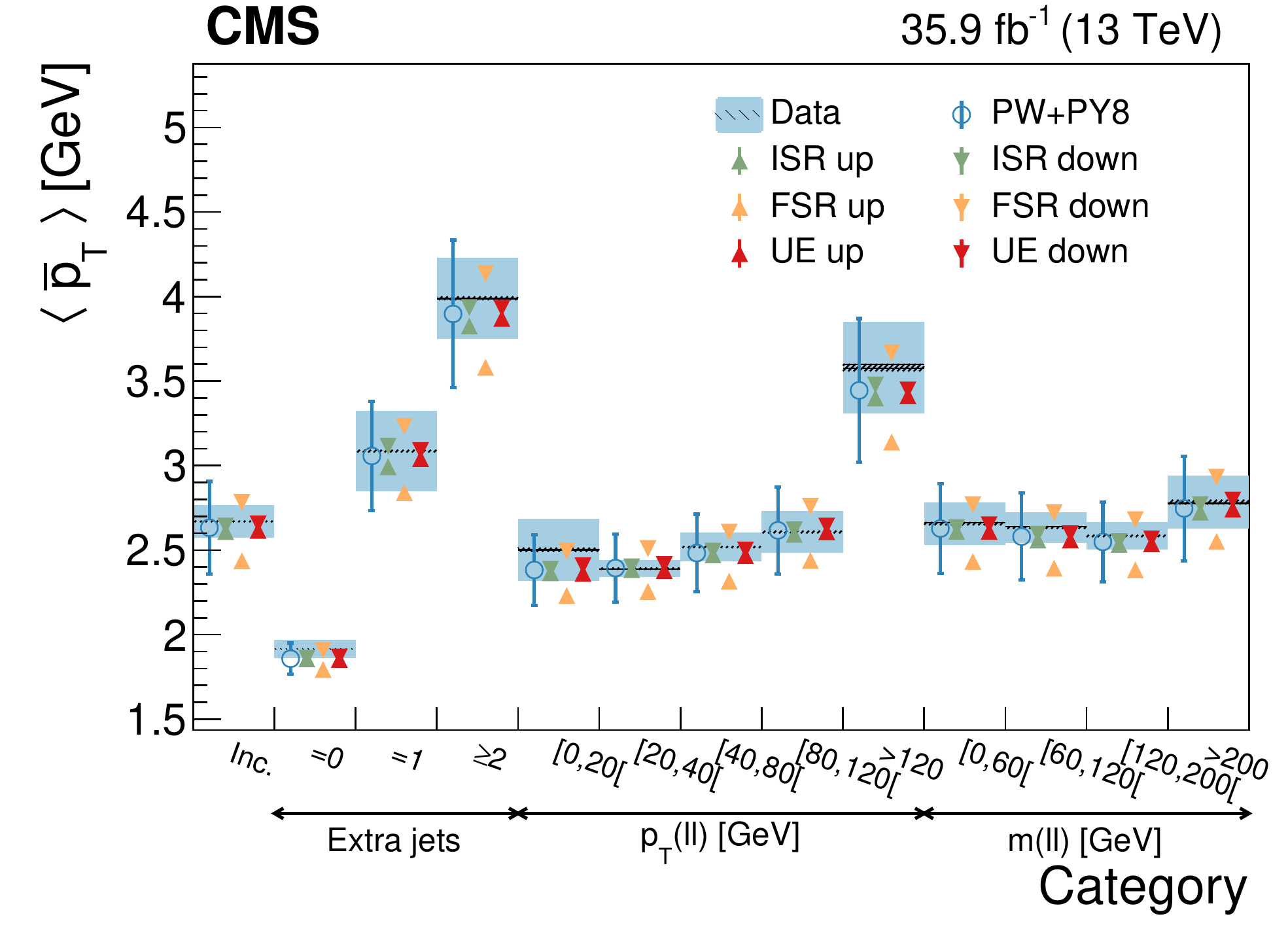}\\
\includegraphics[width=\cmsFigWidthi]{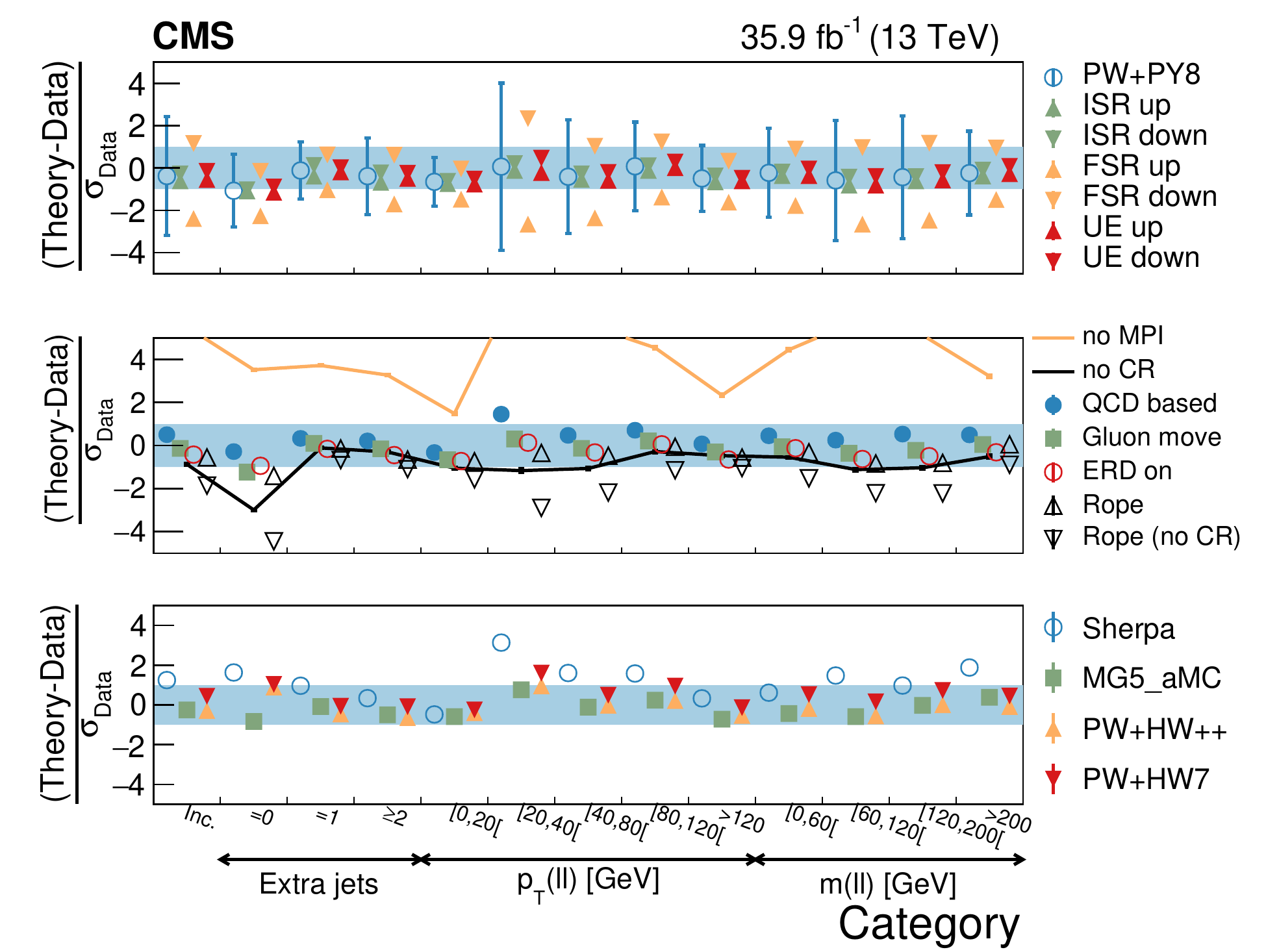}
\caption{
Average {\avgpt} in different categories.
The conventions of Fig.~\ref{fig:ueprofile_chmult} are used.
}
\label{fig:ueprofile_chavgpt}
\end{figure*}

\begin{figure*}[!htp]
\centering
\includegraphics[width=\cmsFigWidthiv]{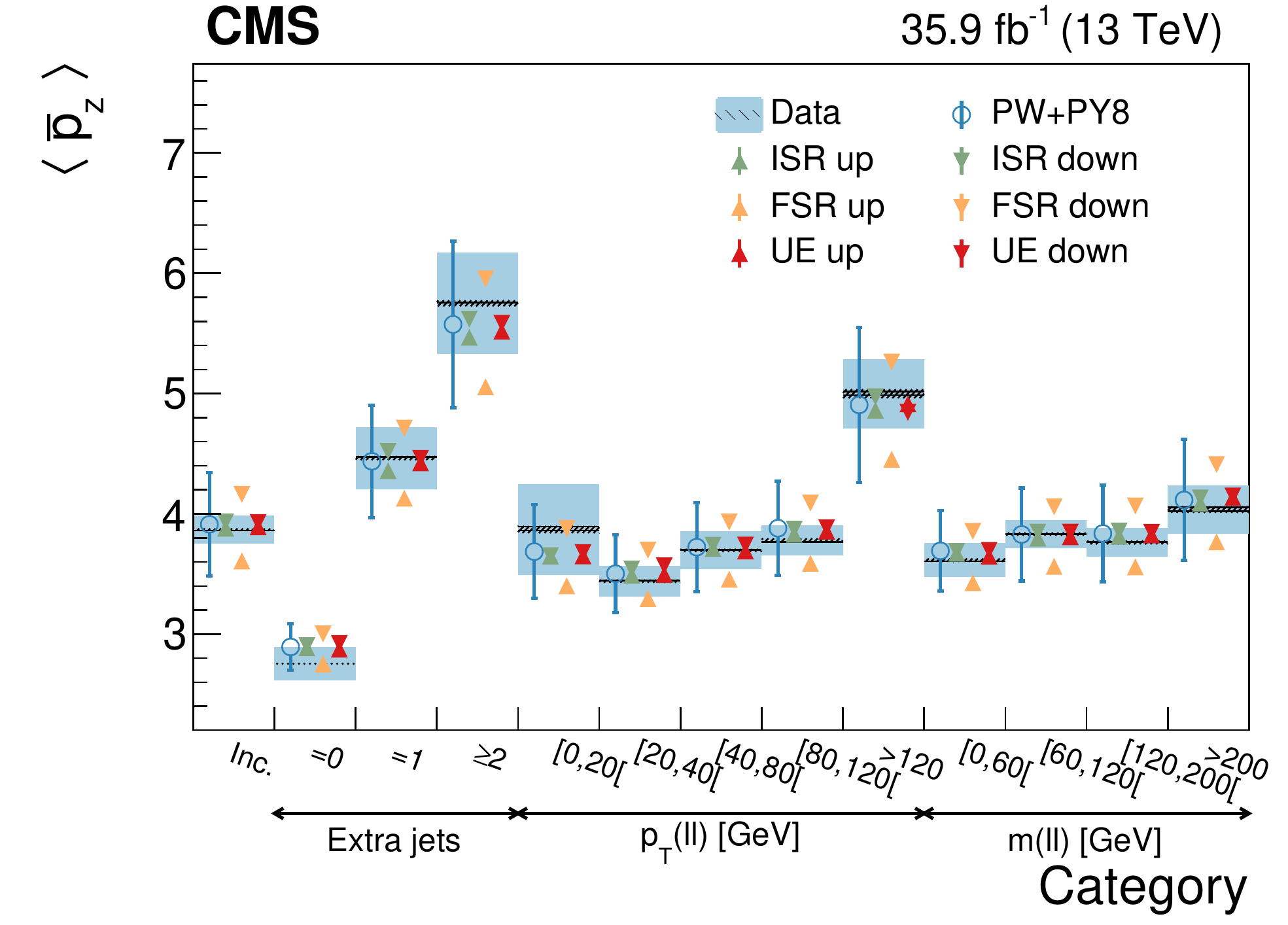}\\
\includegraphics[width=\cmsFigWidthi]{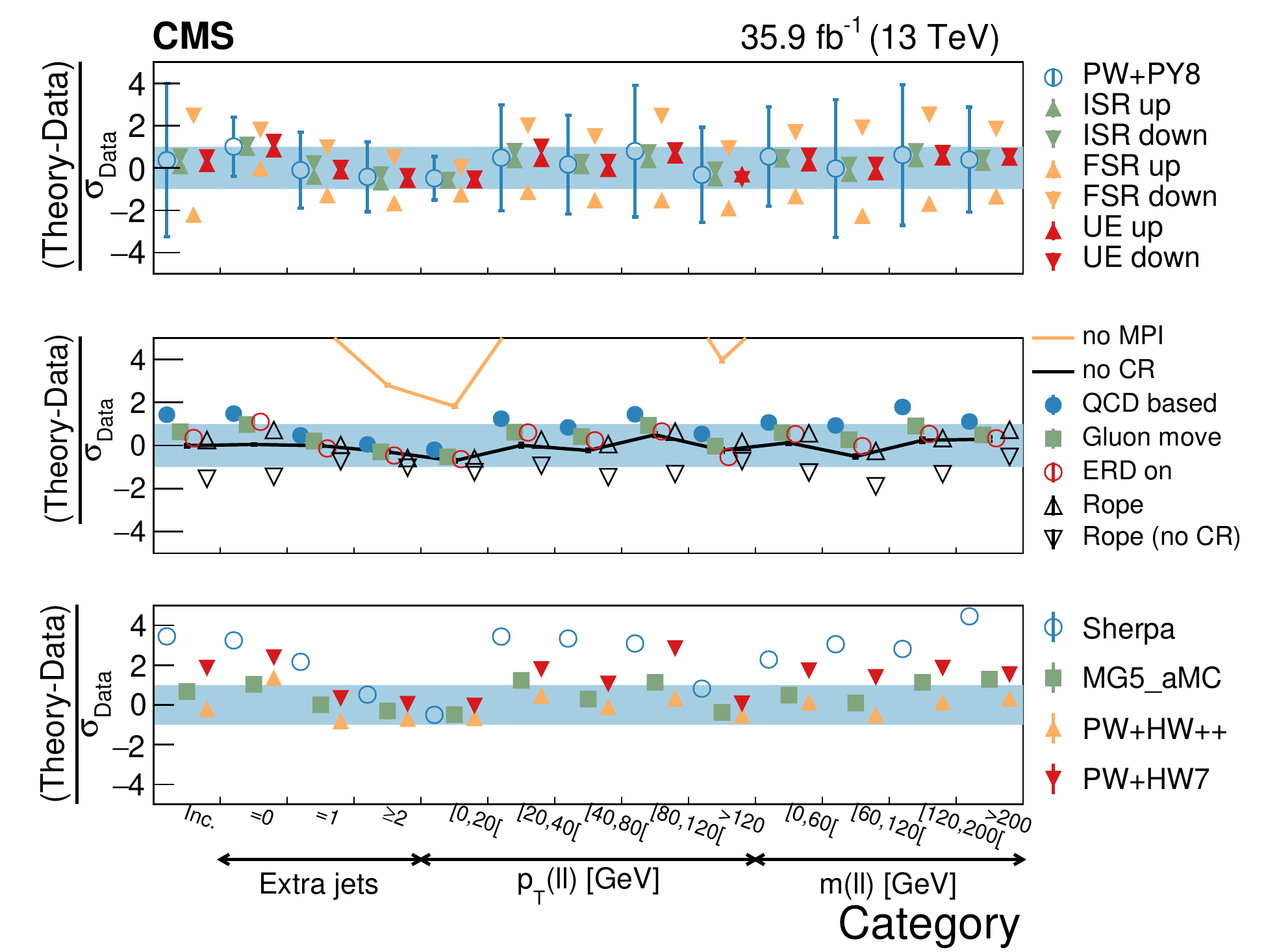}
\caption{
Average {\avgpz} in different categories.
The conventions of Fig.~\ref{fig:ueprofile_chmult} are used.
}
\label{fig:ueprofile_chavgpz}
\end{figure*}

\begin{figure*}[!htp]
\centering
\includegraphics[width=\cmsFigWidthiv]{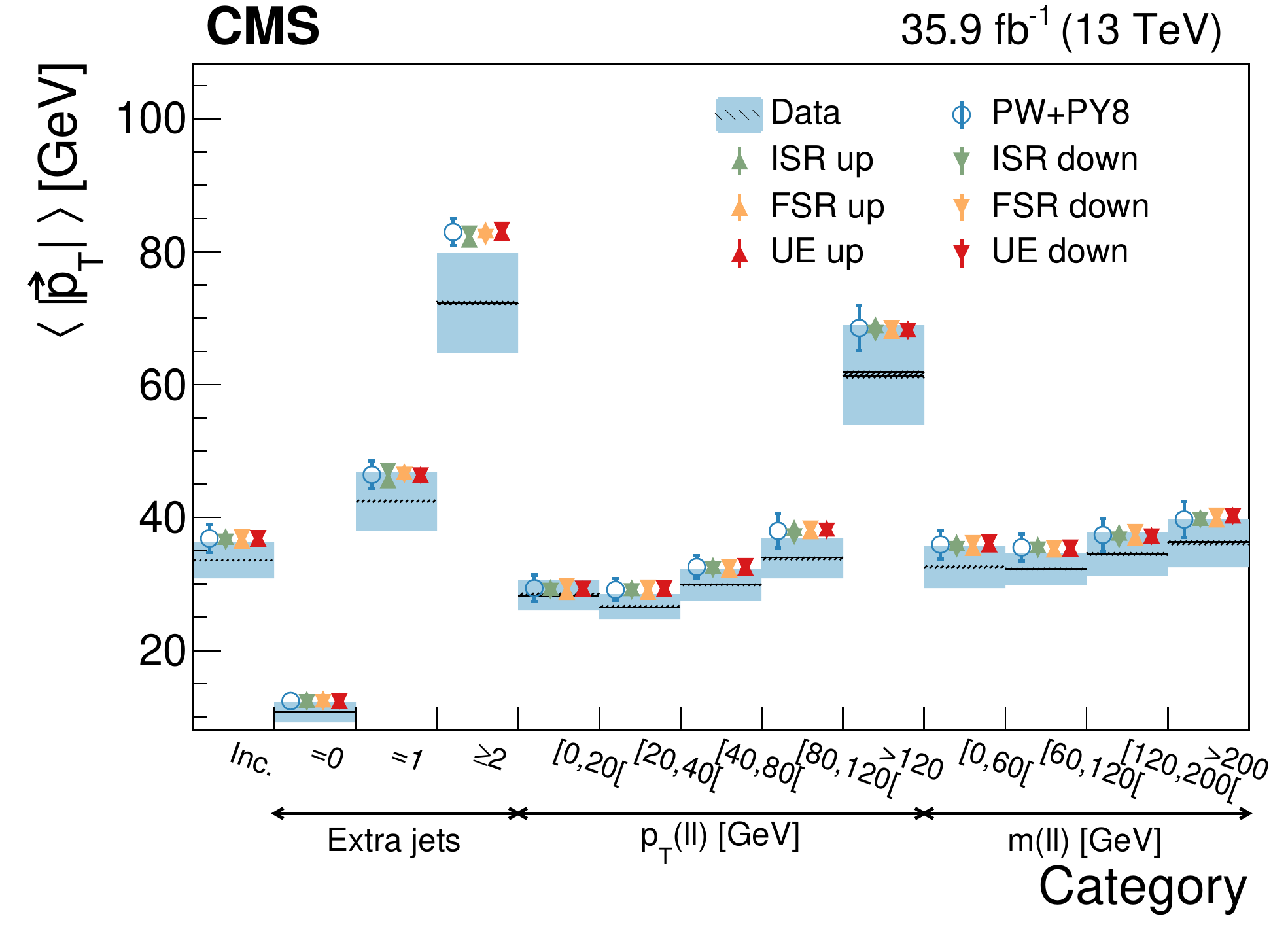}\\
\includegraphics[width=\cmsFigWidthi]{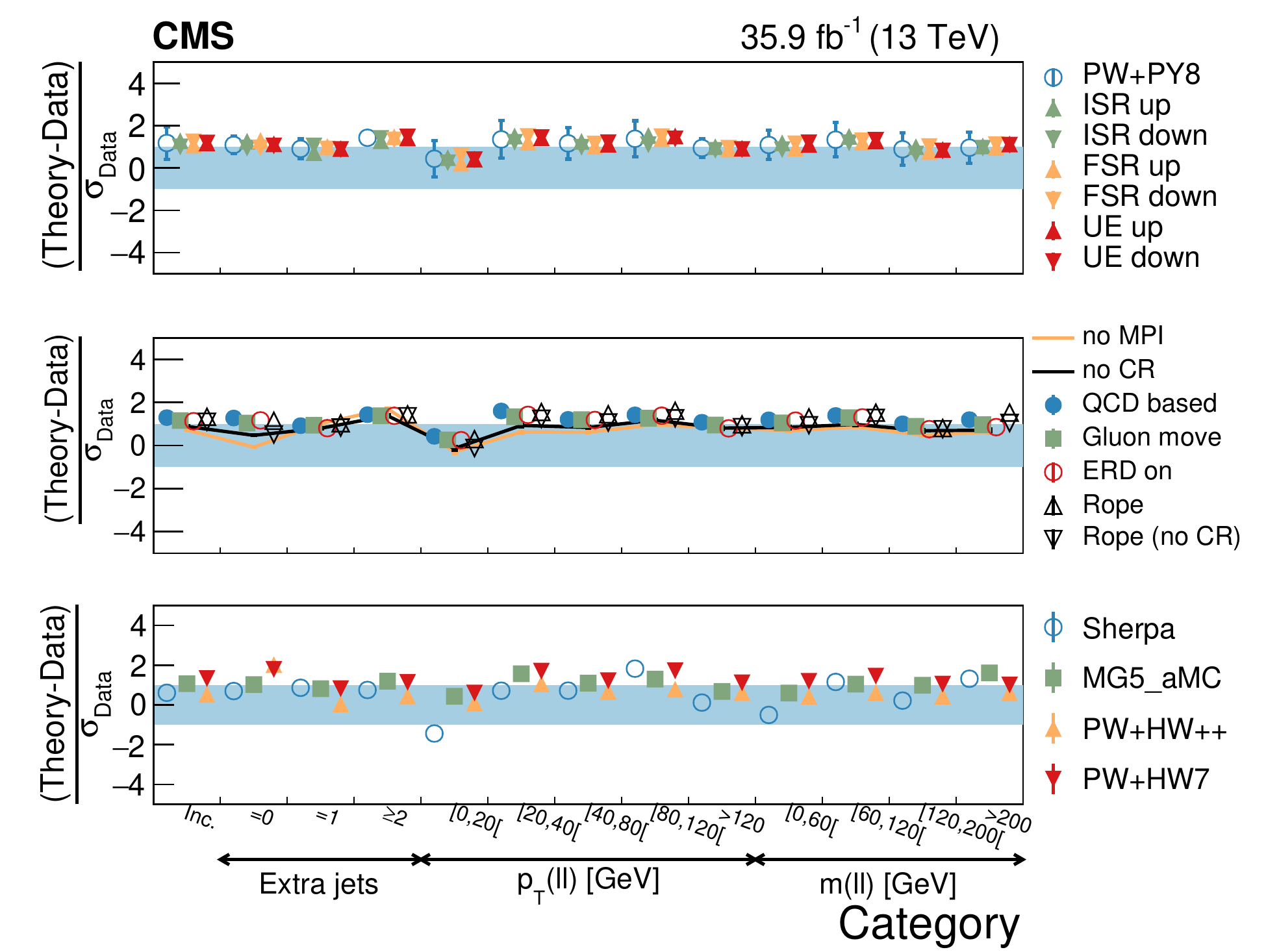}
\caption{
Average \chpt{} in different categories.
The conventions of Fig.~\ref{fig:ueprofile_chmult} are used.
}
\label{fig:ueprofile_chavgpz2}
\end{figure*}

\begin{figure*}[!htp]
\centering
\includegraphics[width=\cmsFigWidthiv]{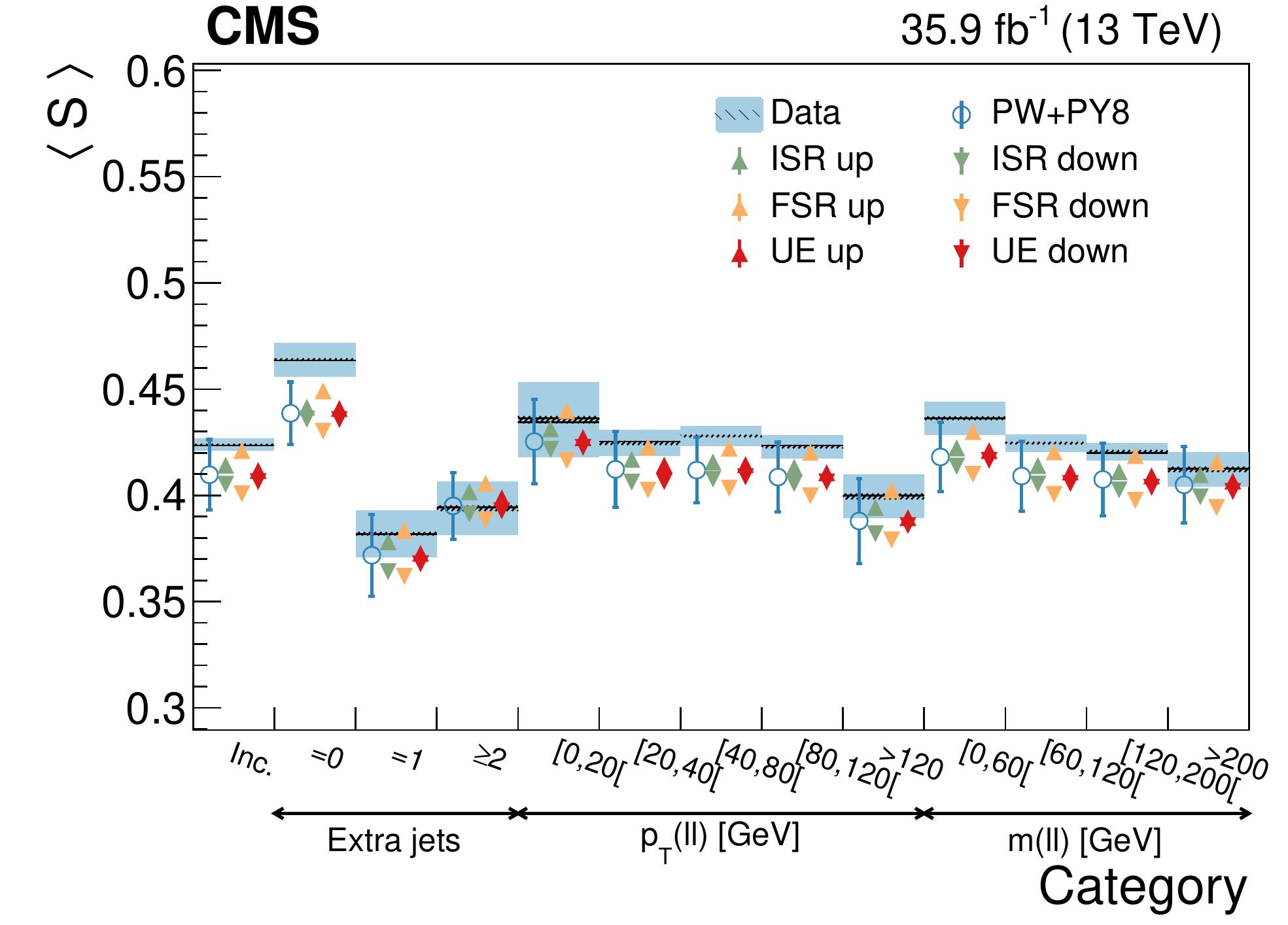}\\
\includegraphics[width=\cmsFigWidthi]{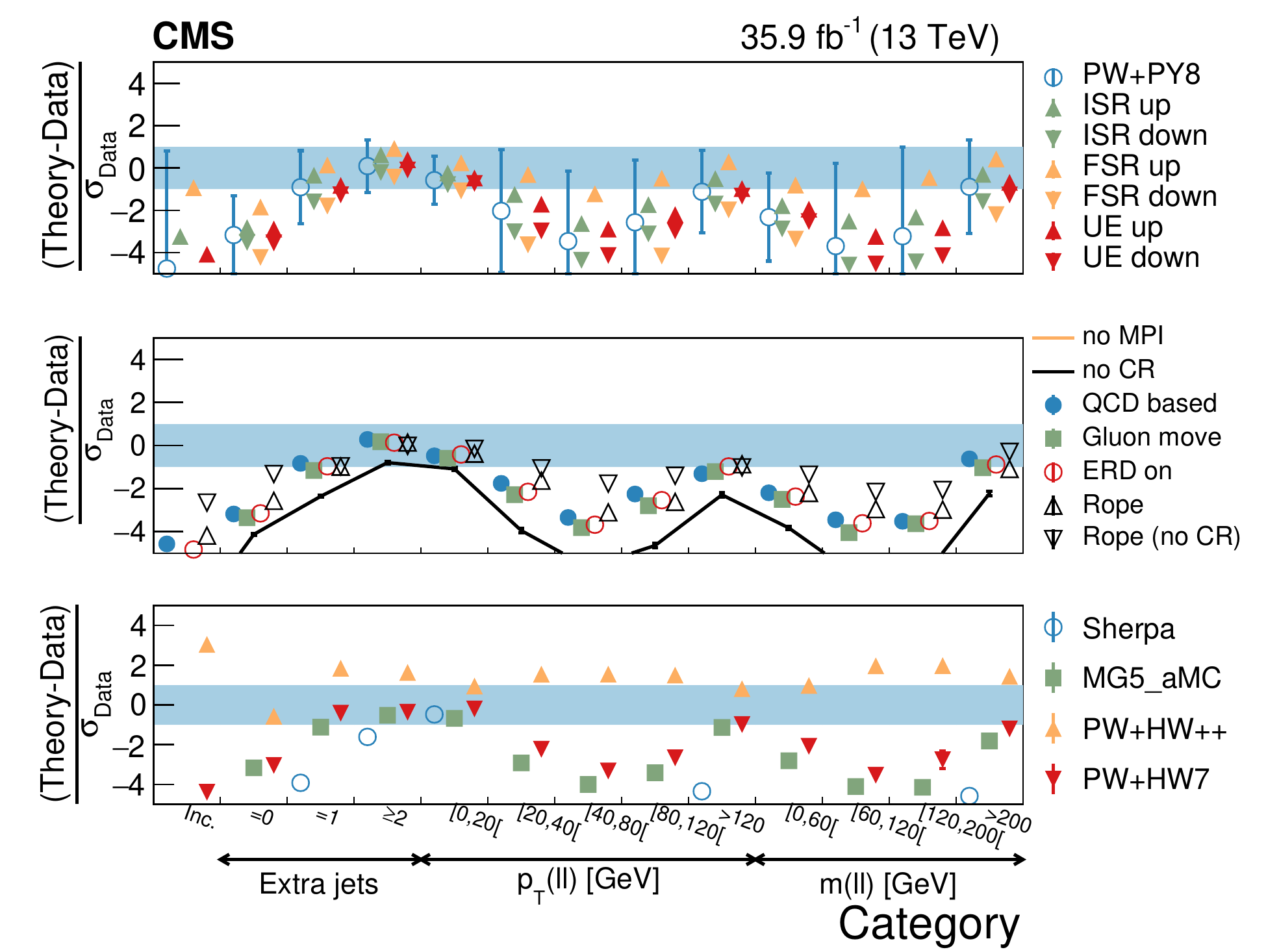}
\caption{
Average sphericity in different categories.
The conventions of Fig.~\ref{fig:ueprofile_chmult} are used.
}
\label{fig:ueprofile_sphericity}
\end{figure*}

\begin{figure*}[!htp]
\centering
\includegraphics[width=\cmsFigWidthiv]{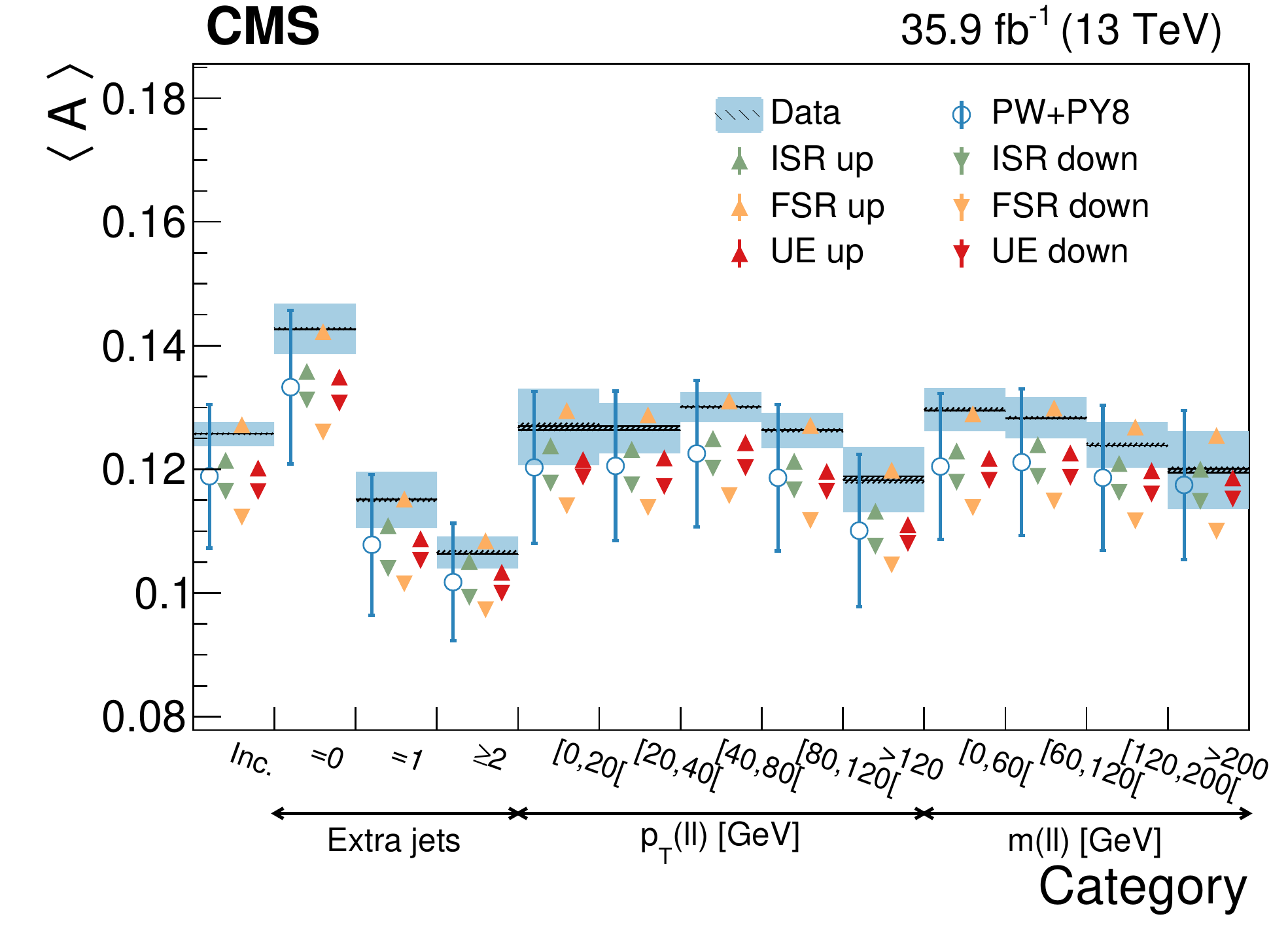}\\
\includegraphics[width=\cmsFigWidthi]{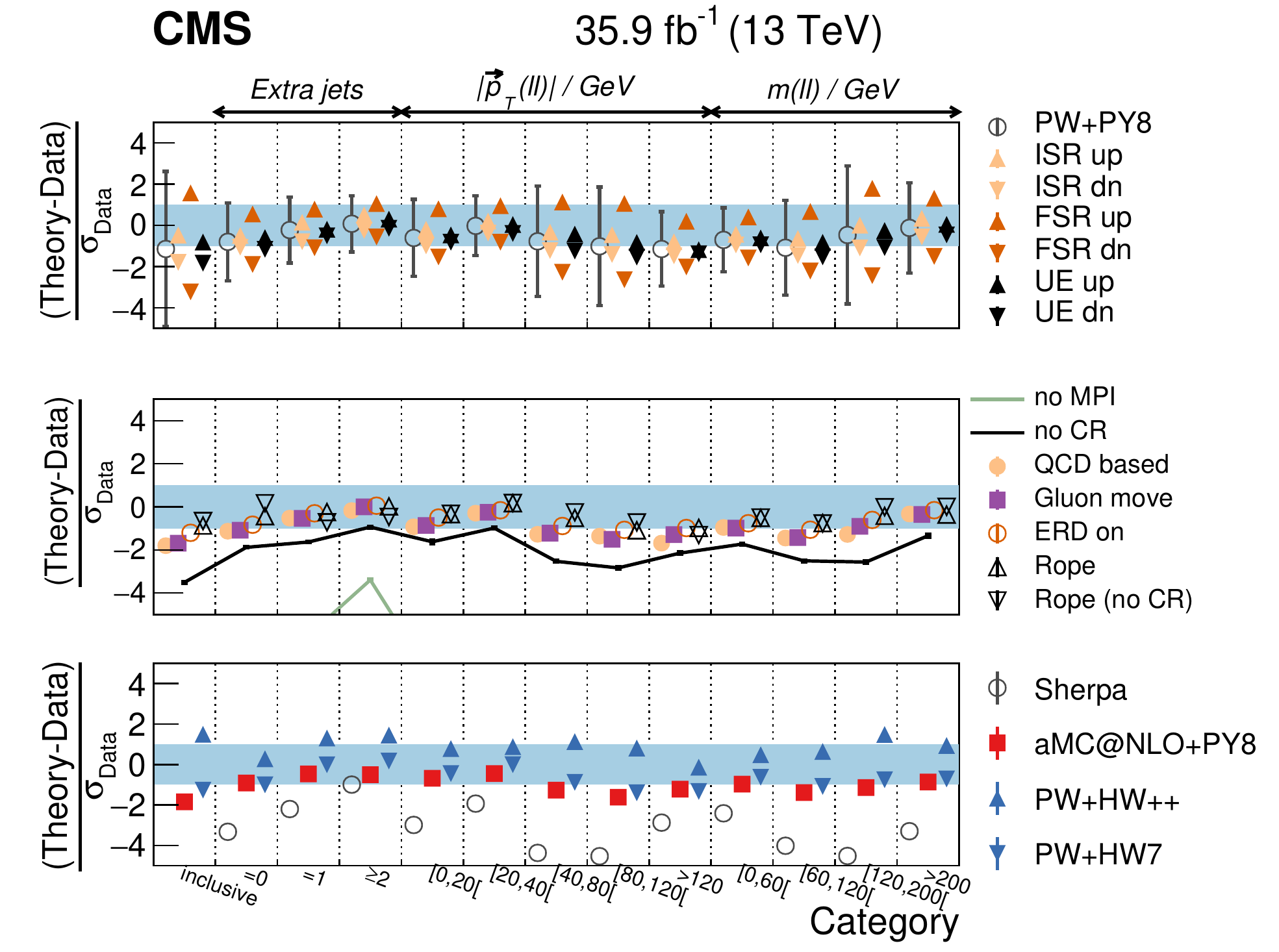}
\caption{
Average aplanarity in different categories.
The conventions of Fig.~\ref{fig:ueprofile_chmult} are used.
}
\label{fig:ueprofile_aplanarity}
\end{figure*}

\begin{figure*}[!htp]
\centering
\includegraphics[width=\cmsFigWidthiv]{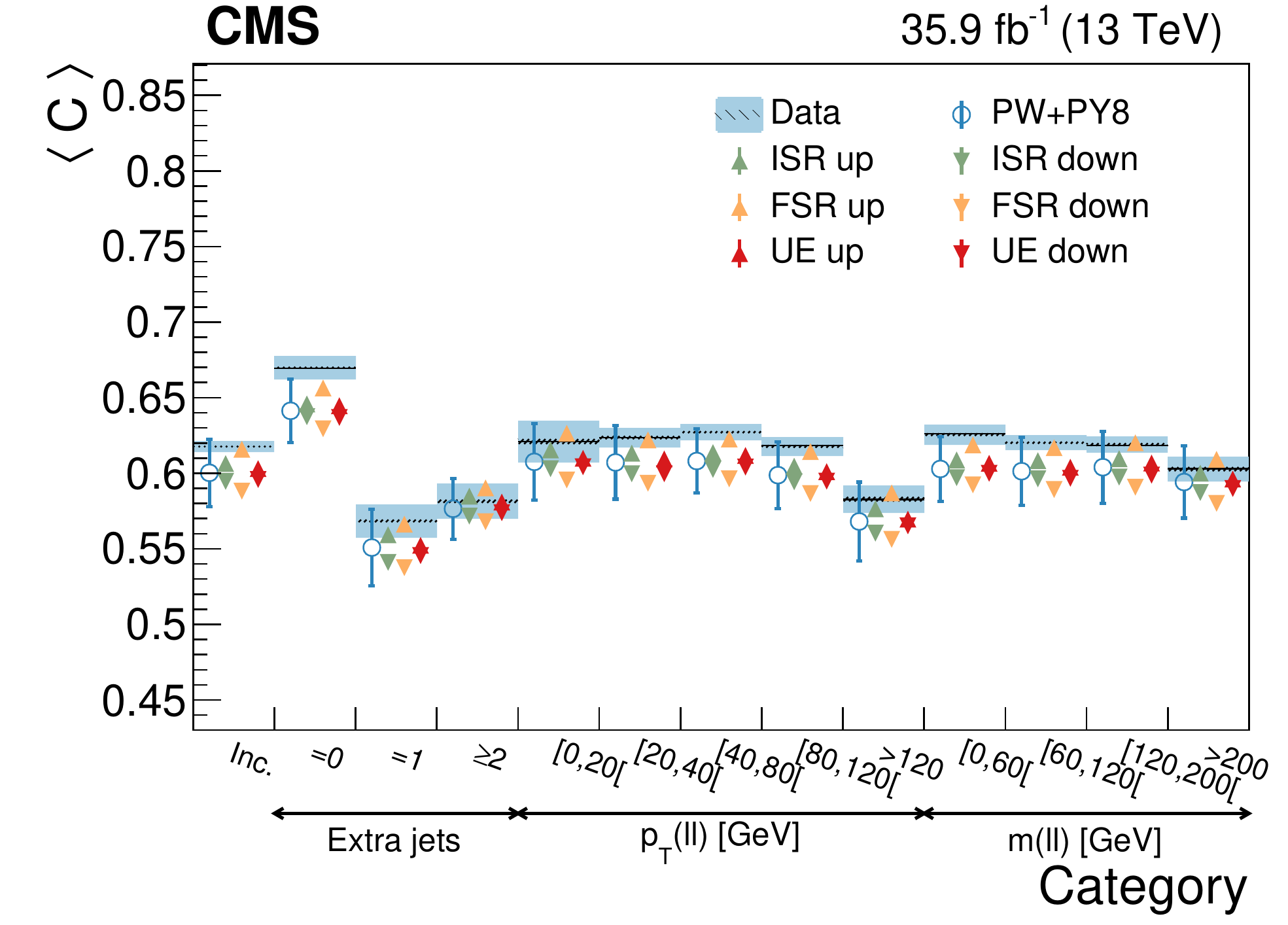}\\
\includegraphics[width=\cmsFigWidthi]{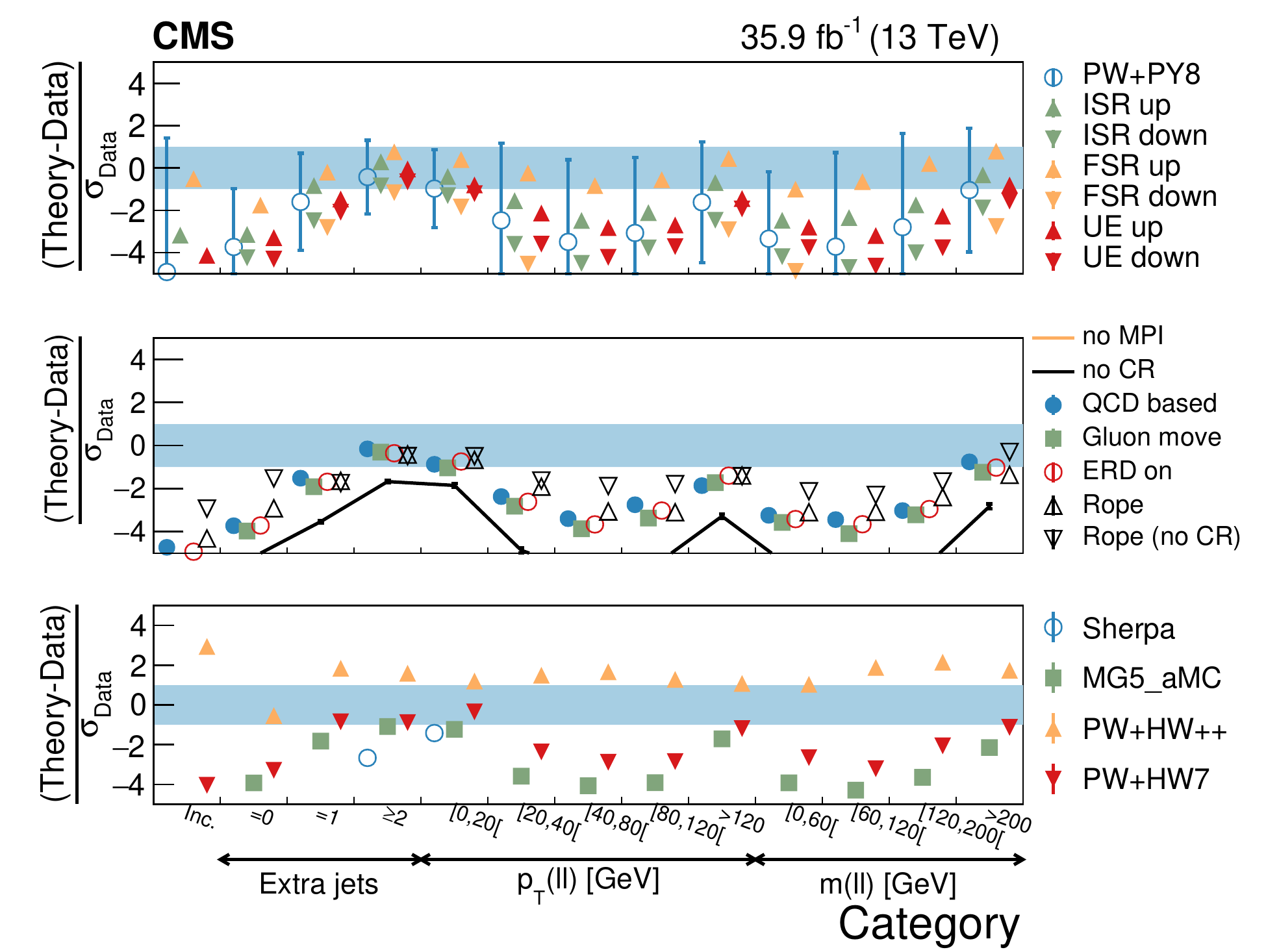}
\caption{
Average $C$ in different categories.
The conventions of Fig.~\ref{fig:ueprofile_chmult} are used.
}
\label{fig:ueprofile_C}
\end{figure*}

\begin{figure*}[!htp]
\centering
\includegraphics[width=\cmsFigWidthiv]{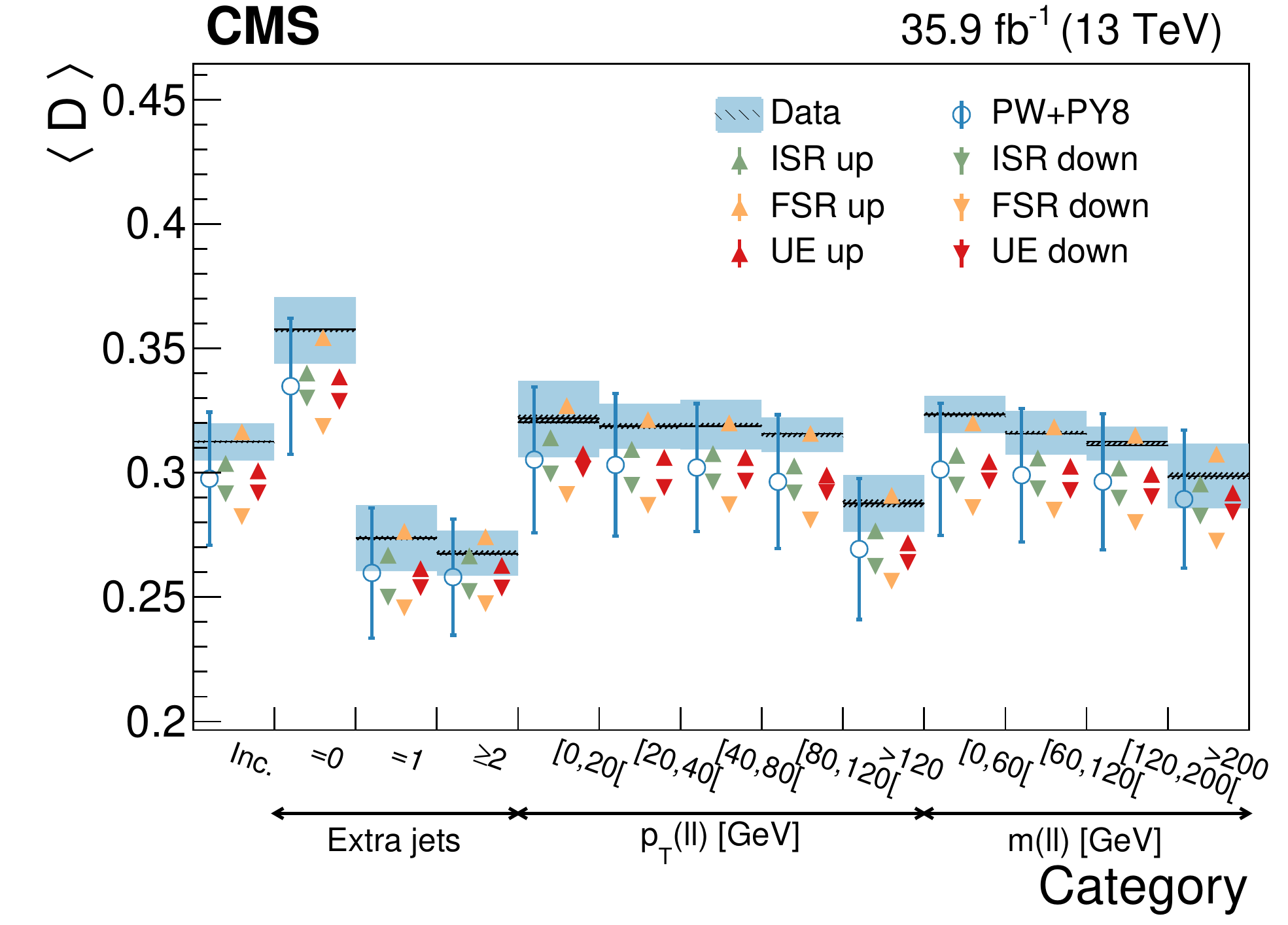}\\
\includegraphics[width=\cmsFigWidthi]{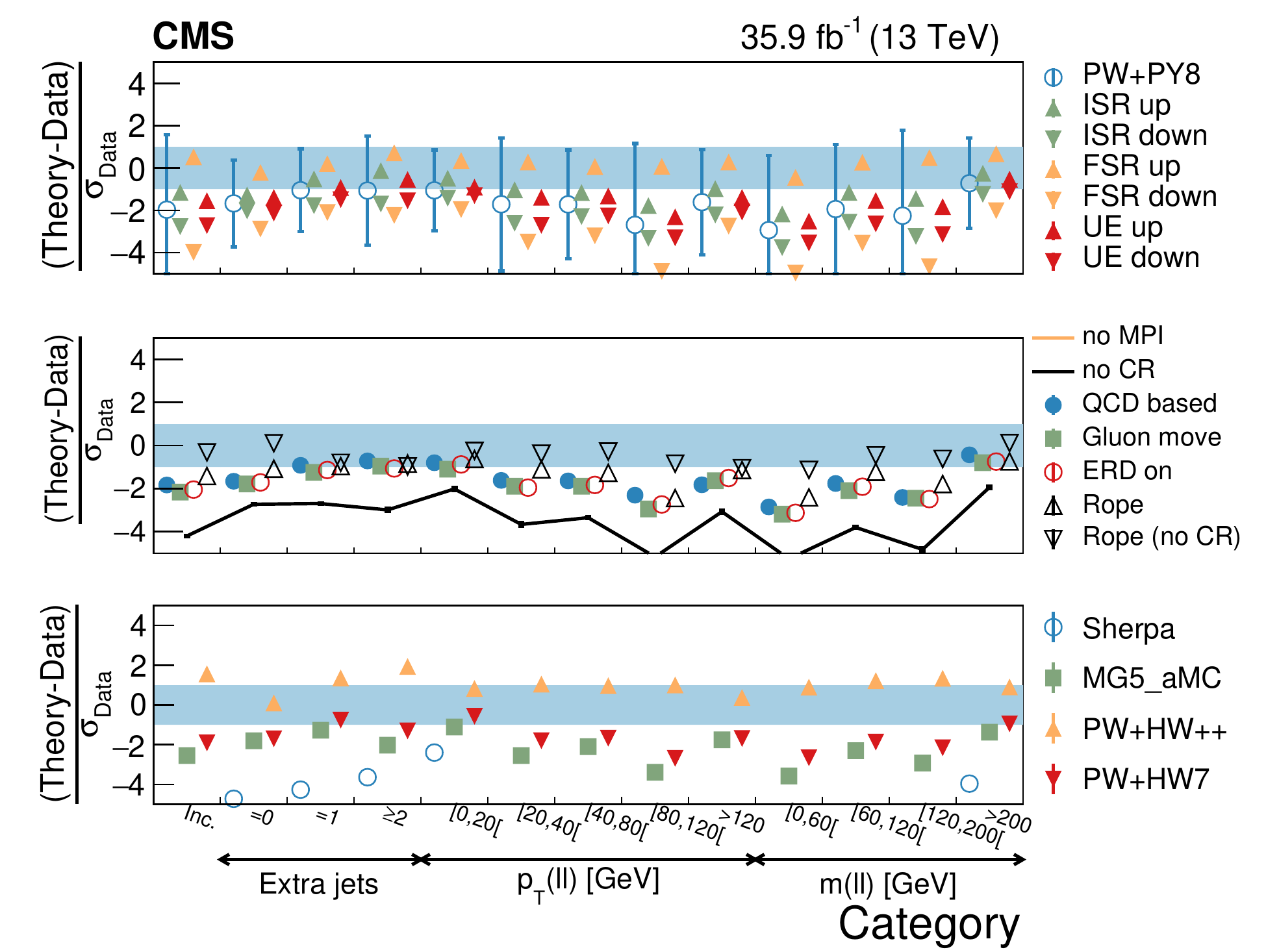}
\caption{
Average $D$ in different categories.
The conventions of Fig.~\ref{fig:ueprofile_chmult} are used.
}
\label{fig:ueprofile_D}
\end{figure*}

\begin{figure*}[!htp]
\centering
\includegraphics[width=\cmsFigWidthiv]{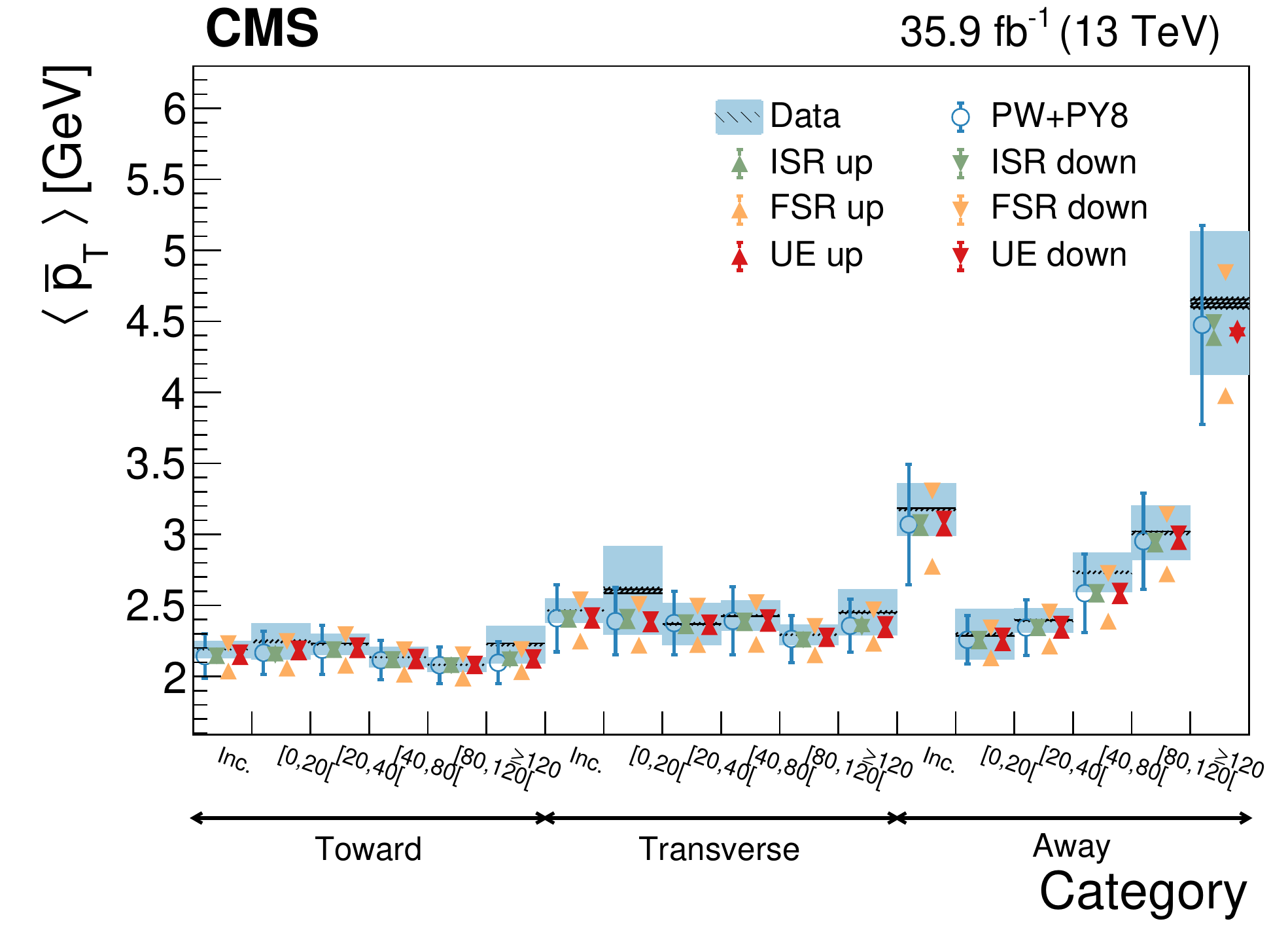}\\
\includegraphics[width=\cmsFigWidthi]{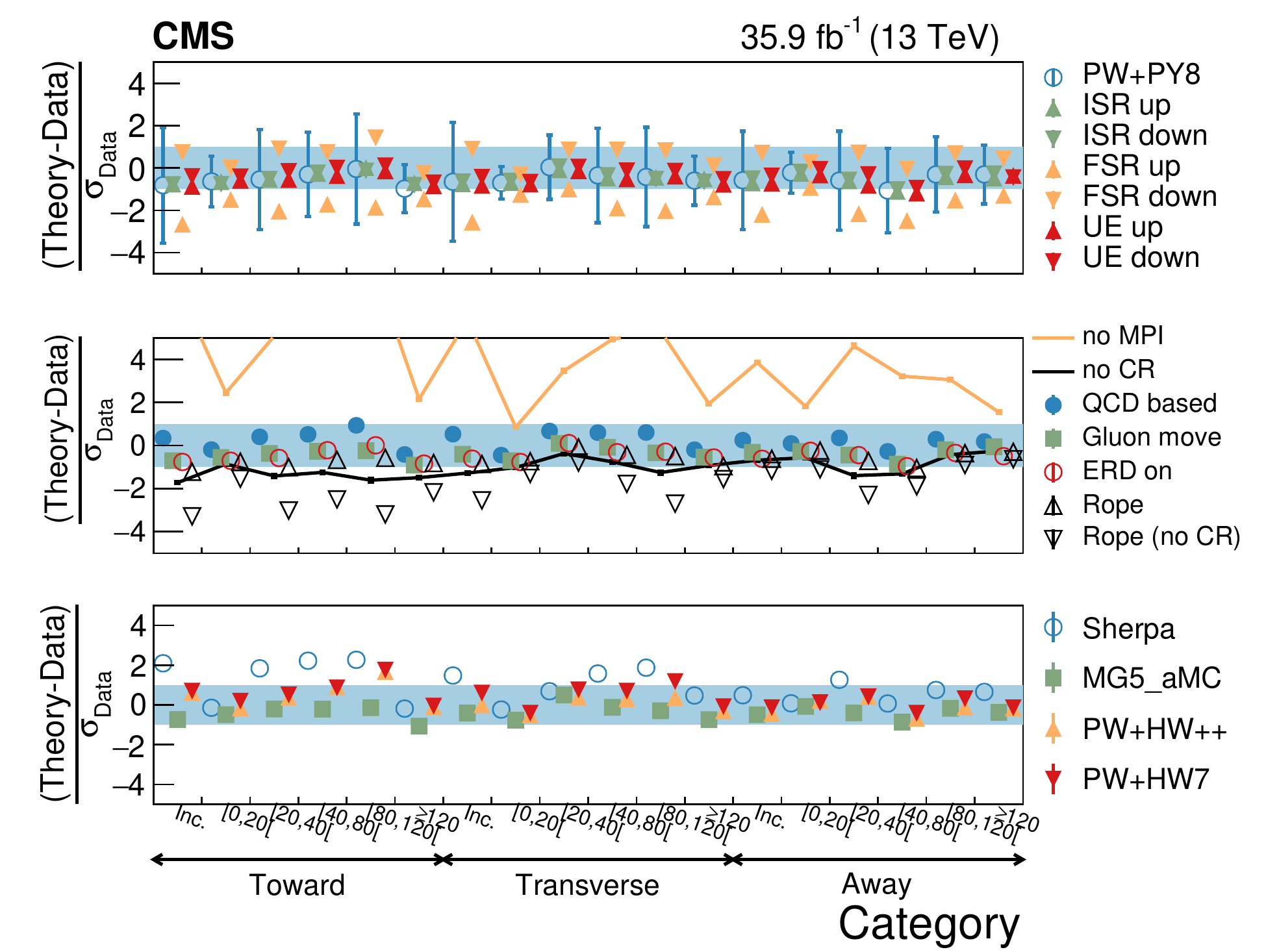}
\caption{
Average {\avgpt} in different \ptll{} categories.
The conventions of Fig.~\ref{fig:ueprofile_chmult} are used.
}
\label{fig:ueprofile_ptll}
\end{figure*}

\begin{figure*}[!htp]
\centering
\includegraphics[width=\cmsFigWidthiv]{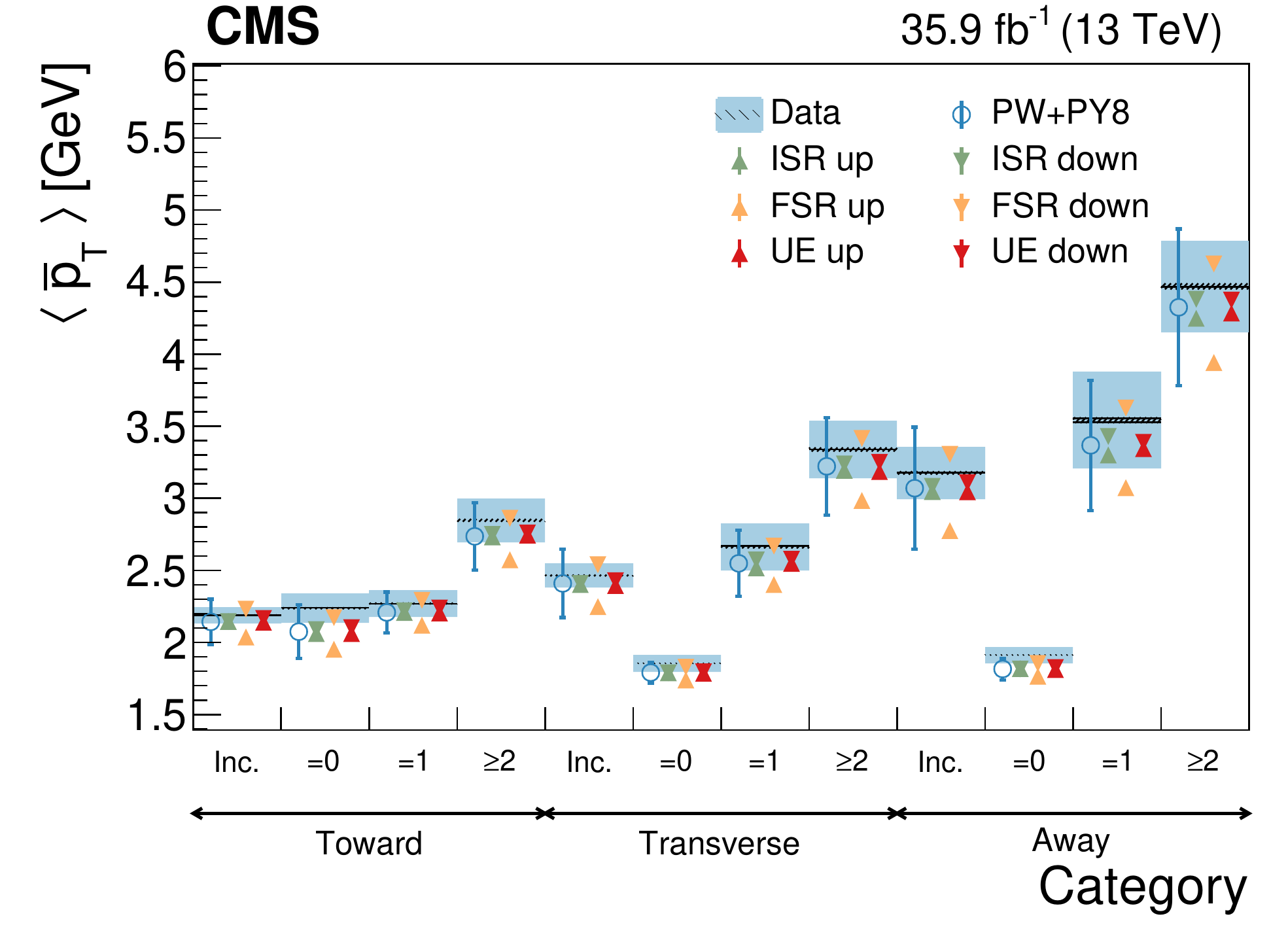}\\
\includegraphics[width=\cmsFigWidthi]{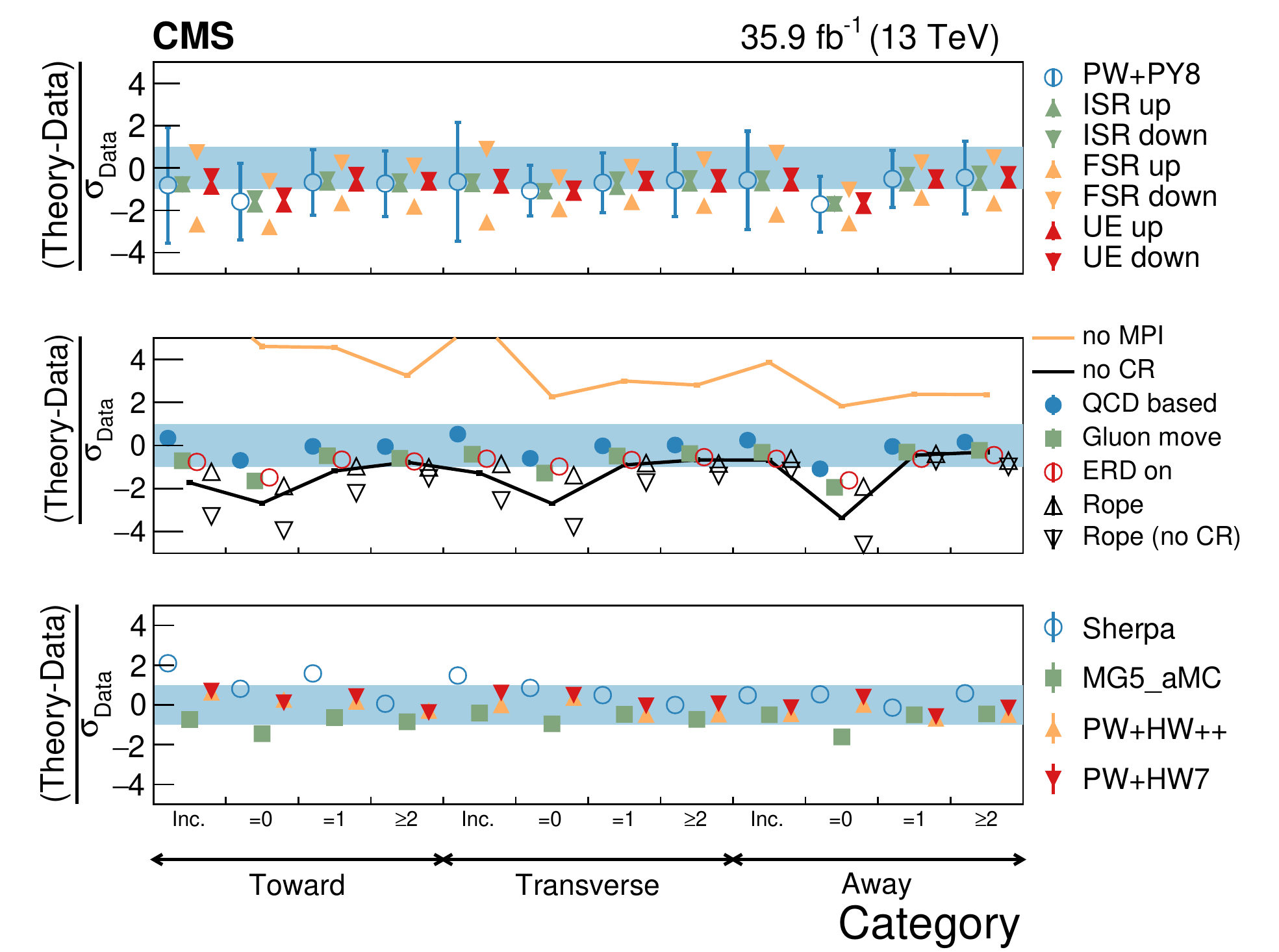}
\caption{
Average {\avgpt} in different jet multiplicity categories.
The conventions of Fig.~\ref{fig:ueprofile_chmult} are used.
}
\label{fig:ueprofile_ptllnj}
\end{figure*}

The differential cross sections as functions of different observables
are measured in different event categories
introduced in Section~\ref{sec:uechar}.
We report the profile, \ie, the average of the measured
differential cross sections in different event categories, and compare it to the expectations
from the different simulation setups.
Figures~\ref{fig:ueprofile_chmult}--\ref{fig:ueprofile_D} summarize the results obtained.
Additional results for \avgpt{}, profiled in different categories of \ptll{} and/or jet multiplicity, are shown in Figs.~\ref{fig:ueprofile_ptll} and \ref{fig:ueprofile_ptllnj}, respectively.
In all figures, the pull of the simulation distributions with respect to data,
defined as the difference between the model and the data divided by the total uncertainty,
is used to quantify the level of agreement.

The average charged-particle multiplicity and
the average of the momentum flux observables vary significantly
when extra jets are found in the event or for higher \ptll{} values.
The same set of variables varies very slowly as a function of \mll{}.
Event shape variables are mostly affected by the presence of extra jets in the event,
while varying slowly as a function of \ptll{} or \mll.
The average sphericity increases significantly when no extra jets are present in the event
showing that the UE is slightly more isotropic in these events.
A noticeable change is also observed for the other event shape variables in the same categories.

For all observables, the MPI contribution is crucial:
most of the pulls are observed to be larger than 5 when MPI is switched off in the simulation.
Color reconnection effects are on the other hand more subtle and are more relevant for \avgpt{},
specifically when no additional jet is present in the event.
This is illustrated by the fact that the pulls of the setup without CR
are larger for events belonging to these categories.
Event shape variables also show sensitivity to CR.
All other variations of the UE and CR models tested
yield smaller variations of the pulls,
compared to the ones discussed.

Although a high pull value of the {\pwpy} simulation is obtained for several categories,
when different theory variations are taken into account, the envelope encompasses the data.
The variations of $\alpS^\text{FSR}(M_\cPZ)$ and $\alpS^\text{ISR}(M_\cPZ)$ account for the largest contribution to this envelope.
As already noted in the previous section,
the {\pwhwpp}, {\pwhw}, and \SHERPA models tend to be in worse agreement with data than {\pwpy},
indicating that further tuning of the first two is needed.

\subsection{Sensitivity to the choice of \texorpdfstring{$\alpS$}{alphaS} in the parton shower}
\label{subsec:alphasSens}

\begin{figure}[!htp]
\centering
\includegraphics[width=\cmsFigWidthMed]{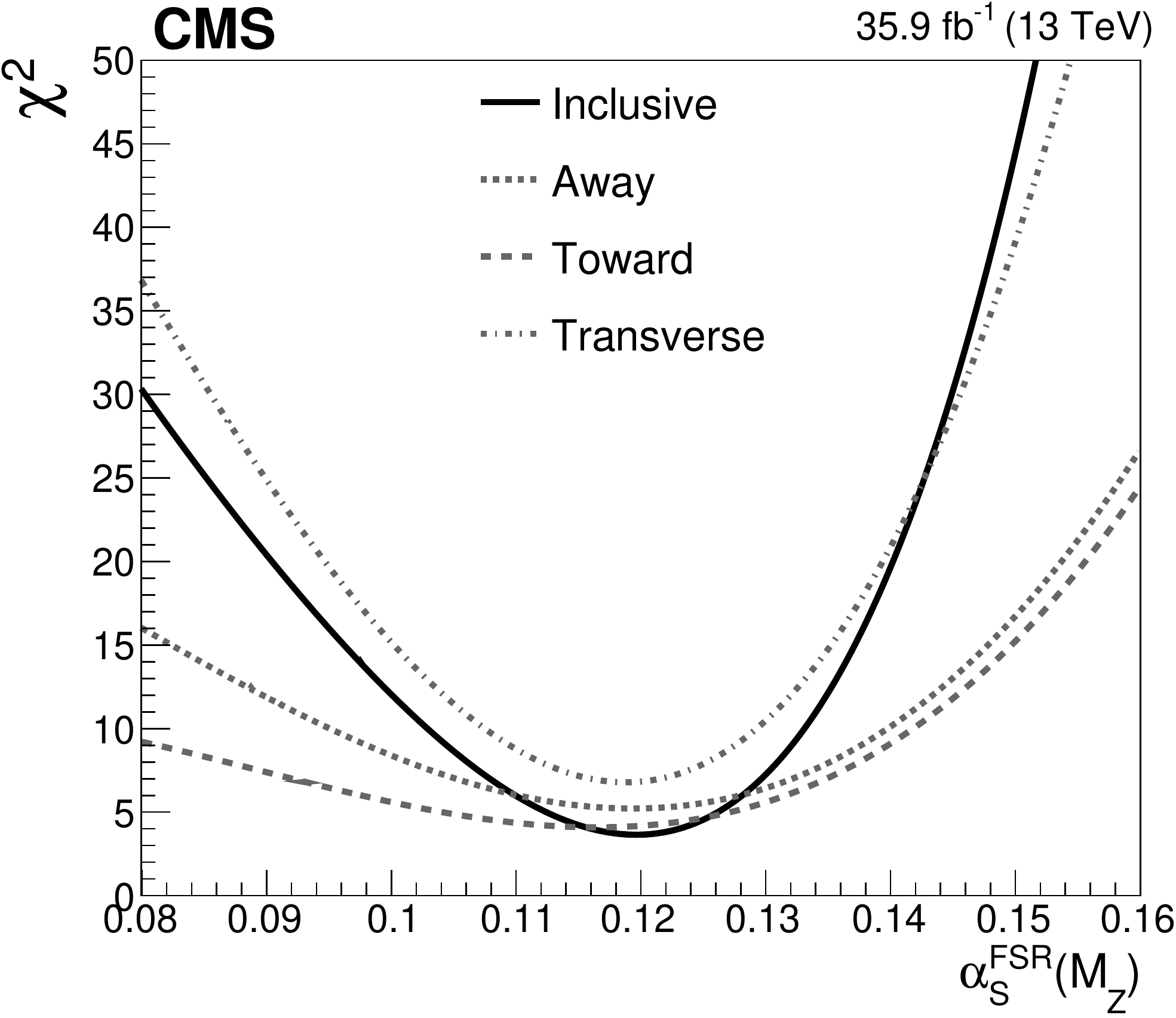}
\caption{
Scan of the $\chi^2$ as a function of the value of $\alpS^\text{FSR}(M_\cPZ)$
employed in the {\pwpy} simulation, when the inclusive {\avgpt} or the {\avgpt} distribution
measured in different regions is used.
The curves result from a fourth-order polynomial interpolation between the simulated $\alpS^\text{FSR}(M_\cPZ)$ points.
}
\label{fig:chi2scan}
\end{figure}

The sensitivity of these results to the choice of $\alpS(M_\cPZ)$ in the parton shower is tested
by performing a scan of the $\chi^2$ value defined by Eq. (\ref{eq:chi2}),
as a function of $\alpS^\text{ISR}(M_\cPZ)$ or $\alpS^\text{FSR}(M_\cPZ)$.
The $\chi^2$ is scanned fixing all the other parameters of the generator.
A more complete treatment could only be achieved with a fully tuned UE,
which lies beyond the scope of this paper.
While no sensitivity is found to $\alpS^\text{ISR}(M_\cPZ)$,
most observables are influenced by the choice of $\alpS^\text{FSR}(M_\cPZ)$.
The most sensitive variable is found to be {\avgpt}
and the corresponding variation of the $\chi^2$ function is reported in Fig.~\ref{fig:chi2scan}.
A polynomial interpolation
is used to determine the minimum of the scan (best fit),
and the points at which the $\chi^2$ function increases by one unit
are used to derive the 68\% confidence interval (CI).
The degree of the polynomial is selected by a stepwise regression based on an F-test statistics~\cite{James:2006zz}.
A value of $\alpS^\text{FSR}(M_\cPZ)=0.120\pm0.006$
is obtained, which is lower than the one assumed in the Monash tune~\cite{Skands:2014pea} and used in the CUETP8M2T4 tune.
The value obtained is compatible with the one obtained from the differential cross sections measured
as a function of {\avgpt} in  different \ptll{} regions
or in events with different additional jet multiplicities.
Table~\ref{tab:chi2_asfsr} summarizes the results obtained.
From the inclusive results, we conclude that the range of the energy scale that corresponds to the
5\% uncertainty attained in the determination
of $\alpS^\text{FSR}(M_\cPZ)$ can be approximated by a $[\sqrt{2},1/\sqrt{2}]$ variation,
improving considerably over the canonical $[2,0.5]$ scale variations.

\begin{table*}[ht]
\centering
\topcaption{
The first rows give the best fit values for $\alpS^\text{FSR}$ for the {\pwpy} setup,
obtained from the inclusive distribution of different observables
and the corresponding 68 and 95\% confidence intervals.
The last two rows give the preferred value of the renormalization scale in units of $M_\cPZ$,
and the associated $\pm 1 \sigma$ interval that
can be used as an estimate of its variation
to encompass the differences between data and the {\pwpy} setup.
}
\cmsTable{
\begin{tabular}{lcccc}
\ptll{} region             & Inclusive        & Away                 & Toward               & Transverse \\
\hline
Best fit $\alpS^\text{FSR}(M_\cPZ)$ & 0.120 & 0.119             & 0.116                & 0.119 \\
68\% CI                 & [-0.006,+0.006]  & [-0.011,+0.010]      & [-0.013,+0.011]      & [-0.006,+0.006] \\
95\% CI                 & [-0.013,+0.011]  & [-0.022,+0.019]      & [-0.030,+0.021]      & [-0.013,+0.012]  \\[\cmsTabSkip]
$\mu_\mathrm{R}/M_\cPZ$    & 2.3              & 2.4                  & 2.9                  & 2.4 \\
68\% CI                 & [1.7,3.3]        & [1.4,4.9]            & [1.6,7.4]            & [1.7,3.5] \\
\end{tabular}
}
\label{tab:chi2_asfsr}
\end{table*}
\clearpage
\section{Summary}
\label{sec:summary}
The first measurement of the underlying event (UE) activity in \ttbar dilepton events produced in hadron colliders has been reported. The measurement makes use of $\sqrt{s}=13\TeV$ proton-proton collision data collected by the CMS experiment in 2016, and corresponding to 35.9\fbinv.
Using particle-flow reconstruction,
the contribution from the UE has been isolated by
removing charged particles associated with the decay products of the \ttbar event candidates
as well as with pileup interactions from the set of reconstructed charged particles per event.
The measurements performed are expected to be valid for other \ttbar final states, and can be used as a reference for complementary studies, \eg,
of how different color reconnection (CR) models
compare to data in the description of the jets
from $\PW\to \PQq\PAQq'$ decays.
The chosen observables and categories enhance the sensitivity to the
modeling of multiparton interactions (MPI),
CR and the choice of
strong coupling parameter at the mass of {\PZ} boson ($\alpS^\text{FSR}(M_\cPZ)$)
in the {\PYTHIA}8 parton shower Monte Carlo simulation.
These parameters have significant impact on the modeling of \ttbar production at the LHC.
In particular, the compatibility of the data with different
choices of the $\alpS^\text{FSR}(M_\cPZ)$ parameter in {\PYTHIA}8 has been quantified,
resulting in a lower value than the one considered in  Ref.~\cite{Skands:2014pea}.

{\tolerance=8000
The majority of the distributions analyzed indicate a fair agreement between the data and
the \POWHEG{}+{\PYTHIA}8 setup with the CUETP8M2T4 tune~\cite{CMS-PAS-TOP-16-021},
but disfavor the setups in which MPI and CR are switched off, or in which $\alpS^\text{FSR}(M_\cPZ)$ is increased.
The data also disfavor the default configurations in \POWHEG{}+\HERWIGpp, \POWHEG{}+{\HERWIG}7, and \SHERPA.
It has been furthermore verified that, as expected,
the choice of the next-to-leading-order matrix-element generator does not impact significantly
the expected characteristics of the UE by comparing predictions from \POWHEG and \MGvATNLO, both interfaced with {\PYTHIA}8.
\par}

The present results test the hypothesis of universality in UE
at an energy scale typically higher than the ones at which models have been studied.
The UE model is tested up to a scale of two times the top quark mass, and the measurements in categories of
dilepton invariant mass indicate that it should be valid at even higher scales.
In addition, they can be used to improve the assessment of systematic uncertainties in future top quark analyses.
The results obtained in this study show that a value of $\alpS^\text{FSR}(M_\cPZ)=0.120\pm0.006$
is consistent with the data. The corresponding uncertainties
translate to a variation of the renormalization scale by a factor of $\sqrt{2}$.

\begin{acknowledgments}
\hyphenation{Bundes-ministerium Forschungs-gemeinschaft Forschungs-zentren} We congratulate our colleagues in the CERN accelerator departments for the excellent performance of the LHC and thank the technical and administrative staffs at CERN and at other CMS institutes for their contributions to the success of the CMS effort. In addition, we gratefully acknowledge the computing centers and personnel of the Worldwide LHC Computing Grid for delivering so effectively the computing infrastructure essential to our analyses. Finally, we acknowledge the enduring support for the construction and operation of the LHC and the CMS detector provided by the following funding agencies: the Austrian Federal Ministry of Science, Research and Economy and the Austrian Science Fund; the Belgian Fonds de la Recherche Scientifique, and Fonds voor Wetenschappelijk Onderzoek; the Brazilian Funding Agencies (CNPq, CAPES, FAPERJ, and FAPESP); the Bulgarian Ministry of Education and Science; CERN; the Chinese Academy of Sciences, Ministry of Science and Technology, and National Natural Science Foundation of China; the Colombian Funding Agency (COLCIENCIAS); the Croatian Ministry of Science, Education and Sport, and the Croatian Science Foundation; the Research Promotion Foundation, Cyprus; the Ministry of Education and Research, Estonian Research Council via IUT23-4 and IUT23-6 and European Regional Development Fund, Estonia; the Academy of Finland, Finnish Ministry of Education and Culture, and Helsinki Institute of Physics; the Institut National de Physique Nucl\'eaire et de Physique des Particules~/~CNRS, and Commissariat \`a l'\'Energie Atomique et aux \'Energies Alternatives~/~CEA, France; the Bundesministerium f\"ur Bildung und Forschung, Deutsche Forschungsgemeinschaft, and Helmholtz-Gemeinschaft Deutscher Forschungszentren, Germany; the General Secretariat for Research and Technology, Greece; the National Scientific Research Foundation, and National Innovation Office, Hungary; the Department of Atomic Energy and the Department of Science and Technology, India; the Institute for Studies in Theoretical Physics and Mathematics, Iran; the Science Foundation, Ireland; the Istituto Nazionale di Fisica Nucleare, Italy; the Ministry of Science, ICT and Future Planning, and National Research Foundation (NRF), Republic of Korea; the Lithuanian Academy of Sciences; the Ministry of Education, and University of Malaya (Malaysia); the Mexican Funding Agencies (CINVESTAV, CONACYT, SEP, and UASLP-FAI); the Ministry of Business, Innovation and Employment, New Zealand; the Pakistan Atomic Energy Commission; the Ministry of Science and Higher Education and the National Science Centre, Poland; the Funda\c{c}\~ao para a Ci\^encia e a Tecnologia, Portugal; JINR, Dubna; the Ministry of Education and Science of the Russian Federation, the Federal Agency of Atomic Energy of the Russian Federation, Russian Academy of Sciences, and the Russian Foundation for Basic Research; the Ministry of Education, Science and Technological Development of Serbia; the Secretar\'{\i}a de Estado de Investigaci\'on, Desarrollo e Innovaci\'on and Programa Consolider-Ingenio 2010, Spain; the Swiss Funding Agencies (ETH Board, ETH Zurich, PSI, SNF, UniZH, Canton Zurich, and SER); the Ministry of Science and Technology, Taipei; the Thailand Center of Excellence in Physics, the Institute for the Promotion of Teaching Science and Technology of Thailand, Special Task Force for Activating Research and the National Science and Technology Development Agency of Thailand; the Scientific and Technical Research Council of Turkey, and Turkish Atomic Energy Authority; the National Academy of Sciences of Ukraine, and State Fund for Fundamental Researches, Ukraine; the Science and Technology Facilities Council, UK; the US Department of Energy, and the US National Science Foundation.

Individuals have received support from the Marie-Curie program and the European Research Council and EPLANET (European Union); the Leventis Foundation; the A. P. Sloan Foundation; the Alexander von Humboldt Foundation; the Belgian Federal Science Policy Office; the Fonds pour la Formation \`a la Recherche dans l'Industrie et dans l'Agriculture (FRIA-Belgium); the Agentschap voor Innovatie door Wetenschap en Technologie (IWT-Belgium); the Ministry of Education, Youth and Sports (MEYS) of the Czech Republic; the Council of Science and Industrial Research, India; the HOMING PLUS program of Foundation for Polish Science, cofinanced from European Union, Regional Development Fund; the Compagnia di San Paolo (Torino); the Consorzio per la Fisica (Trieste); MIUR project 20108T4XTM (Italy); the Thalis and Aristeia programs cofinanced by EU-ESF and the Greek NSRF; and the National Priorities Research Program by Qatar National Research Fund.
\end{acknowledgments}
\bibliography{auto_generated}

\appendix

\section{Appendix: Variations of the \POWHEG{}+{\PYTHIA}8 setup}
\label{tableofvariations}
\begin{table*}[!htp]
\centering
\topcaption{
Variations of the {\pwpy} setup used for the comparison with the measurements.
The values changed with respect to the CUETP8M2T4 tune are given in the columns corresponding to each model.
Further details on parameters or specificities of the models can be found in Ref.~\cite{Sjostrand:2014zea,Skands:2014pea,Sjostrand:2013cya,Christiansen:2015yqa,Christiansen:2015yqa,Argyropoulos:2014zoa,Bierlich:2014xba,Bierlich:2015rha}.
For the Rope hadronization model two variations are considered: one with no CR and the other with the default CR model.
The settings for the former are denoted in parenthesis in the last column.
}
\resizebox{\textwidth}{!}{
\begin{tabular}{ l|c|c|c|c|c|c|c|c|c|c }
\multicolumn{1}{c}{\multirow{5}{*}{Parameter}} & \multicolumn{10}{c}{{\pwpy} simulation setups}\\\cline{2-11}
                           & \multirow{4}{*}{CUETP8M2T4} & \multicolumn{2}{c|}{Extreme}    & \multicolumn{7}{c}{Fine grain variations}\\\cline{5-11}
                           &                      & \multicolumn{2}{c|}{variations} & MPI/CR          & \multicolumn{2}{c|}{Parton shower scale} & \multicolumn{4}{c}{CR including \ttbar} \\\cline{3-11}
                           & & no  & no & UE          & ISR         & FSR         & ERD & QCD & Gluon  & Rope (no CR)\\
                           & & MPI & CR & \small up/down & \small up/down & \small up/down & on  & based~\cite{Christiansen:2015yqa} & move~\cite{Argyropoulos:2014zoa} & \cite{Bierlich:2014xba,Bierlich:2015rha} \\
\hline
PartonLevel                &        &        &       &                 &                &                &        &           &      &\\
~~~~~MPI                   & on     & off    &       &                 &                &                &        &           &      &\\\hline
SpaceShower                &        &        &       &                 &                &                &        &           &      &\\
~~~~~renormMultFac         & 1.0    &        &       &                 & 4/0.25         &                &        &           &      &\\
~~~~~alphaSvalue           & 0.1108 &        &       &                 &                &                &        &           &      &\\\hline
TimeShower                 &        &        &       &                 &                &                &        &           &      &\\
~~~~~renormMultFac         & 1.0    &        &       &                 &                & 4/0.25         &        &           &      &\\
~~~~~alphaSvalue           & 0.1365 &        &       &                 &                &                &        &           &      &\\\hline
MultipartonInteractions    &        &        &       &                 &                &                &        &           &      &\\
~~~~~pT0Ref                & 2.2    &        &       & 2.20/2.128      &                &                &        & 2.174     & 2.3  &\\
~~~~~ecmPow                & 0.2521 &        &       &                 &                &                &        & 0.2521    &      &\\
~~~~~expPow                & 1.6    &        &       & 1.711/1.562     &                &                &        & 1.312     & 1.35 &\\\hline
ColorReconnection          &        &        &       &                 &                &                &        &           &      &\\
~~~~~reconnect             & on     &        & off   &                 &                &                &        &           &      & (off) \\
~~~~~range                 & 6.59   &        &       & 6.5/8.7         &                &                &        &           &      &\\
~~~~~mode                  & 0      &        &       &                 &                &                &        & 1         & 2    &\\
~~~~~junctionCorrection    &        &        &       &                 &                &                &        & 0.1222    &      &\\
~~~~~timeDilationPar       &        &        &       &                 &                &                &        & 15.86     &      &\\
~~~~~m0                    &        &        &       &                 &                &                &        & 1.204     &      &\\
~~~~~flipMode              &        &        &       &                 &                &                &        &           & 0    &\\
~~~~~m2Lambda              &        &        &       &                 &                &                &        &           & 1.89 &\\
~~~~~fracGluon             &        &        &       &                 &                &                &        &           & 1    &\\
~~~~~dLambdaCut            &        &        &       &                 &                &                &        &           & 0    &\\\hline
PartonVertex               &        &        &       &                 &                &                &        &           &      &\\
~~~~~setVertex             &        &        &       &                 &                &                &        &           &      & on \\\hline
Ropewalk                   &        &        &       &                 &                &                &        &           &      &\\
~~~~~RopeHadronization     &        &        &       &                 &                &                &        &           &      & on \\
~~~~~doShoving             &        &        &       &                 &                &                &        &           &      & on \\
~~~~~doFlavour             &        &        &       &                 &                &                &        &           &      & on \\\hline
PartonLevel                &        &        &       &                 &                &                &        &           &      & \\
~~~~~earlyResDec           & off    &        &       &                 &                &                & on     & on        & on   & on \\
\end{tabular}
}
\label{tab:mcsetups_detailed}
\end{table*}
\cleardoublepage \section{The CMS Collaboration \label{app:collab}}\begin{sloppypar}\hyphenpenalty=5000\widowpenalty=500\clubpenalty=5000\vskip\cmsinstskip
\textbf{Yerevan Physics Institute, Yerevan, Armenia}\\*[0pt]
A.M.~Sirunyan, A.~Tumasyan
\vskip\cmsinstskip
\textbf{Institut f\"{u}r Hochenergiephysik, Wien, Austria}\\*[0pt]
W.~Adam, F.~Ambrogi, E.~Asilar, T.~Bergauer, J.~Brandstetter, E.~Brondolin, M.~Dragicevic, J.~Er\"{o}, A.~Escalante~Del~Valle, M.~Flechl, R.~Fr\"{u}hwirth\cmsAuthorMark{1}, V.M.~Ghete, J.~Hrubec, M.~Jeitler\cmsAuthorMark{1}, N.~Krammer, I.~Kr\"{a}tschmer, D.~Liko, T.~Madlener, I.~Mikulec, N.~Rad, H.~Rohringer, J.~Schieck\cmsAuthorMark{1}, R.~Sch\"{o}fbeck, M.~Spanring, D.~Spitzbart, A.~Taurok, W.~Waltenberger, J.~Wittmann, C.-E.~Wulz\cmsAuthorMark{1}, M.~Zarucki
% ...
\vskip\cmsinstskip
% ...
1:  Also at Vienna University of Technology, Vienna, Austria\\
% ...
\end{sloppypar}
\end{document}